\documentclass[12pt,aps,prd,longbibliography]{article}
\usepackage{graphicx,tabularx,epsfig}
\usepackage{color}
\usepackage{enumitem}
\usepackage{caption}
\usepackage[labelformat = simple]{subcaption}
\usepackage{amsmath}
\usepackage{amssymb}
\usepackage{amsthm}
\usepackage{physics}
\usepackage{dsfont, mathtools}
\usepackage{setspace}
\usepackage{psvectorian}
\usepackage{blkarray}
\usepackage{bm}
\usepackage{url}
\usepackage{setspace}

\usepackage{xcolor}
\colorlet{linkequation}{blue}

\usepackage[colorlinks = true,
            linkcolor = blue,
            urlcolor  = blue,
            citecolor = blue,
            anchorcolor = blue]{hyperref}

\usepackage{floatrow}

\usepackage{amsthm}

\usepackage{amssymb}

\newlength{\abstractwidth}
\renewcommand{\thefootnote}{\fnsymbol{footnote}}
\renewcommand{\thanks}[1]{\footnote{#1}} 
\newcommand{\starttext}{
\setcounter{footnote}{0}
\renewcommand{\thefootnote}{\arabic{footnote}}}

\makeatletter
\g@addto@macro\normalsize{%
  \setlength\abovedisplayskip{10pt}
  \setlength\belowdisplayskip{20pt}
  \setlength\abovedisplayshortskip{10pt}
  \setlength\belowdisplayshortskip{20pt}
}
\makeatother

\usepackage{color}

\renewcommand{\title}[1]{\vbox{\center\LARGE{#1}}\vspace{5mm}}
\renewcommand{\author}[1]{\vbox{\center#1}\vspace{5mm}}
\newcommand{\address}[1]{\vbox{\center\em#1}}
\newcommand{\email}[1]{\vbox{\center\tt#1}\vspace{5mm}}

\theoremstyle{definition}
\newtheorem{assumption}{}

\begin{document}

\singlespacing

\begin{titlepage}
\rightline{}
\bigskip
\bigskip\bigskip\bigskip\bigskip
\bigskip
\centerline{\Large \bf {Complexity and Boost Symmetry}}

\bigskip \noindent

\bigskip
\begin{center}


\author{Ying Zhao}

\address{Stanford Institute for Theoretical Physics and Department of Physics, \\
Stanford University, Stanford, CA 94305-4060, USA }

\email{zhaoying@stanford.edu}

\bigskip

\end{center}

\begin{abstract}

We find that the time dependence of holographic complexity is controlled by the Rindler boost symmetry across the horizon. By studying the collision energy experienced by an infalling object, we see the breaking of this boost symmetry is closely related to firewalls, which in turn shows the connection between the time dependence of complexity and firewalls.
We further identify the black and white hole interiors as two tapes storing different parts of the minimal circuit preparing the state. Depending on whether the quantum gates are being laid on the tape at a particular moment, each tape can be in two states: working, or locked. We interpret the existence of firewalls as the locking of tapes.

\medskip
\noindent
\end{abstract}

\end{titlepage}

\starttext \baselineskip=17.63pt \setcounter{footnote}{0}

{\hypersetup{hidelinks}
\tableofcontents
}

\setcounter{equation}{0}
\section{Introduction}\label{Introduction}

The growth of the black hole interior is a mystery. From the field theory point of view, thermal equilibrium is quickly achieved; but from the bulk geometry, one can see that the interior keeps expanding. It is conjectured that this reflects the increase of the state complexity, i.e. the minimal number of simple gates needed to prepare a state starting from some simple reference state \cite{Susskind:2014rva}.

The picture of a tensor network lying in spacetime \cite{Swingle:2009bg} \cite{Swingle:2012wq} offers an intuitive explanation about the duality between boundary state complexity and bulk spacetime. It also makes explicit the idea of a black hole interior as emergent spacetime \cite{Hartman:2013qma} \cite{Roberts:2014isa} \cite{Brown:2015lvg}. We can view the tensor network as a quantum circuit lying inside the horizon, recording the dynamics that creates the state starting from some simple state. The complexity is one important parameter characterizing this dynamics. In particular, if the state is made by inserting various precursors to the thermofield double, complexity keeps track of the dynamics of the CFT disturbed at various points.

In \cite{Stanford:2014jda} \cite{Roberts:2014isa}, complexity of a holographic state is conjectured to be dual to the volume of maximal surface anchored at boundary. In \cite{Brown:2015bva} \cite{Brown:2015lvg}, it is pointed out that the action in the Wheeler-DeWitt patch offers a better candidate for complexity. Both prescriptions give certain features of time dependence of complexity, like linear increase for sub-exponential time, switchback effect, e.t.c., which match quite well with what one expects from the quantum circuit picture.

\begin{figure}[H]
\begin{center}
\includegraphics[scale=.4]{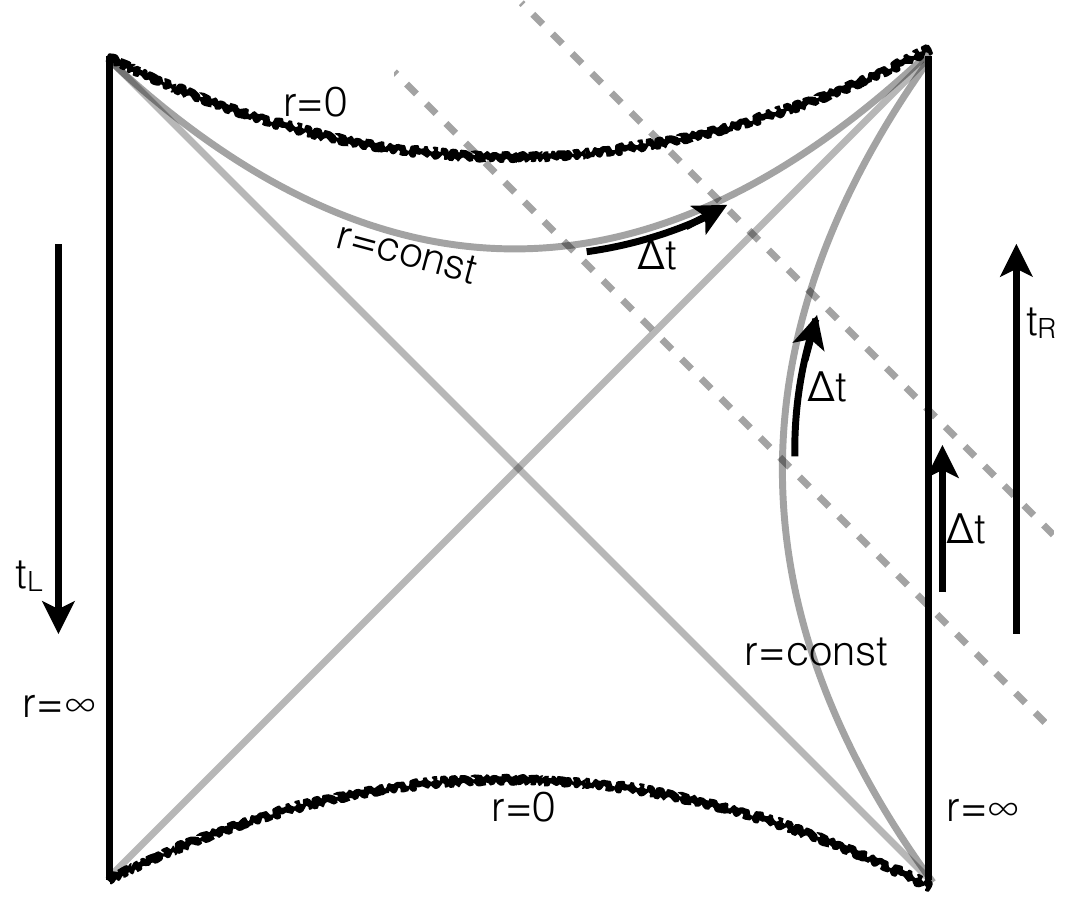}
\caption{Boost symmetry across horizon. Schwarzschild time inside the horizon synchronizes with Schwarzschild time outside.}
\label{symmetryacrosshorizon}
\end{center}
\end{figure}

We'll explore the origin of these various features, see that the time dependence of complexity is basically a consequence of boost symmetry across the horizon as indicated in Figure \ref{symmetryacrosshorizon}. Notice that the Schwarzschild time inside the horizon is synchronized with the Schwarzschild time outside. This boost symmetry is the origin of the linear time dependence of complexity. With the insertion of various precursors, the uniform dynamics is disrupted, and the bulk symmetry is broken. However, it's broken in a particularly simple pattern which results from the Rindler-like nature of near horizon geometry. We'll see that events near the horizon can play a crucial role. If they affect the synchronization, there will be abnormal time dependence of complexity, and vise versa. \footnote{In this thesis, we are talking about sub-exponential time. We don't really know what will happen after the complexity saturates. } 

The quantum circuit description of the black hole interior was given by Hartman and Maldacena \cite{Hartman:2013qma}. The idea of a tape storing the quantum circuit behind the future horizon appeared in \cite{Susskind:2014moa}. We'll see that in order to incorporate the possibility of complexity decrease and the existence of firewalls \cite{Susskind:2015toa}, we need two tapes: future and past tapes, which are stored in the future and past interiors respectively \cite{Zhao:2017iul}. There is a family of examples of black hole, in which the minimal quantum circuit preparing the state is explicitly known and is composed of segments of forward and backward Hamiltonian evolutions \cite{Shenker:2013pqa} \cite{Shenker:2013yza}. The forward Hamiltonian evolution part is stored in the future tape, while the backward evolution part is stored in the past tape. When the quantum gates preparing the state are laid on one particular tape, we say that tape is working, and the corresponding horizon is smooth. When there are no gates laid down on that tape, the entry to the interior is forbidden. From the point of view of an exterior observer, a black hole is just a particle with certain properties \cite{tHooft:1984kcu}. What's special about it is that there is spacetime behind the horizon, which an infalling observer can enter. In that sense, the horizon is like a door to the interior region. One learns that whether the door is open or not depends on the working status of the corresponding tape; one can cross the horizon only when the corresponding tape is working. \\

In this paper we'll explore the time dependence of the complexity using the following assumptions:\footnote{In these four assumptions, assumption 2 essentially implies assumptions 1, 3, and 4. I thank Daniel Harlow for pointing this out. }

\begin{assumption}
\label{assumption1}
In holographic duality, the complexity of a boundary state is dual to some bulk geometric quantity inside the Wheeler-DeWitt patch. 
\end{assumption}

\begin{assumption}
\label{assumption2}
Tensor networks support spacetime. They have locality down to AdS scale, and respect the symmetry of spacetime, in particular, time translation symmetry and rotational symmetry. 
\end{assumption}

\begin{assumption}
\label{assumption3}
Complexity counts the number of gates in tensor network.
\end{assumption}

\begin{assumption}
\label{assumption4}
Contributions to complexity from very small regions are also small. 
\end{assumption}

The most important assumption we use throughout this paper is probably the validity of classical general relativity, at least before very late time. \\

By doing a more detailed matching between the picture of the quantum circuit preparing a state and the dual black hole geometry, we then identify the different roles played by future and past horizons. In the minimal circuit preparing the state, there can be both forward and backward Hamiltonian evolutions. They are stored in two different tapes. The future interior serves as the future tape storing the forward Hamiltonian evolution, while the past interior is the past tape storing backward Hamiltonian evolution. In this sense, the existence of past horizons means complexity can increase towards the past, and is a manifestation of time reversibility. 


Here we should also emphasize the limit of this picture. The examples we studied are special. We know the minimal circuit preparing them, and their minimal circuits are made of Hamiltonian evolutions perturbed at various points. We don't know what will happen for more general states.

For an exterior observer, a black hole is just a particle with certain properties \cite{tHooft:1984kcu}. What's particularly special about it is, there is spacetime behind the horizon, and an infalling observer can enter. In this sense, the horizon is like a door to the interior region. What we learn is, whether the door is open or not depends on the working status of the corresponding tape. You can only enter an interior when the corresponding tape is working. When the tape is not working, the entry is forbidden by firewalls.

This paper is organized as follows: In section \ref{calculation} we give a brief review of some known results about holographic complexity, and then illustrate the idea of symmetry determining complexity by explicit calculations in shockwave geometries. In particular, without a detailed prescription, we'll recover the earlier results about the time dependence of complexity in \cite{Stanford:2014jda} \cite{Roberts:2014isa} \cite{Brown:2015lvg} from the above assumptions. In section \ref{diagnosis} we use this idea to show the connection between the time dependence of complexity and smoothness of horizons \cite{Susskind:2015toa}. We study the connection between the boost symmetry breaking and collision energy in general relativity. In section \ref{circuit} we introduce the two-tape picture, and make a connection between the working status of the tapes and the smoothness of horizon. \\

For the convenience of the reader we'll list some conventions used throughout this chapter.

\begin{itemize}
\item
$D$ refers to the spacetime dimension of the bulk.
\item
Our convention of time will be that time flows upward on the right side of Penrose diagram, and downward on the left side. See Figure \ref{symmetryacrosshorizon}. So a precursor $W_L(t_w)=e^{iH_Lt_w}W_Le^{-iH_Lt_w}$ acting on the left CFT from the remote past would have large positive $t_w$. 
\item
We'll use boost symmetry and time translation symmetry interchangeably. 
\end{itemize}

\section{Time dependence of complexity and boost symmetry}
\label{calculation}

\subsection{Known results about holographic complexity}
\label{knownresults}


Given a set of simple gates and a simple reference state, the complexity of a state is defined as the minimum number of gates needed to prepare the state. For a continuum field theory, we still don't have a rigorous definition of complexity, but we expect there exists a similar quantity which characterizes how hard it is to prepare a state starting from a particular simple reference state (thermofield double for a two-sided black hole). On the bulk side, it was proposed that complexity is dual to the volume of the maximal surface anchored on the boundary, with some prefactor involving the cosmological constant \cite{Stanford:2014jda} \cite{Roberts:2014isa}. Later, it was proposed that complexity equals the action of the Wheeler-DeWitt patch in the bulk \cite{Brown:2015bva} \cite{Brown:2015lvg}, and a universal rate of complexity increase was found. Both proposals were tested in various examples. 

For example, take the time evolution of the thermofield double:
\begin{align}
\label{TFD}
|\psi(t_R,t_L)\rangle=\frac{1}{\sqrt Z}\sum_ne^{-\frac{\beta}{2}E_n}|n\rangle_L|n\rangle_R e^{-iE_n(t_R-t_L)}, 
\end{align}
whose holographic dual is a two-sided black hole in Anti-de Sitter space \cite{Maldacena:2001kr} (Figure \ref{symmetryacrosshorizon}). Since black hole dynamics is chaotic, we don't expect shortcuts (fast forwarding in \cite{Atia:2016sax}) before exponentially long time. The complexity will increase linearly with time after an initial transient period of order the thermal time \cite{Stanford:2014jda} \cite{Brown:2015lvg}.
\begin{align}
\label{linear}
\mathcal{C}(t_R, t_L)=\mathcal{C}(|\text{TFD}\rangle)+C|t_R-t_L|
\end{align}
We can also consider more complicated states by perturbing the system at various times. If we act on the state with a precursor $W_L(t_w)=e^{iH_Lt_w}W_Le^{-iH_Lt_w}$, where $W_L$ is some thermal scale perturbation, it amounts to throwing in a thermal quantum at early time $t_w$. In the dual geometry there is a shockwave lying close to the horizon \cite{Shenker:2013pqa}. 
What's the effect of such a perturbation? For a black hole of size $l_{AdS}$, we expect this quantum takes scrambling time $t_*=\frac{\beta}{2\pi}\log S$ to affect the entire system, during which the perturbation has little effect on complexity. After that the complexity will increase linearly. Bulk calculations indeed give \cite{Stanford:2014jda} \cite{Brown:2015lvg}
\begin{align}
\label{sphericaloneshockresult}
\mathcal{C}(t_R, t_L)=
\begin{cases}
\mathcal{C}(|\text{TFD}\rangle)+C|t_R-t_L| & t_w-t_R<t_*\  \text{or}\ t_w-t_L<t_*\\
\mathcal{C}(|\text{TFD}\rangle)+C(2(t_w-t_*)-t_R-t_L) & t_w-t_R>t_*\  \text{and}\ t_w-t_L>t_*,
\end{cases}
\end{align}
where in the first line it behaves the same as the unperturbed system. The subtraction of $2t_*$ in the second line is called the switchback effect \cite{Stanford:2014jda} \cite{Susskind:2014jwa} \cite{Brown:2016wib}. The switchback effect is due to cancellations between forward and backward time evolutions during which the perturbation has not yet affected much of the system. Also, as was pointed out in \cite{Stanford:2014jda} \cite{Susskind:2015toa}, in the range $t_R<t_w-t_*,\ t_L<t_w-t_*$, the complexity always decreases with $t_R$. In contrast, without any perturbations the complexity will increase with $t_R$ as long as $t_R>t_L$. This is what we mean by abnormal time dependence. We will use this to diagnose transparency of horizons in section \ref{diagnosis} \cite{Susskind:2015toa}.

\begin{figure}[H]
\begin{center}
\includegraphics[scale=.4]{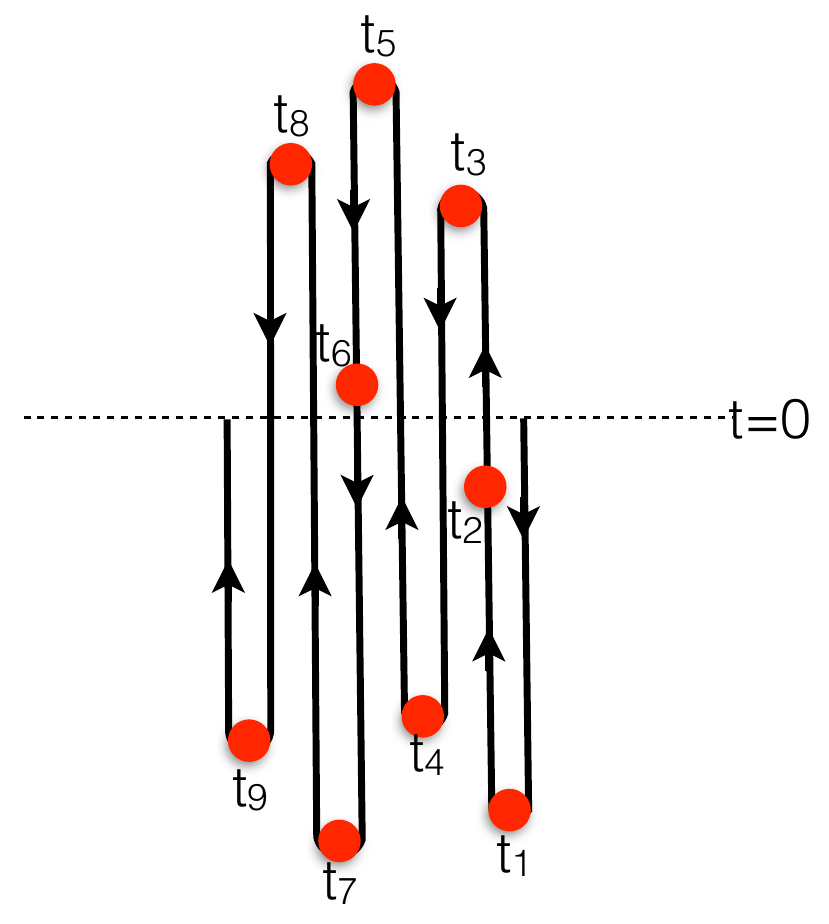}
\caption{Product of multiple precursors}
\label{multiprecursor}
\end{center}
\end{figure}

We can also consider insertion of multiple precursors:
\begin{align*}
&W_n(t_n)W_{n-1}(t_{n-1})...W_2(t_2)W_1(t_1)\\
=&e^{-iHt_n}W_ne^{iH(t_n-t_{n-1})}W_{n-1}e^{iH(t_{n-1}-t_{n-2})}...W_2e^{iH(t_2-t_1)}W_1e^{iHt_1},
\end{align*}
where the perturbations are well separated, $|t_i-t_{i+1}|>2 t_*$, so that the spreadings of different precursors do not interfere. See Figure \ref{multiprecursor}. The arrows indicate the direction of time evolution.
From considerations of quantum circuits, we expect one switchback from each turn, so the complexity of this operator will be
\begin{align}
\label{multiswitchback}
\mathcal{C}=C(t_f-2n_{sb}t_*),
\end{align}
where $t_f$ is total folded time, and $n_{sb}$ is the number of switchbacks. On the gravity side the system is dual to black hole geometries perturbed by multiple shockwaves separated by large time \cite{Shenker:2013yza}. The result \eqref{multiswitchback} is reproduced in gravity calculations \cite{Stanford:2014jda} \cite{Brown:2015lvg}.

In previous examples we considered systems dual to black holes of size $l_{AdS}$. We can also consider a bigger system with position-dependent perturbations originating at position $x_w$. Now the perturbation also spreads ballistically. In \cite{Dray:1984ha} \cite{Sfetsos:1994xa} \cite{Roberts:2014isa} geometries with localized shocks were studied in detail, and it was shown that in the complexity there will be more complicated position-dependent cancellations \cite{Roberts:2014isa}. Roughly speaking, the perturbation spreads with butterflly velocity $v_B$, so the degrees of freedom at position $x$ feel the effect of the perturbation at time $t_w-\frac{|x-x_w|}{v_B}$. This means the contribution to complexity from site $x$ will be proportional to $2\left(t_w-\frac{|x-x_w|}{v_B}-t_*\right)$.

In all the above calculations, the time dependence of complexity shows certain robust features, like linear increase at large time, switchback delay time of $2t_*$, and abnormal decrease when the interior contains shockwaves. In the rest of this section, we'll rederive these results just from general assumptions \ref{assumption1} to \ref{assumption4}. 
We'll see that, in the bulk, these features are essentially consequences of symmetry and a particular pattern of symmetry breaking which results from the Rindler nature of horizons. 

\subsection{Linear increase of complexity: Boost symmetry across the horizon}

Let's first look at the time evolution of the thermofield double without perturbations \eqref{TFD}. 
%
%
The state $|\psi(t_R, t_L)\rangle$ describes the Wheeler-DeWitt patch at time $(t_R, t_L)$, i.e., the bulk region that is spacelike separated from the chosen boundary time \cite{Maldacena:2013xja}. From assumption \ref{assumption1}, complexity is dual to a certain geometric quantity defined within the WDW patch. 

Earlier we mentioned boost symmetry across the horizon. Here we'll make this statement precise. In Figure \ref{symmetryacrosshorizon}, if $t_R$ increases by $\Delta t$, the right boundary of WDW patch will move by $\Delta t$ along any $r=$ constant curve. This is more easily seen in Eddington-Finkelstein coordinates:
Define ingoing time: $dv^*=dt+\frac{dr}{f(r)}$.
Then the metric becomes
\begin{align}
ds^2=-f(r)dv^{*2}+2drdv^*+r^2 d\Omega_{D-2}^2
\end{align}
These coordinates cover right exterior and future interior regions in the Penrose diagram, and $v^*$ is a good time coordinate on the right future horizon. The boost symmetry in these two quadrants can now be written as
\begin{align}
v^*\rightarrow v^*+\Delta t.
\end{align} 
We can similarly define $du^*=dt-\frac{dr}{f(r)}$ and it covers the right exterior region as well as the white hole interior.

We can intuitively view $v^*$ as a trajectory of a light beam sent from boundary at time $t_R$ along radial directions. The beam will stay on the boundary of the WDW patch $v^*=t_R$. As $t_R$ increases, the spacetime region between $v^*=t_R$ and $v^*=t_R+\Delta t$ is scanned by the beam, and the time dependence of complexity is seen in this scan. As we'll make precise later, the scan can tell us whether the interior is growing uniformly, which shows whether the dynamics of making the state is going well without being disrupted. We should emphasize here, the scanning beam is merely a geometric analogy. All we want is something representing boundary of WDW patch. We don't mean physically throwing in any quanta.

Also note that, even though we work with thermofield double as an example, what we mean by boost symmetry in this paper is a property near the horizon and does not rely on the existence of the other side. 

Let's fix $t_L$ and just evolve $t_R$, when $t_R-t_L$ is large, we see a pattern emerge. 
 \begin{figure}[H]
 \begin{center}
  \begin{subfigure}[b]{0.4\textwidth}
  \begin{center}
    \includegraphics[scale=0.5]{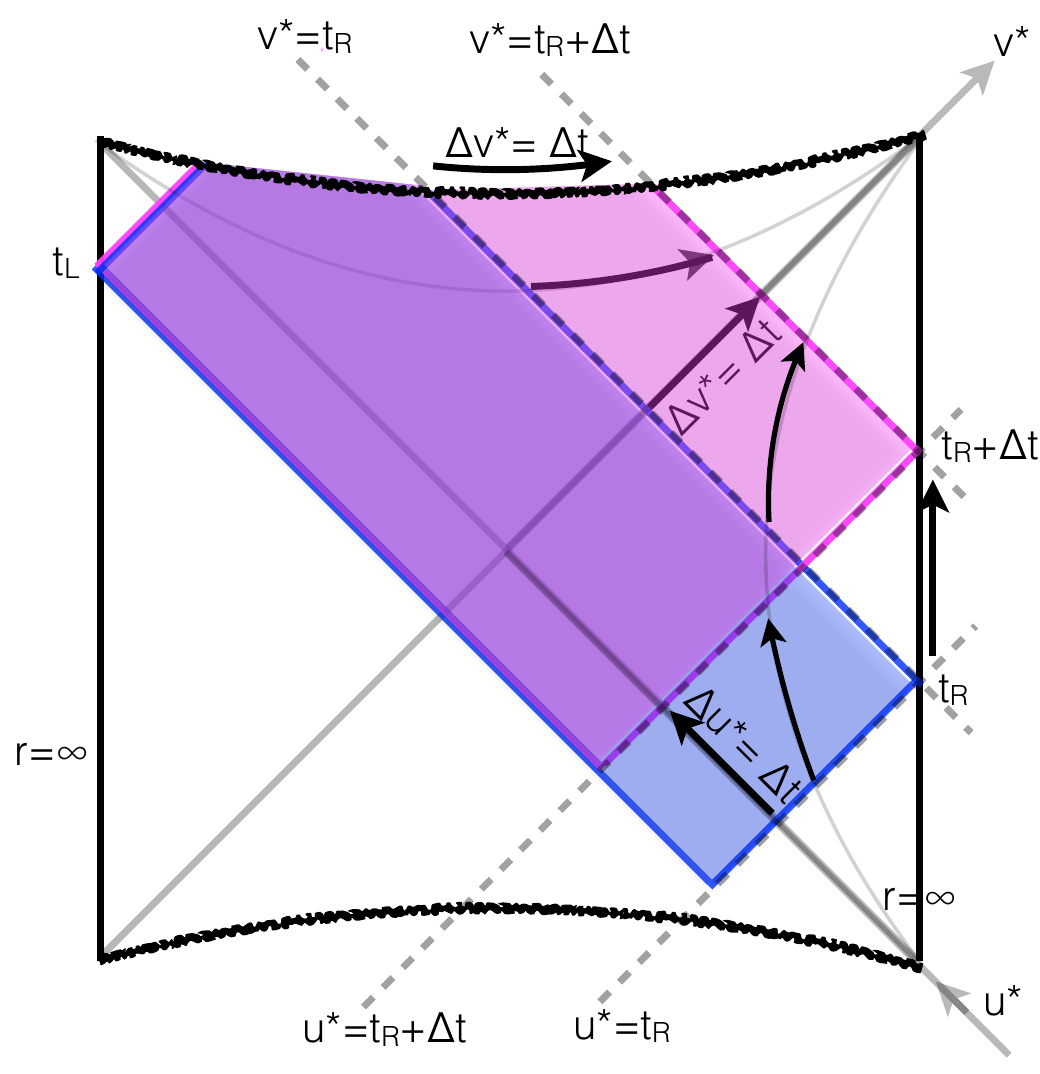}
    \caption{Two WDW patches at different $t_R$ with $t_L$ fixed.}
    \label{TFDsymmetryWDW1}
    \end{center}
  \end{subfigure}
  \hspace{+2em}
  \begin{subfigure}[b]{0.4\textwidth}
  \begin{center}
    \includegraphics[scale=0.5]{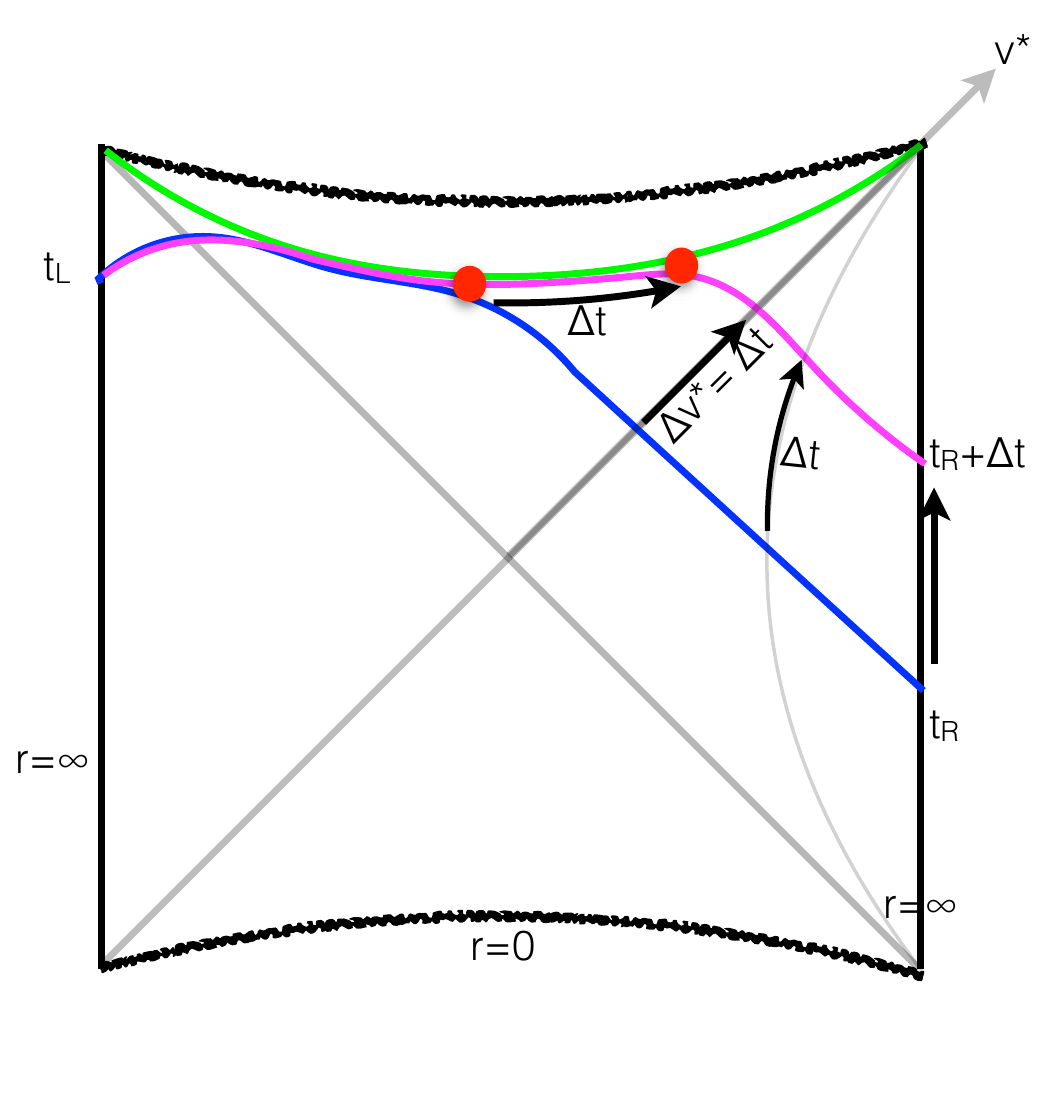}
    \caption{Two maximal surfaces at different $t_R$ with $t_L$ fixed.}
    \label{TFDsymmetrymaxsurface1}
    \end{center}
  \end{subfigure}
  \caption{Bulk duals of $|\psi(t_R,t_L)\rangle$ and $|\psi(t_R+\Delta t, t_L)\rangle$.}
  \label{TFDsymmetry1}
  \end{center}
\end{figure}
Figure \ref{TFDsymmetryWDW1} shows WDW patches at two times. As we increase $t_R$, we move the upper right boundary of WDW patch from $v^*=t_R$ to $v^*=t_R+\Delta t$, and the lower right boundary from $u^*=t_R$ to $u^*=t_R+\Delta t$.  The black hole interior is expanding, while the white hole interior is shrinking. With large $t_R-t_L$, the white hole interior is small, and we expect it to have less and less effect on the complexity (assumption \ref{assumption4}). The change in the WDW patch is basically to add a piece of Figure \ref{increasedregion1} to the interior region. Regions like this will be the basic building blocks for our following discussions. They have the feature of being uniform in the Schwarzschild time direction. Look at Figure \ref{increasedregion2}. We divided it into two parts. The two parts are related by a symmetry transformation $v^*\rightarrow v^*+\Delta t$, so we expect them to give equal contributions to complexity. As a consequence, the contribution of a region like this to complexity will be proportional to $\Delta v^*$.\footnote{In complexity-action prescription, additivity of action is a subtle issue, which was discussed in \cite{Lehner:2016vdi}.} From now on we will denote it by $C\Delta v^*$:
\begin{align}
\label{buildingblockcomplexity}
\Delta \mathcal{C}=C\Delta v^*
\end{align}
This symmetry is the reason for which we find linear increase of complexity. But of course, without a specific prescription we cannot fix the coefficient $C$. 

 \begin{figure}[H]
 \begin{center}
  \begin{subfigure}[b]{0.4\textwidth}
  \begin{center}
    \includegraphics[scale=0.5]{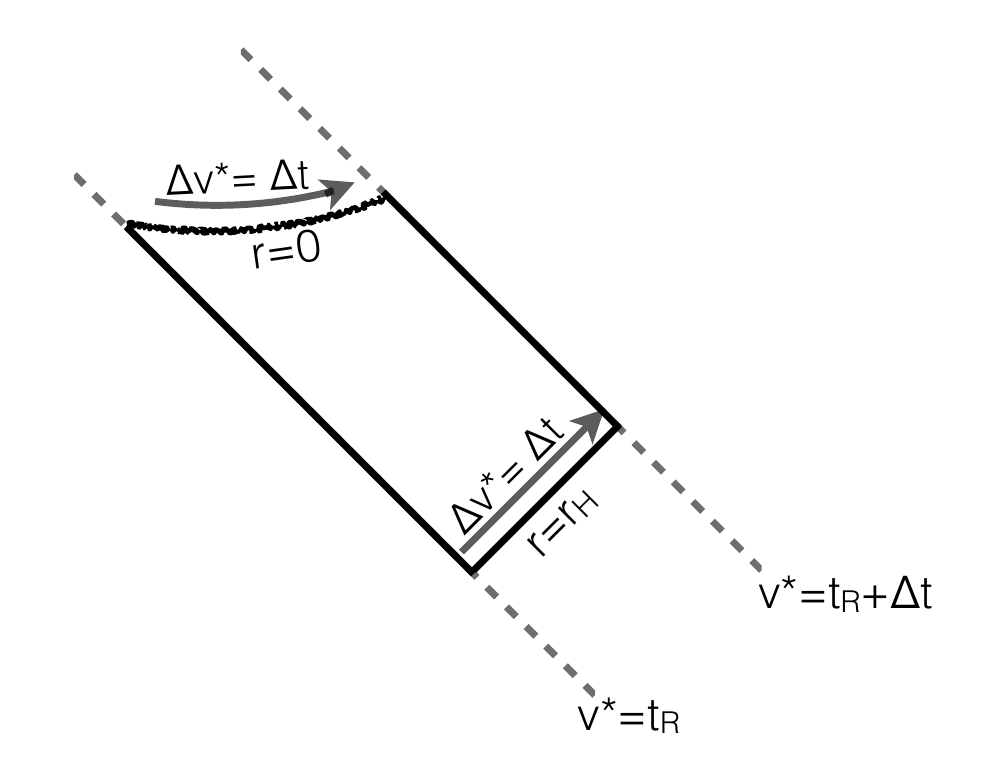}
    \caption{}
    \label{increasedregion1}
    \end{center}
  \end{subfigure}
  \hspace{-1em}
  \begin{subfigure}[b]{0.4\textwidth}
  \begin{center}
    \includegraphics[scale=0.5]{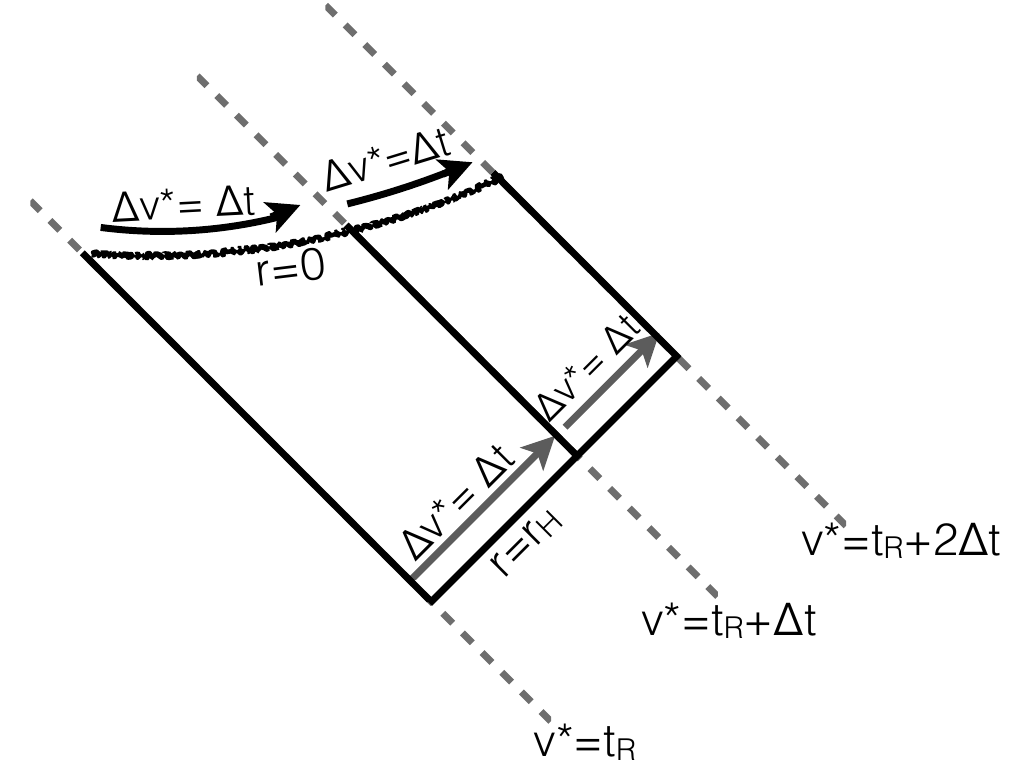}
    \caption{}
    \label{increasedregion2}
    \end{center}
  \end{subfigure}
  \caption{Increase of WDW patch as right time increases from $t_R$ to $t_R+\Delta t$.}
  \label{increasedregion}
  \end{center}
\end{figure}
The above argument does not depend on the detailed prescriptions of the bulk dual of complexity. It only requires that the prescription respects time translation symmetry, i.e., we don't use different prescriptions at different times! For example, we draw maximal volume surfaces in Figure \ref{TFDsymmetrymaxsurface1}. The maximal volume surfaces hug the limiting surface (green line), until they separate (red dots). The volume increases linearly with the separation time, and that time is synchronized (up to an additive constant) with boundary time $t_R$ as a consequence of the above mentioned symmetry. 

There will be a transient period of order the thermal time, during which the above symmetry argument does not apply. It happens when the tip of the WDW wedge leaves the singularity and the corresponding piece of the interior starts to shrink. We don't expect this period to give big contributions to complexity. Also, we are interested in time scales much bigger than the thermal time, so we will ignore this transient. The transient period ends when the shrinking interior piece becomes very small, or the maximal surface touches the limiting surface. After that the robust linear increase kicks in, and the corrections continue to die off exponentially. Symmetry dictates the linear time dependence in \eqref{linear}.

\subsection{Breaking boost symmetry: Abnormal time dependence of complexity}
\label{homogeneousperturbation}
\begin{figure}[H]                      
      \includegraphics[scale=.5]{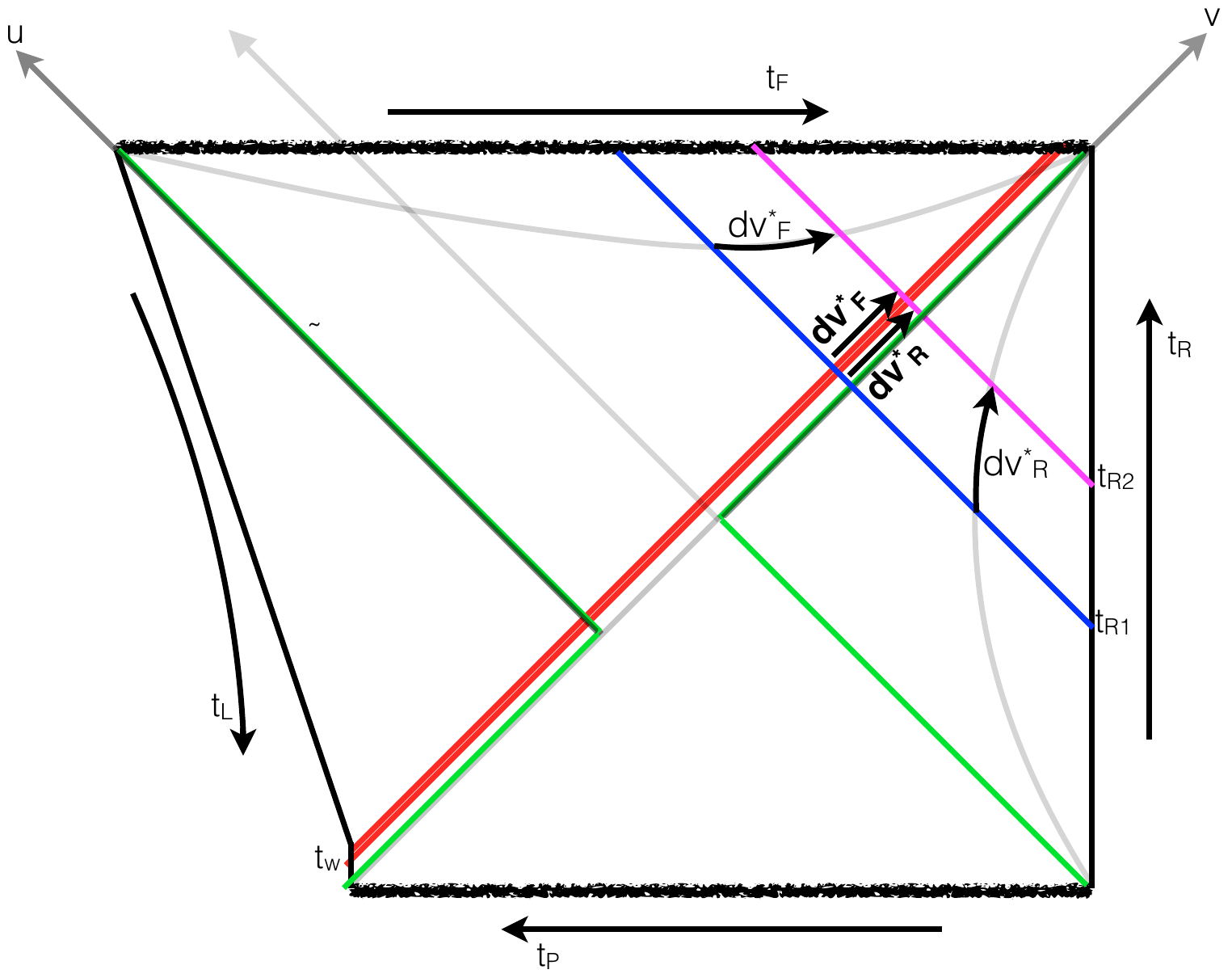}
  \caption{Two sided black hole with one spherically symmetric shockwave}
  \label{sphericaloneshock}
\end{figure}

What if we perturb the state by inserting a precursor $W_L(t_w)=e^{iHt_w}W_Le^{-iH t_w}$ with $t_w$ large? The geometry was studied in \cite{Shenker:2013pqa}. Now there is a shockwave lying close to the horizon, and it would break the boost symmetry in a specific pattern, see Figure \ref{sphericaloneshock}.
From earlier discussions, we see that what determines the time dependence of complexity is how the interior portion of the WDW patch changes as we vary the boundary time, i.e., as the detecting beam scans the interior it would keep a record of what's going on. Now with a shockwave present, the scanning beam will be pushed forward, by an amount depending on the position where the beam meets the shockwave. Mathematically, $\partial_{v^*}$ is no longer a Killing vector. The time separation $\Delta v^*$ between the two beams inside horizon will no longer be same as that outside. In Figure \ref{sphericaloneshock}, consider two ingoing null geodesics from the right boundary with time separation $\Delta t_R=\Delta v^*_{R}$. Across the shockwave, there will be relative time dilation, and $\Delta v^*$ will jump when the two geodesics pass the shell. 

To quantify this effect, note that near the horizons, the geometry is Rindler-like on both sides of the shockwave:
\begin{align*}
ds^2=-f(r)dv^{*2}+2drdv^*+r^2d\Omega_{D-2}^2
\end{align*}
where $f(r)=\frac{4\pi}{\beta}(r-r_H)$. Across the shell, there is discontinuity in $v^*$ coordinate:
\begin{align*}
\frac{dv^*_F}{dv^*_R}(t_R)=\frac{f(r_w)}{\tilde f(r_w)}=\frac{r_H-r_w}{\tilde r_H-r_w}=\frac{r_H-r_w}{\delta r_H+r_H-r_w},
\end{align*}
where $r_w$ is the radius at which the ingoing null line crosses the shockwave, and $\delta r_H$ is the increase of horizon radius due to the extra quantum. If we throw in a thermal scale quantum, we have
\begin{align*}
\delta S\sim 1,\ \ \frac{\delta r_H}{r_H}=\frac{1}{D-2}\frac{\delta S}{S}.
\end{align*}
Rindler geometry also gives 
\begin{align*}
r_w-r_H=-\frac{\beta}{4\pi}e^{-\frac{2\pi}{\beta}(t_w-t_R-2R^*)},
\end{align*}
where $\frac{R^*}{\beta}$ is an order one constant depending on the asymptotic geometry and dimension. 
So we see the dilation factor is given by
\begin{align}
\label{timedilation}
\frac{dv^*_F}{dv^*_R}(t_R)=\frac{1}{1+ce^{\frac{2\pi}{\beta}(t_w-t_R-t_*)}},
\end{align}
where $c=\frac{1}{D-2}\frac{4\pi r_H}{\beta}e^{-\frac{4\pi}{\beta}R^*}\delta S$ is an order one constant. 

\begin{figure}[H] 
 \hspace{1cm}                    
      \includegraphics[scale=.45]{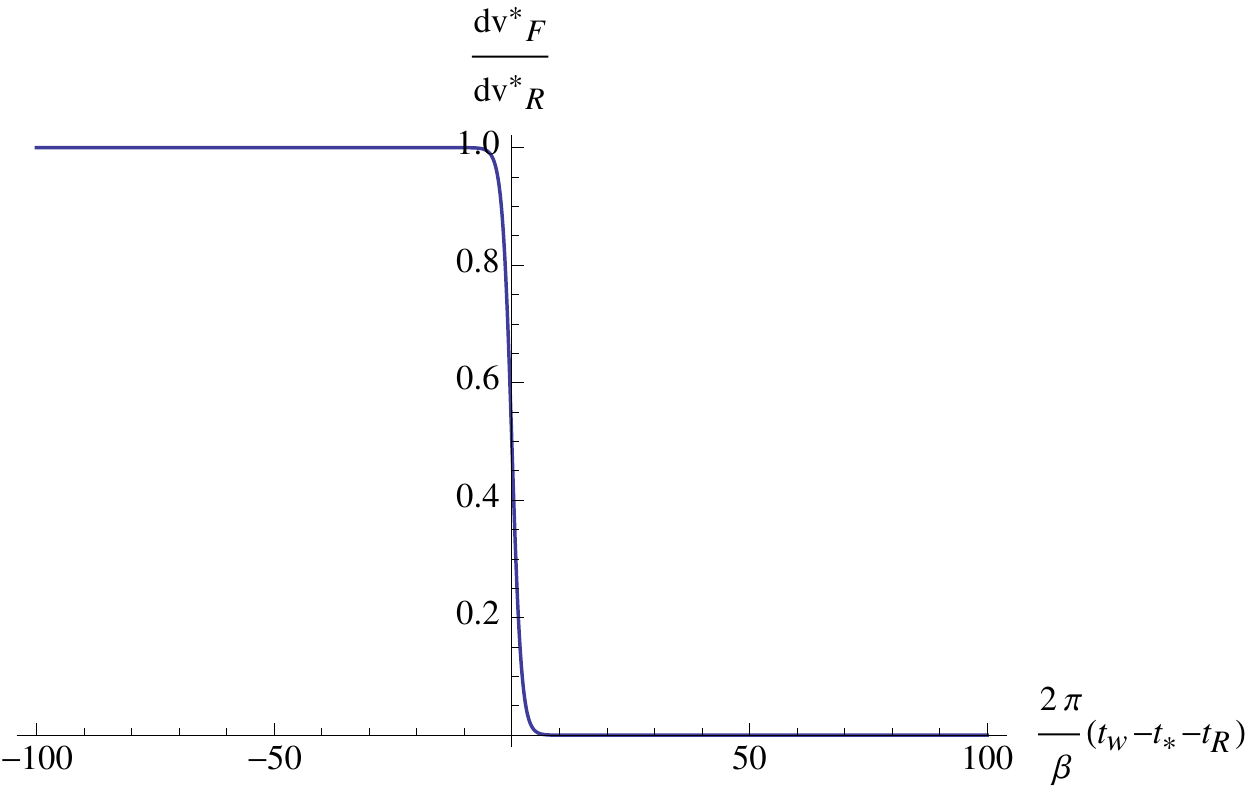}
  \caption{Eddington-Finkelstein time dilation factor as we go across the shockwave, as a function of boundary time.}
  \label{dilation}
\end{figure}

\begin{figure}[H]
 \begin{center}
  \begin{subfigure}[b]{0.3\textwidth}
  \begin{center}
    \includegraphics[scale=0.32]{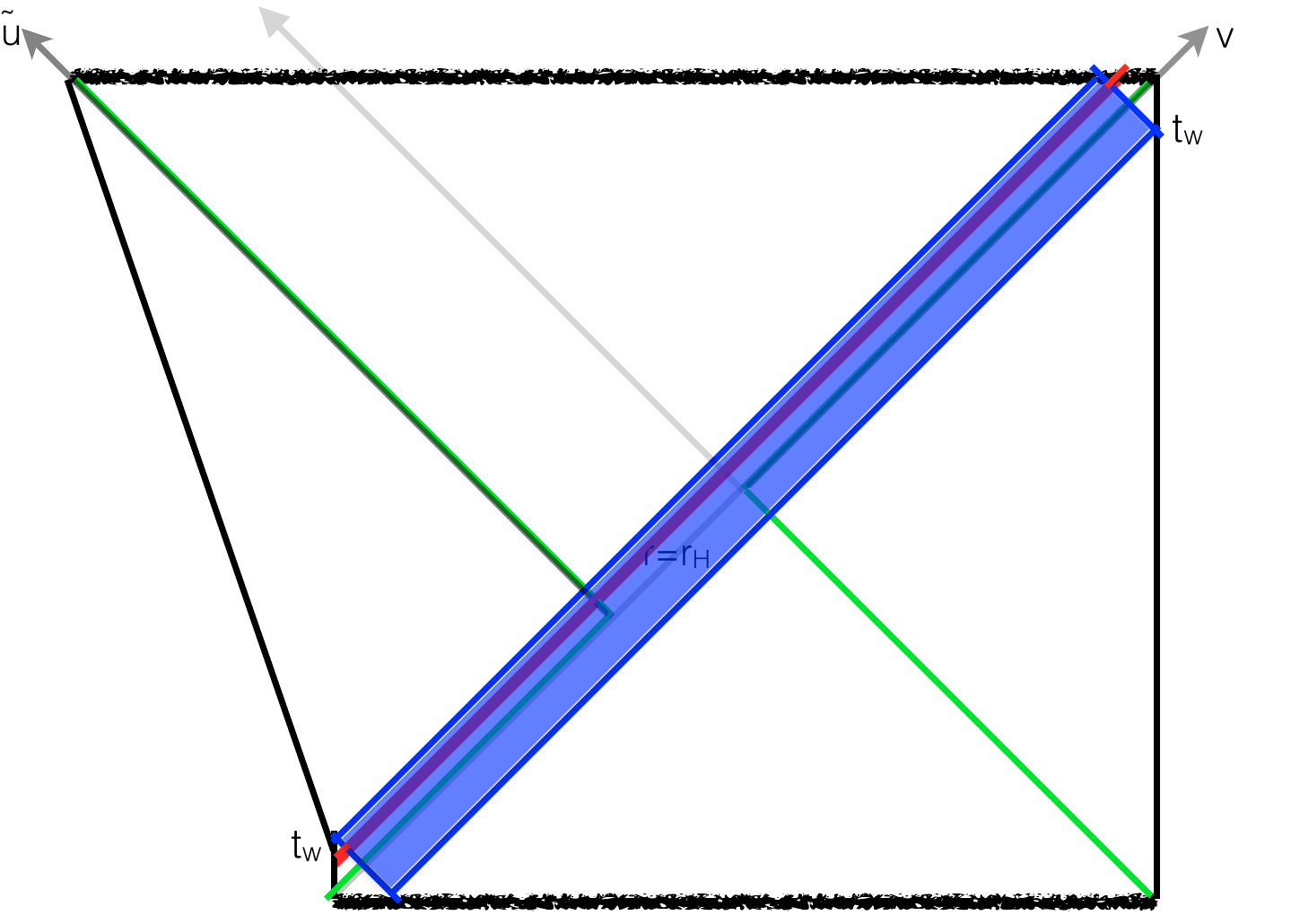}
    \caption{Right after perturbation}
    \label{sphericaloneshocktime1a}
    \end{center}
  \end{subfigure}
  \hspace{-.5cm}
  \begin{subfigure}[b]{0.4\textwidth}
  \begin{center}
    \includegraphics[scale=0.32]{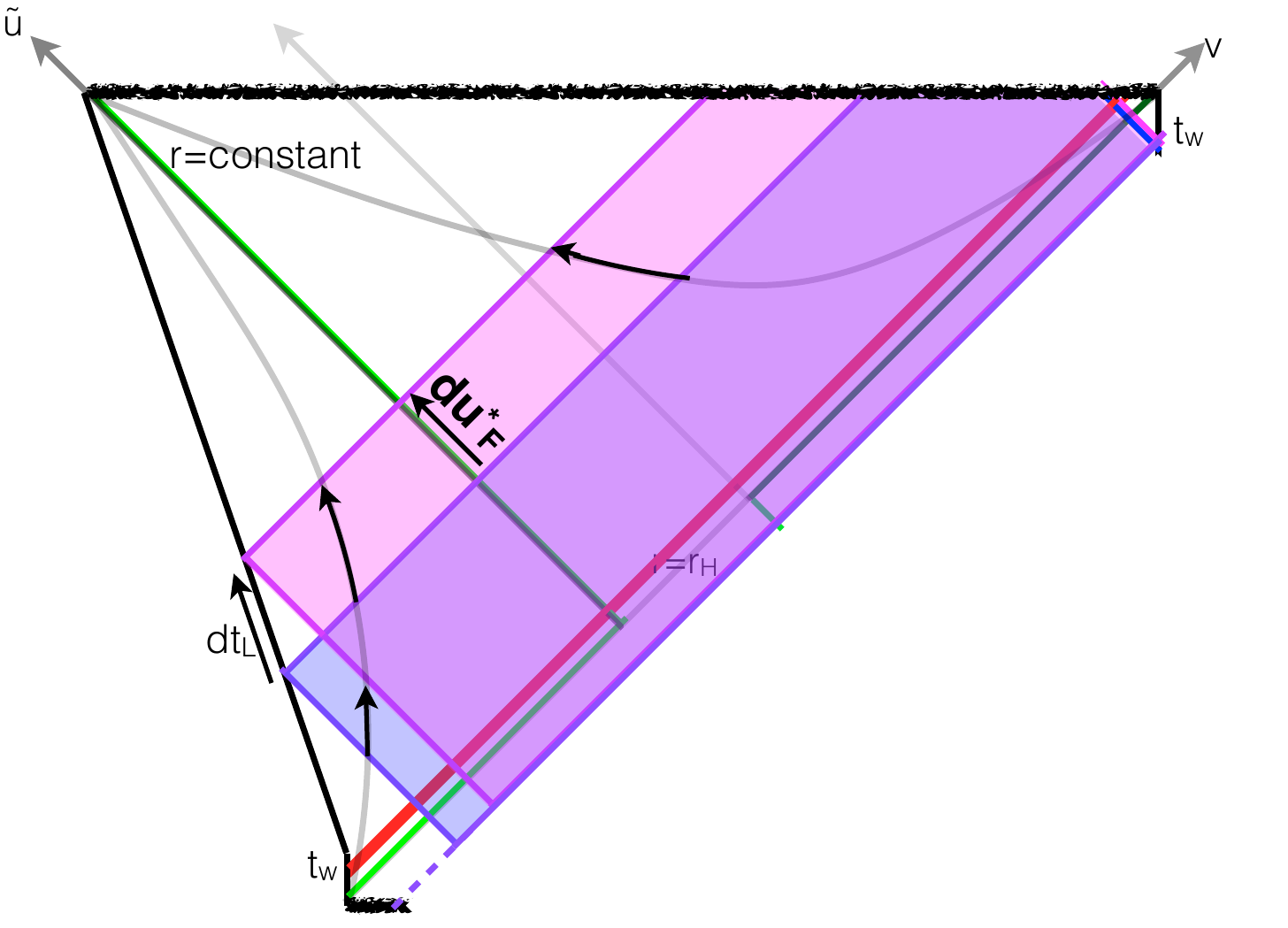}
    \caption{$t_L$ goes up, $t_R$ fixed}
    \label{sphericaloneshocktime1b}
    \end{center}
  \end{subfigure}
  \hspace{-1.6cm}
  \begin{subfigure}[b]{0.4\textwidth}
  \begin{center}
    \includegraphics[scale=0.32]{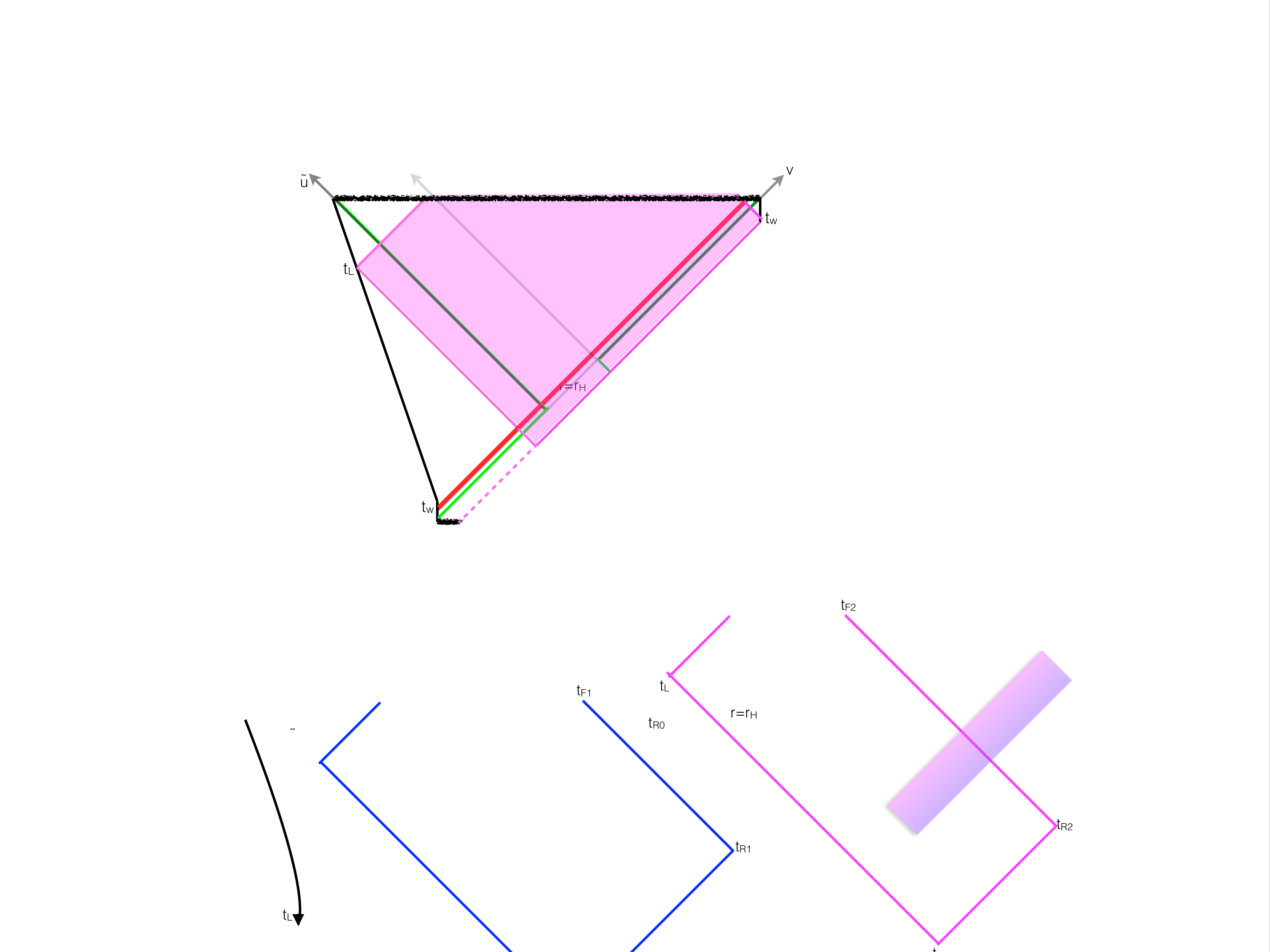}
    \caption{$t_w-t_L>t_*$}
    \label{sphericaloneshocktime1c}
    \end{center}
  \end{subfigure}
  \end{center}
    \label{sphericaloneshocktime1}
    \caption{Evolution of WDW patch as we increase left time.}
\end{figure}

This time dilation factor quantitatively characterizes to what extent the boost symmetry is broken. Note that to arrive at this factor \eqref{timedilation} we only used the Rindler-like geometry of the horizon. Its functional form is completely robust. Different dimensions or different asymptotic boundary conditions only change the constant $c$. Let's look at the behavior of this function (Figure \ref{dilation}). When $t_w-t_R<t_*$, $dv^*_F=dv^*_R$, i.e., there is still good symmetry across the horizon. But when $t_w-t_R>t_*$, $dv^*_F=e^{-\frac{2\pi}{\beta}(t_w-t_*-t_R)}dv^*_R$ is exponentially smaller than $dv^*_R$, i.e., as we increase $t_R$ the future interior does not grow accordingly. We've seen that the linear increase of complexity is a result of the steady expansion of the black hole interior. Now this symmetry is broken, and it gives rise to abnormal time dependence of complexity.\footnote{In this type of perturbation we considered here (low energy, early time), the symmetry is broken in a particularly simple pattern. The only effect of the shockwave is to give a relative time dilation across the horizon. The change of interior always has the same shape as in Figure \ref{increasedregion}. The only difference is, $\Delta v^*$ inside horizon can be smaller than $\Delta v^*$ outside, which will give rise to a shift to the time dependence of complexity. } From here we already see that the time dependence of complexity is closely related to the transparency of the horizon, as proposed by Susskind in \cite{Susskind:2015toa}. We'll explore this more in sections \ref{diagnosis} and \ref{circuit}. 
Because the perturbation at time $t_w$ has very low energy, it causes very little change to the system right after the perturbation. In order to explore the time dependence of complexity, let's begin with boundary time $t_L$ right above $t_w$. (In our convention this means $t_L$ is slightly smaller than $t_w$.) Let $t_R=t_L=t_w$. 
In Figure \ref{sphericaloneshocktime1a}, the dual WDW patch is shown in blue.

Next, let's push $t_L$ upward with $t_R$ fixed (Figure \ref{sphericaloneshocktime1b}), i.e., decrease $t_L$, until $t_w-t_L>t_*$ (Figure \ref{sphericaloneshocktime1c}). 
The change in the white hole interior will only show up as some transient behavior for thermal time, so we ignore it. The black hole interior steadily expands. We have
\begin{align}
&\Delta \mathcal{C}_1=C\Delta u^*_F\nonumber\\
&\frac{d\mathcal{C}_1(t_L, t_R=t_w)}{dt_L}=C\frac{d u^*_F(t_L)}{dt_L}=-C\nonumber\\
\label{complexitytime1}
&\mathcal{C}_1(t_L,t_R=t_w)=\mathcal{C}(|\text{TFD}\rangle)+C(t_w-t_L)
\end{align}
Thus we see the complexity increases as we push $t_L$ upward.\\

\begin{figure}[H]
 \begin{center}
  \begin{subfigure}[b]{0.3\textwidth}
  \begin{center}
    \includegraphics[scale=0.32]{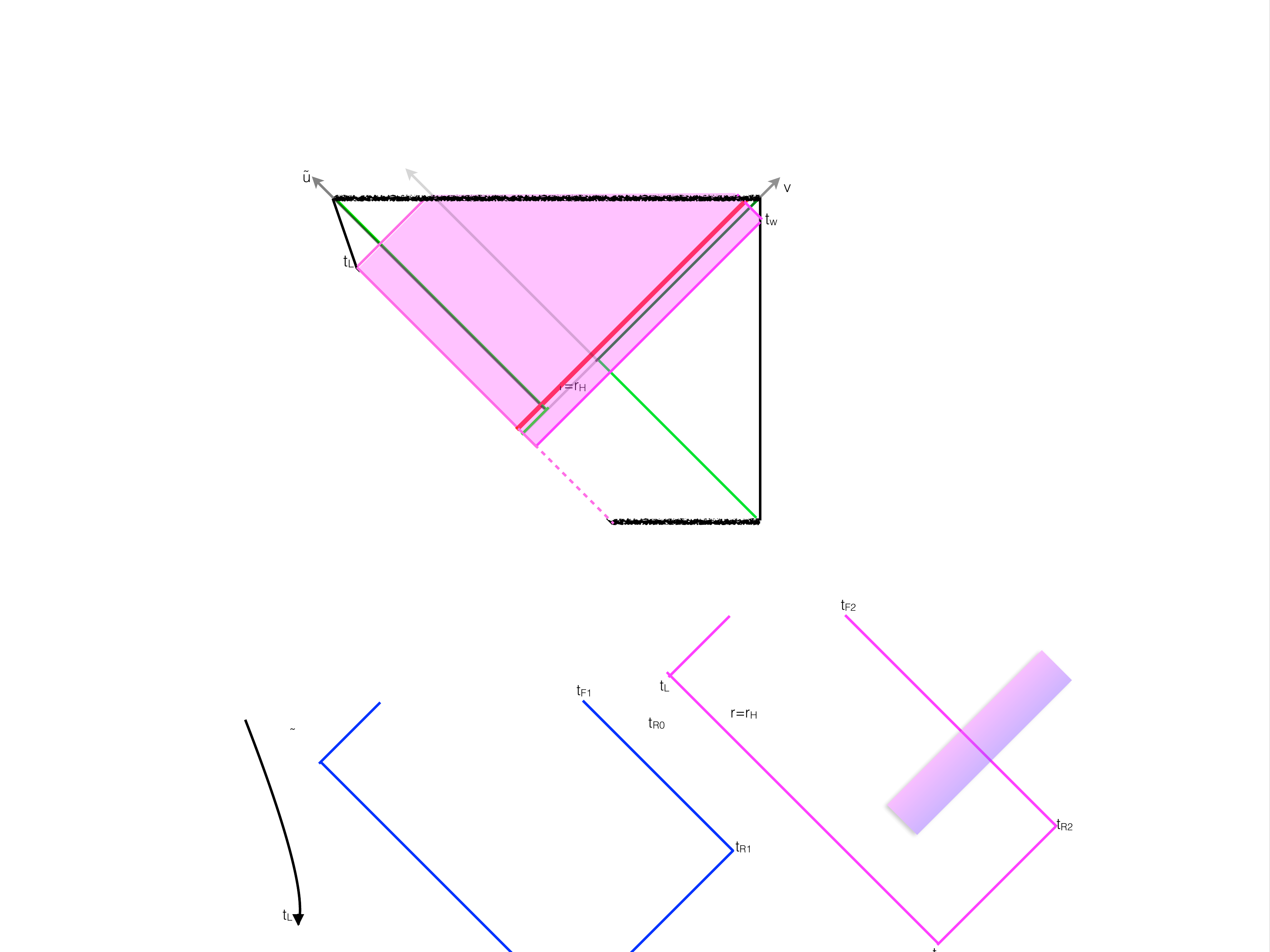}
    \caption{$t_R=t_w$}
    \label{sphericaloneshocktime2a}
    \end{center}
  \end{subfigure}
  \hspace{-0.5cm}
  \begin{subfigure}[b]{0.4\textwidth}
  \begin{center}
    \includegraphics[scale=0.32]{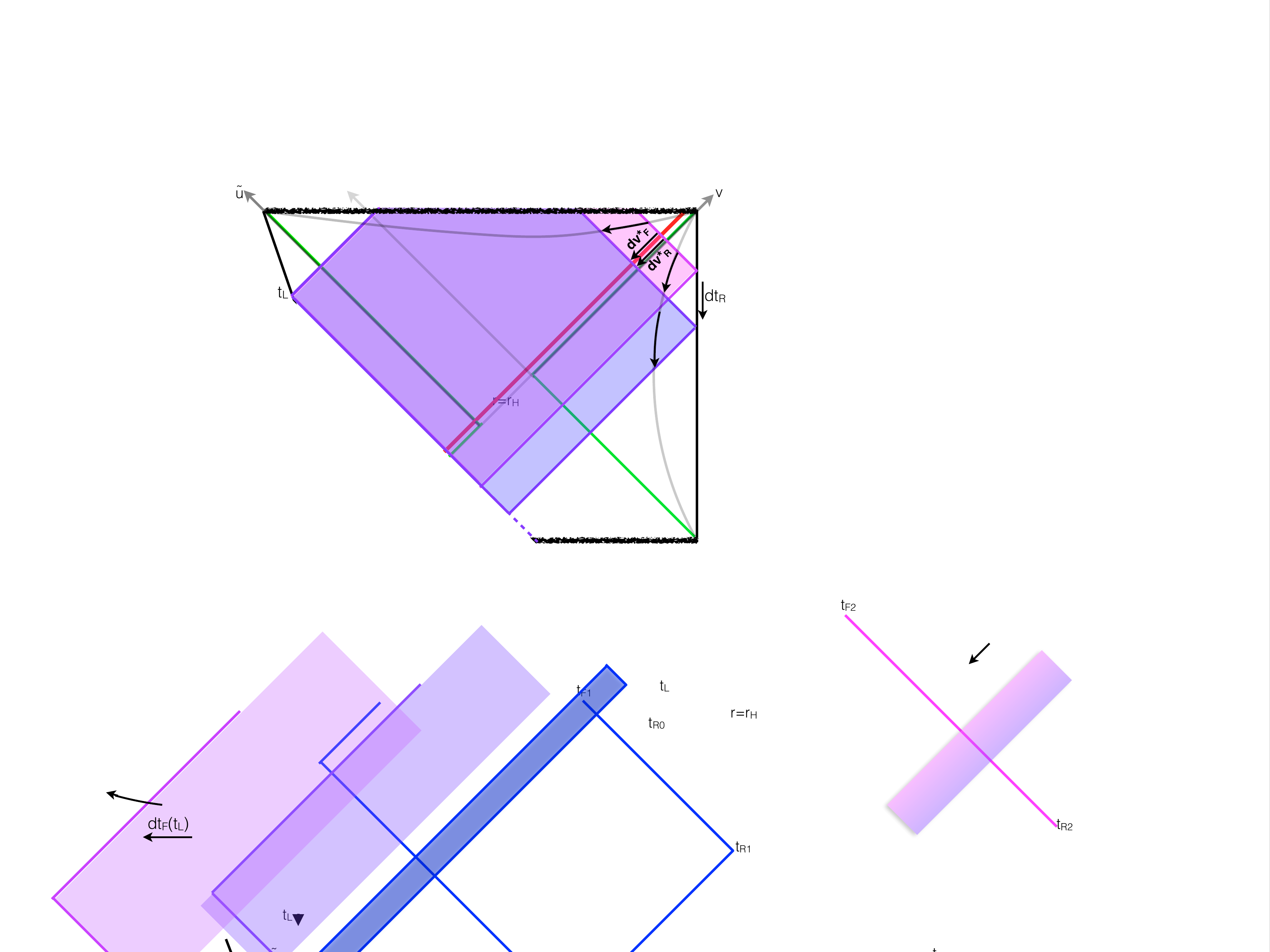}
    \caption{$t_R$ goes down, $t_L$ fixed}
    \label{sphericaloneshocktime2b}
    \end{center}
  \end{subfigure}
  \hspace{-1.6cm}
    \begin{subfigure}[b]{0.4\textwidth}
  \begin{center}
    \includegraphics[scale=0.32]{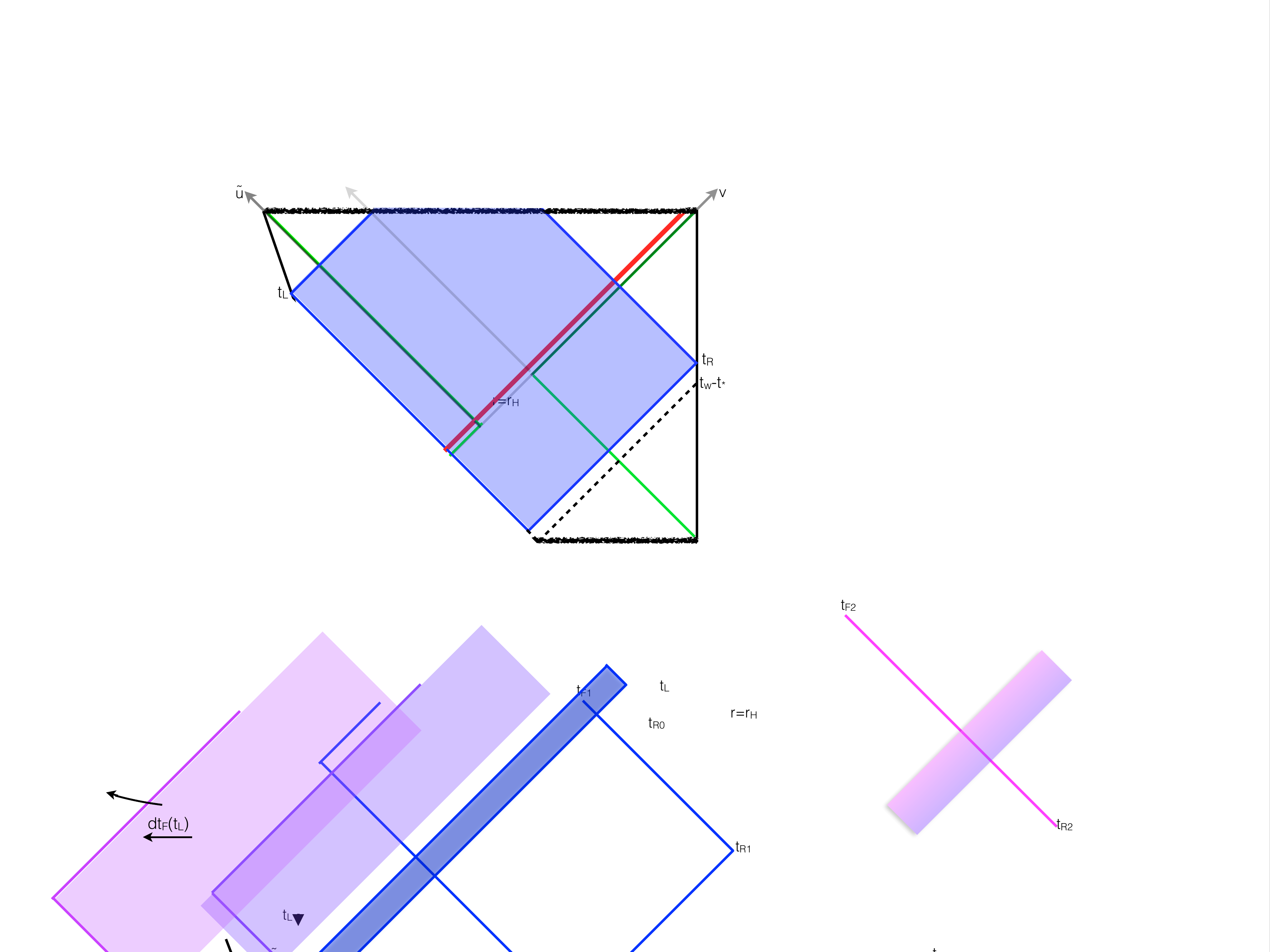}
    \caption{$t_w-t_R<t_*$}
    \label{sphericaloneshocktime2c}
    \end{center}
  \end{subfigure}
   \end{center}
  \caption{Evolution of WDW patch as we decrease right time.}
  \label{sphericaloneshocktime2}
 \end{figure}

Now, we fix the left time at $t_L$, and move $t_R$ downward. See Figure \ref{sphericaloneshocktime2}.

We start from Figure \ref{sphericaloneshocktime2a}, where the complexity is given by \eqref{complexitytime1}. With $t_w-t_R<t_*$ (Figure \ref{sphericaloneshocktime2b}), we can still ignore the white hole interior. The black hole interior is shrinking, but here, we encounter the shockwave, so we need to take into account the time dilation by \eqref{timedilation}.
\begin{align}
\Delta \mathcal{C}_1 =  C\Delta v^*_F&,\ \ dv^*_R = dt_R\nonumber\\
\frac{d\mathcal{C}_1(t_L, t_R)}{dt_R}=&C\frac{d v^*_F}{dv^*_R}(t_R)=\frac{C}{1+c e^{\frac{2\pi}{\beta}(t_w-t_*-t_R)}}\nonumber\\
\mathcal{C}_1(t_L,t_R)=&\mathcal{C}_1(t_L, t_w)+C\left(t_R+\frac{\beta}{2\pi}\log\left(1+ce^{\frac{2\pi}{\beta}(t_w-t_*-t_R)}\right)\right)-Ct_w\nonumber\\
=&\mathcal{C}(|\text{TFD}\rangle)+C(t_R-t_L)\label{complexitytime2}
\end{align}
In this regime ($t_w-t_R<t_*$, Figure \ref{sphericaloneshocktime2c}), the symmetry across the horizon is still approximately satisfied, and the complexity increases linearly with the right boundary time. So far we haven't encountered any abnormal behaviors of complexity. \\

\begin{figure}[H]
 \begin{center}
  \begin{subfigure}[b]{0.3\textwidth}
  \begin{center}
    \includegraphics[scale=0.32]{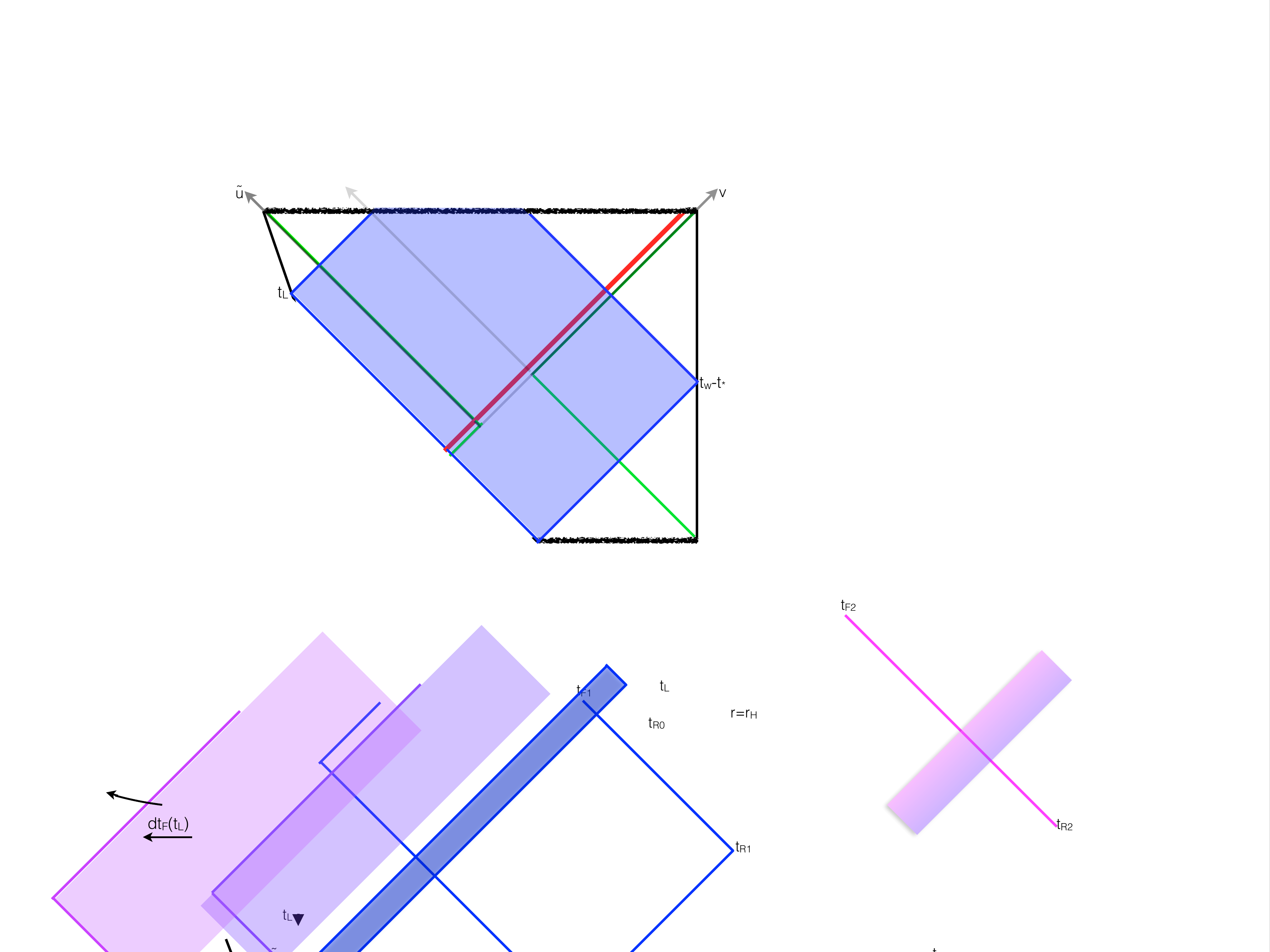}
    \caption{$t_R=t_w-t_*$}
    \label{sphericaloneshocktime3a}
    \end{center}
  \end{subfigure}
  \hspace{-0.6cm}
  \begin{subfigure}[b]{0.4\textwidth}
  \begin{center}
    \includegraphics[scale=0.32]{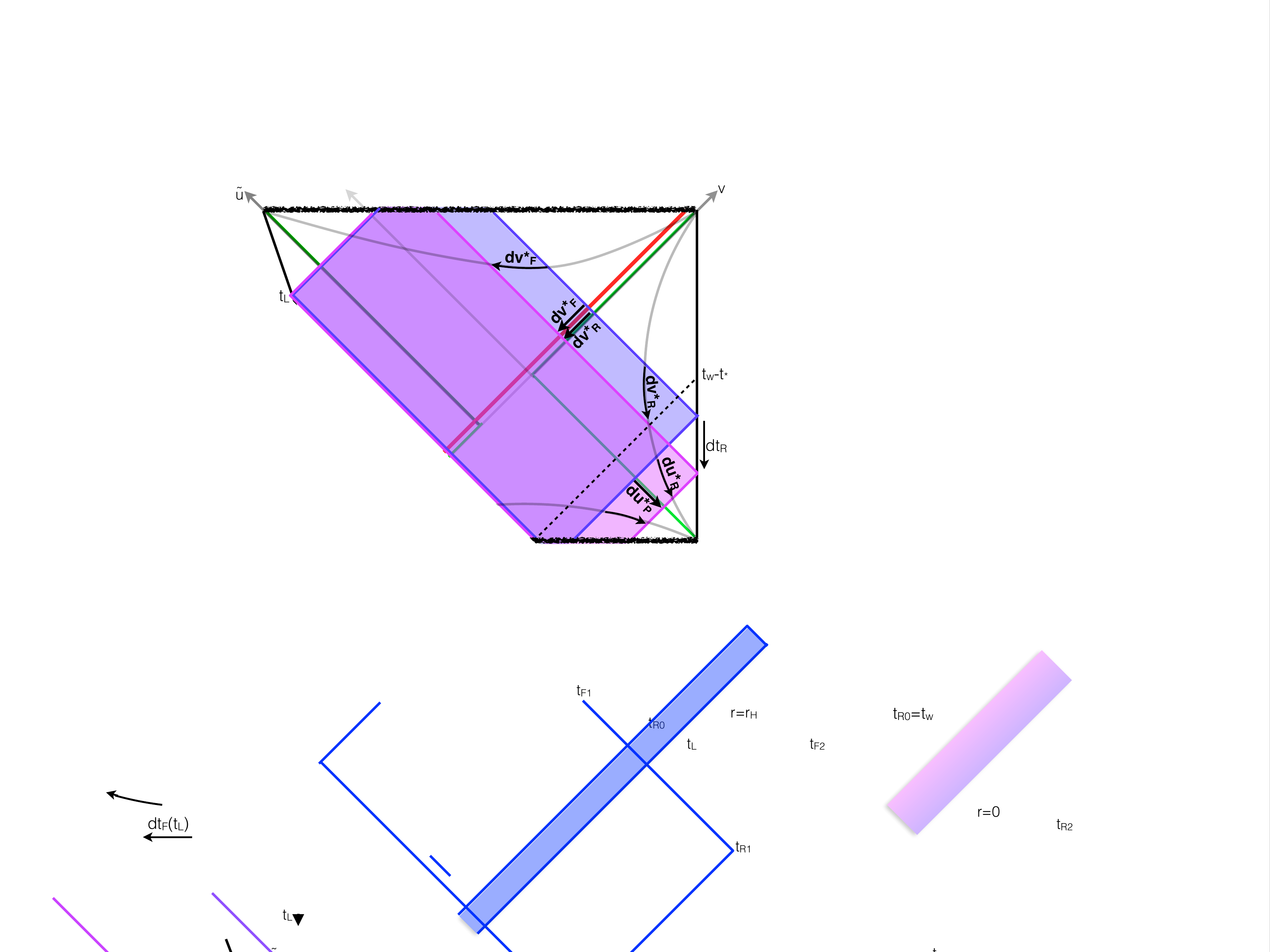}
    \caption{$t_R$ goes down, $t_L$ fixed}
    \label{sphericaloneshocktime3b}
    \end{center}
  \end{subfigure}
  \hspace{-1.6cm}
    \begin{subfigure}[b]{0.4\textwidth}
  \begin{center}
    \includegraphics[scale=0.32]{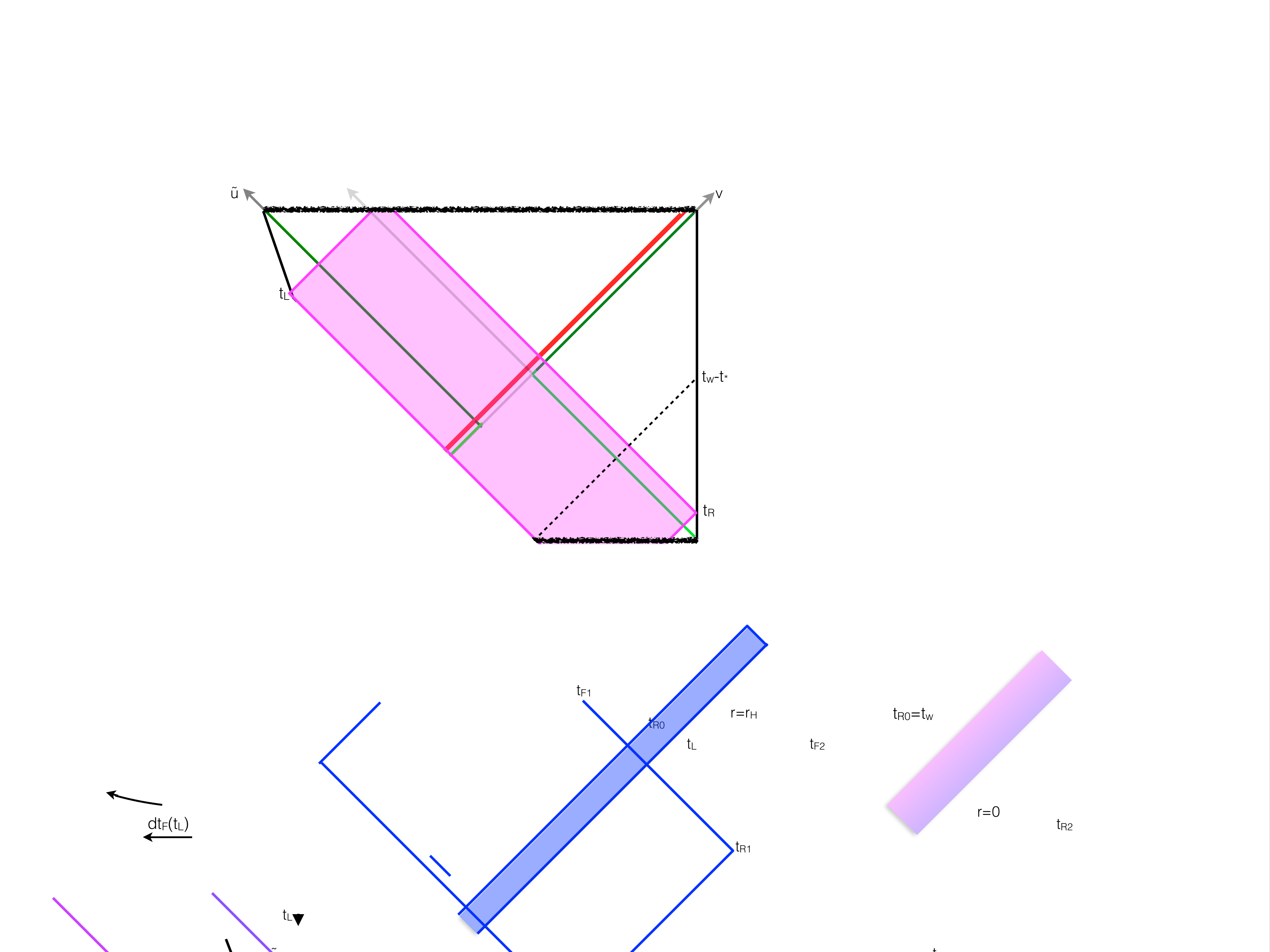}
    \caption{$t_w-t_R>t_*$}
    \label{sphericaloneshocktime3c}
    \end{center}
  \end{subfigure}
  \caption{Evolution of WDW patch as we decrease right time.}
  \label{sphericaloneshocktime3}
  \end{center}
\end{figure}

Next, let's push $t_R$ down further, i.e., consider when $t_L<t_w-t_*$ and $t_R<t_w-t_*$. See Figure \ref{sphericaloneshocktime3}.

Now interiors of both the black hole and the white hole change (Figure \ref{sphericaloneshocktime3b}). We need to consider the contributions to the complexity from both regions. Here, when we talk about the contributions to the complexity from different regions, we are implicitly using assumption \ref{assumption2} that the tensor network has locality down to AdS scale. We have
\begin{align}
\Delta\mathcal{C}_1 = C\Delta v^*_F& - C\Delta u^*_P\nonumber \\
\frac{d\mathcal{C}_1(t_L, t_R)}{dt_R}=&\ C\left(\frac{dv^*_F}{dv^*_R}(t_R)-\frac{du^*_P}{dt_R}\right)=-C\frac{ce^{\frac{2\pi}{\beta}(t_w-t_*-t_R)}}{1+ce^{\frac{2\pi}{\beta}(t_w-t_*-t_R)}}\nonumber\\
\mathcal{C}_1(t_L, t_R)=&\ \mathcal{C}_1(t_L, t_w-t_*)+C\frac{\beta}{2\pi}\log\left(1+ce^{\frac{2\pi}{\beta}(t_w-t_*-t_R)}\right)\nonumber\\
&\ -C\frac{\beta}{2\pi}\log\left(1+ce^{\frac{2\pi}{\beta}(t_w-t_*-(t_w-t_*))}\right)\nonumber\\
=&\ \mathcal{C}(|\text{TFD}\rangle)+C(2(t_w-t_*)-t_L-t_R)\label{abnormalregime}
\end{align}



Without perturbations, the answer would be $\mathcal{C}(|\text{TFD}\rangle)+C|t_R-t_L|$. We see that the time dependence becomes quite different in this regime. For example, let's set $t_L=-\infty$\footnote{This is just a technical choice to avoid transient period. We'll give more explanations about this in section \ref{diagnosis}.}. Without perturbations, the complexity will always increases with $t_R$. Now with a shockwave present, the complexity will decrease with $t_R$ when $t_w-t_R>t_*$. This abnormal time dependence is a consequence of breaking of boost symmetry as we go across the shockwave. 

Combining \eqref{complexitytime1}, \eqref{complexitytime2}, and \eqref{abnormalregime}, we recover the time dependence in \eqref{sphericaloneshockresult}.

In $D>3$, because the asymptotic boundaries bend outward, there will be time of thermal length when the complexity is not changing. We don't understand the reason. In this paper we'll consider it as part of the transient behavior and ignore it in our calculations.

In the above calculations, we ignored contributions from transient periods. It will depend on specific geometries and cannot be completely determined by symmetry argument. However, in calculating the complexity of the precursor $W(t_w)$, the transient behavior has little effect. We carry out the calculation in Appendix \ref{complexityoneprecursor}, and see a smooth exponential-to-linear transition as one expects from analysis of quantum circuits.

The idea that boost symmetry determines the time dependence of complexity should be clear from the above calculations. Following the same procedure we can consider more complicated examples. For states perturbed by multiple precursors, we follow the time evolution of preparing the state. Each time we make a perturbation, it takes a scrambling time for the memory of the previous perturbation to disappear, and the new perturbation takes a scrambling time to be squeezed up against the horizon. This is the origin of the $2t_*n_f$ subtractions in \eqref{multiswitchback}. The detailed calculations are in Appendix \ref{multipleperturbation}.

For states with localized perturbations, the time dilation factors will be position dependent. From a symmetry point of view, not only the boost symmetry is broken, but the translation symmetry in the spatial transverse direction is also broken. But again, at a coarse-grained level it's broken in a particularly simple pattern. More details are in Appendix \ref{localperturbation}. When we vary the transverse positions of the scanning beam\footnote{For localized perturbations, scanning light beam is not a perfect analogy. With localized shockwaves, null geodesics do not stay on constant spatial positions. Nevertheless we want to detect the interior along slices with constant spatial positions in order to compare with the case without perturbations.  }, we will also detect the transverse variations of geometry caused by the propagation of perturbations. Intuitively, one can imagine a tensor network lying at the stretched horizon. It evolves as the dynamics goes on. When we look at the complexity history, we scan this network. In particular, from the profile of complexity we can see some part of the network is normal while some part is disrupted by high energy. This will be the topic of the next section: diagnosing the horizons by complexity.

\subsection{Time dependence of complexity as a diagnosis of good interior}
\label{diagnosis}

From earlier discussions it should be already clear why we can use the complexity to detect the interiors of black holes or white holes. When there is nothing abnormal, the interior steadily expands as the boundary time evolves, and this will give linear increase of complexity. An abnormal decrease of complexity means this evolution is disrupted. 

\begin{figure}[H]
\begin{center}
    \begin{subfigure}[b]{0.4\textwidth}
    \begin{center}
    \hspace{-1.86cm}
    \includegraphics[scale=0.388]{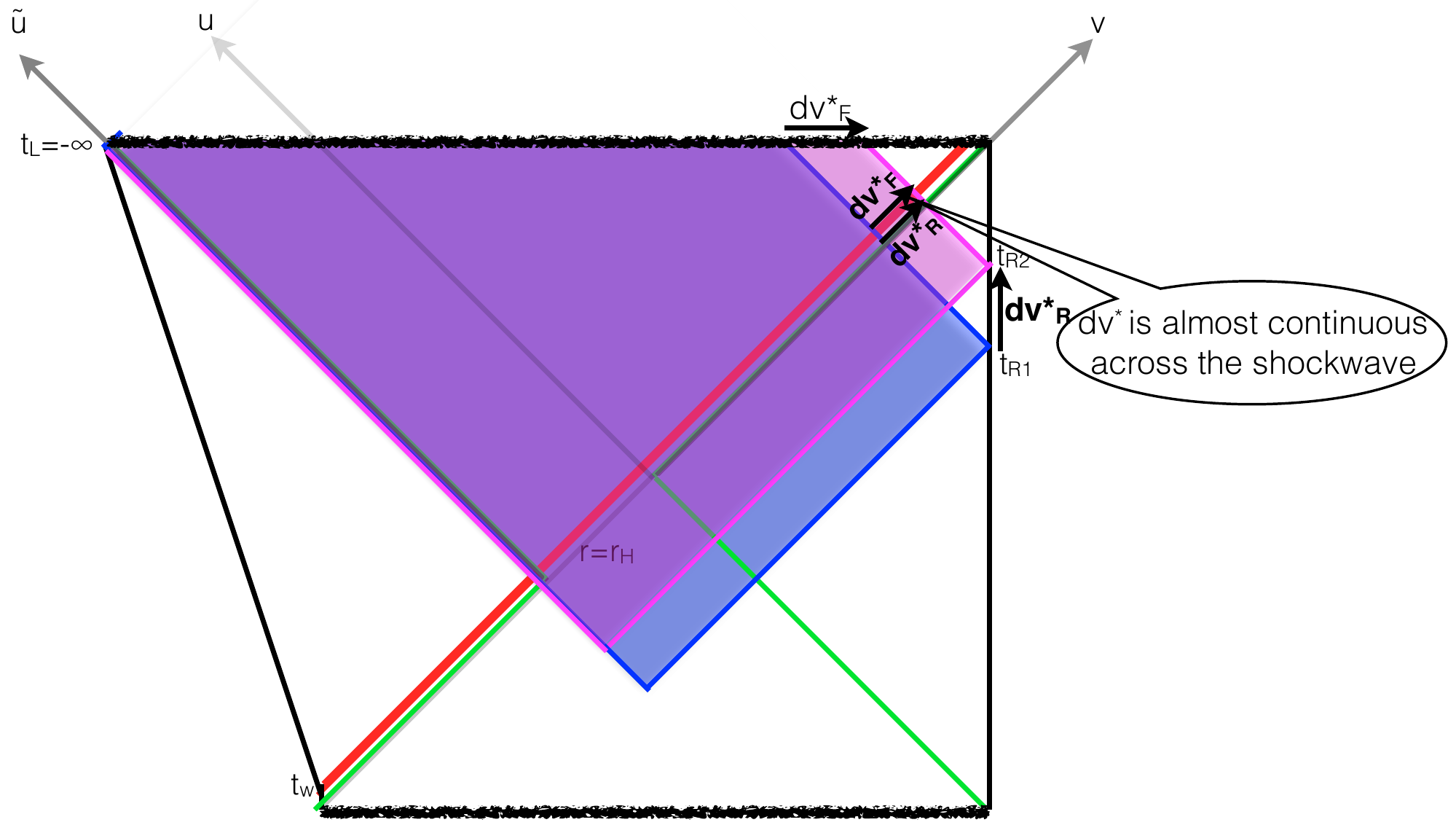}
    \caption{When the energy of matter is mild, $dv^*$ is continuous and the complexity increases.}
    \label{smooth}
    \end{center}
  \end{subfigure}
  \hspace{1.4cm}
  \begin{subfigure}[b]{0.4\textwidth}
    \includegraphics[scale=0.388]{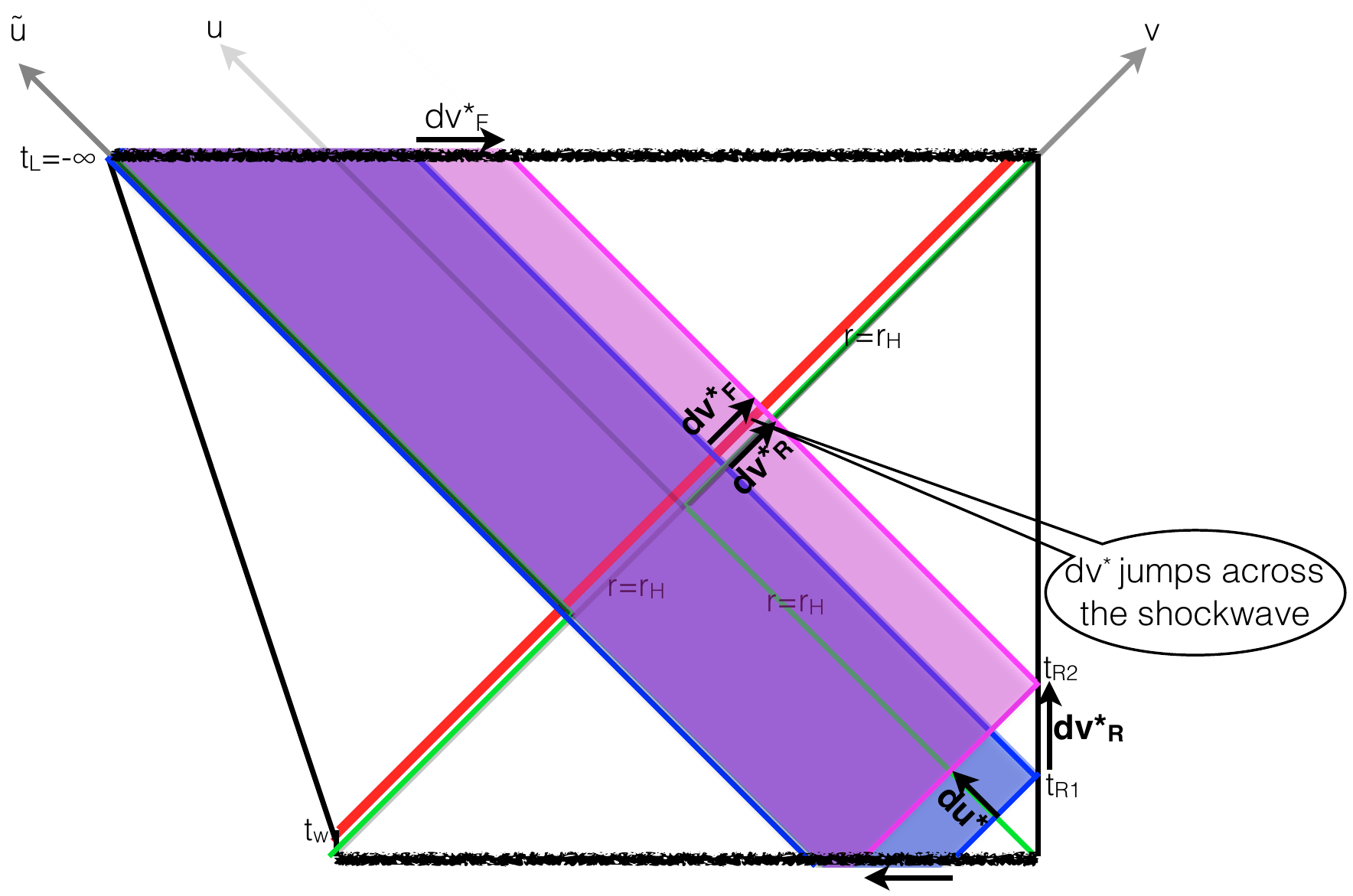}
    \caption{When there is high energy matter, $d v^*$ jumps and the complexity decreases.}
    \label{nonsmooth}
  \end{subfigure}
   \caption{Diagnose the smoothness of right future horizon. Set $t_L$ at the top and scan the interior by increasing Einddington-Finkelstein coordinates $v^*=$ constant.}
  \label{detectsmoothnessRU}
  \end{center}
\end{figure}

There is one technical point here. There are four horizons in the example we discussed, right / left future and past horizons. If we want to detect the transparency of one particular horizon, we don't want to be in transient period, i.e., we want the boundary of WDW patch to touch the singularity. To achieve this, we send the other time to infinity. For example, if we want to detect the transparency of the right future horizon, we set $t_L=-\infty$, i.e., push the left time up to the upper corner. See Figure \ref{detectsmoothnessRU}. As we increase the right time, a linear increase in complexity means the part just scanned has a good interior (Figure \ref{smooth}). If the complexity decreases somewhere, it means there is something wrong in that part of the interior (Figure \ref{nonsmooth}). 

\begin{figure}[H]  
      \includegraphics[scale=0.405]{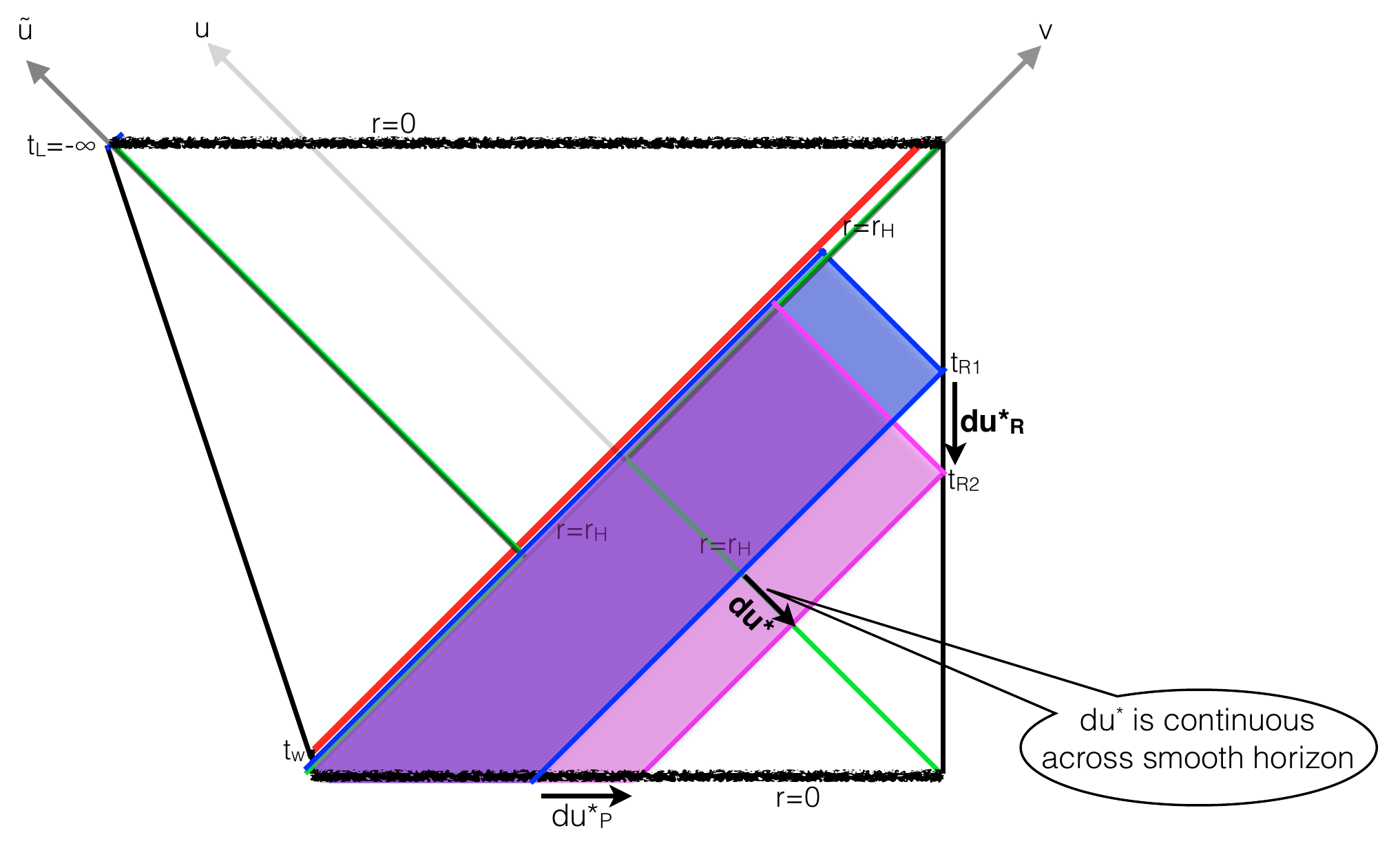}
  \caption{Diagnose the smoothness of right past horizon by setting $t_L$ at the bottom and scan the interior by Eddington-Finkelstein coordinates $u^*=$ constant.}
  \label{detectsmoothnessRP}
\end{figure}

If we want to detect the transparency of the right past horizon, we instead put the left time at the lower corner ($t_L=\infty$), and scan the interior with Eddington-Finkelstein time $u^*$= constant, see Figure \ref{detectsmoothnessRP}.

Similarly, one can detect the smoothness of the left horizons by looking at the dependence of complexity on the left time. If we want to detect the upper left horizon, we set $t_R=\infty$. If we want to look at the lower left horizon, we set $t_R=-\infty$.

In the above examples of two-sided black holes, we anchored time of the other side at infinity to avoid transient period. But for black holes with wide enough Penrose diagrams, this is unnecessary (Figure \ref{sphericaltwoshocktime1}, \ref{sphericaltwoshocktime2}). It's certainly unnecessary for one-sided black holes. In those cases, if we move time up, we are detecting the transparency of the upper horizon, while if we move time down, we are detecting the transparency of the lower horizon. Then a natural conjecture is, for a state which is not maximally complex and has both future and past horizons, at a fixed time you cannot have firewalls on both of them. This is true in the examples we considered here. It's also natural from complexity considerations. The interiors store the quantum circuit making the state. Firewalls lock them up. But the complexity always changes before it reaches the maximum, so there must always be some horizon free of firewalls at work. More discussions on quantum circuit and firewalls are in section \ref{TapeLocking}.\\

To further study the connection between the time dependence of complexity and the smoothness of horizon, we can see look at the collision energy experienced by an infalling thermal quantum. 

Assume there is a homogeneous perturbation, as in section \ref{homogeneousperturbation}, Figure \ref{sphericaloneshock}. Let's say we throw in another thermal scale quantum from right boundary at time $t_R$. It meets the matter shell at radius $r_w$. We have
\begin{align*}
\frac{dv^*_F}{dv^*_R}(r_w)=\frac{r_H-r_w}{\delta r_H+r_H-r_w}=\frac{(r_H-r_w)/\delta r_H}{1+(r_H-r_w)/\delta r_H}
\end{align*}
where for a thermal scale quantum, $\delta r_H\sim\frac{r_H}{S}\sim \frac{l_p}{S^{(D-3)/(D-2)}}$ which is sub-Planck scale. 
On the other hand, the collision energy will be
\begin{align}
\label{collisionenergydilation}
 \frac{E_c^2}{m_p^2} &=\frac{4l_p^2 E_1E_2}{f(r_w)}\sim\frac{l_p^2T}{r_H-r_w}\sim\frac{l_p^2TS}{r_H}\left[\left(\frac{dv^*_F}{dv^*_R}\right)^{-1}-1\right]\nonumber\\
&\sim \frac{r_H}{\beta}S^{\frac{D-4}{D-2}}\left[\left(\frac{dv^*_F}{dv^*_R}\right)^{-1}-1\right]
\propto \left(\frac{dv^*_F}{dv^*_R}\right)^{-1}-1
\end{align}
We see that with significant time dilation: $\frac{dv^*_F}{dv^*_R}\ll 1$, the collision-energy-squared is inversely proportional to the dilation factor and can reach the Planck scale. 

More generally, in appendix \ref{Raychaudhuriequation} we use Raychaudhuri equation to relate the time dilation factors to the stress-energy tensors. Stress-energy tensors cause geodesics to focus. This fact can be translated into time dilation factors which control the time dependence of complexity. 

There is one subtlety here. What do we mean by firewalls? Stress-energy tensors surely signal firewalls. They characterize energy from matter. But gravitons also carry energy. In fact, in presence of localized perturbations, there can be nontrivial time dilation factors at places where the energy momentum tensor $T_{\mu\nu}$ vanishes. Someone who tries to cross the horizon there will still be hit hard, not by matter, but by gravitational shockwaves. We should also consider them as firewalls even though the local stress-energy tensors vanish. To quantitively describe this effect, we can use the Landau-Lifshitz pseudotensor \cite{Landau:1982dva}. 

The Landau-Lifshitz pseudotensor characterizes the energy-momentum flows carried by gravitons. It is only a tensor under linear coordinate transformations, in particular, Lorentz transformations. To describe the experience of an infalling observer, we should evaluate it in the local inertial frame right before the observer crosses the horizon.  In appendix \ref{Landau-LifshitzTensor}, we show that in the geometries we considered in this chapter, the collision energy is always controlled by the time dilation factors as in \eqref{collisionenergydilation}, no matter it's from collisions with matters or with gravitons.

\section{Quantum circuit and the smoothness of horizons}
\label{circuit}

\subsection{Future and past tapes}

\begin{figure}[H]
\begin{center}
\includegraphics[scale=.4]{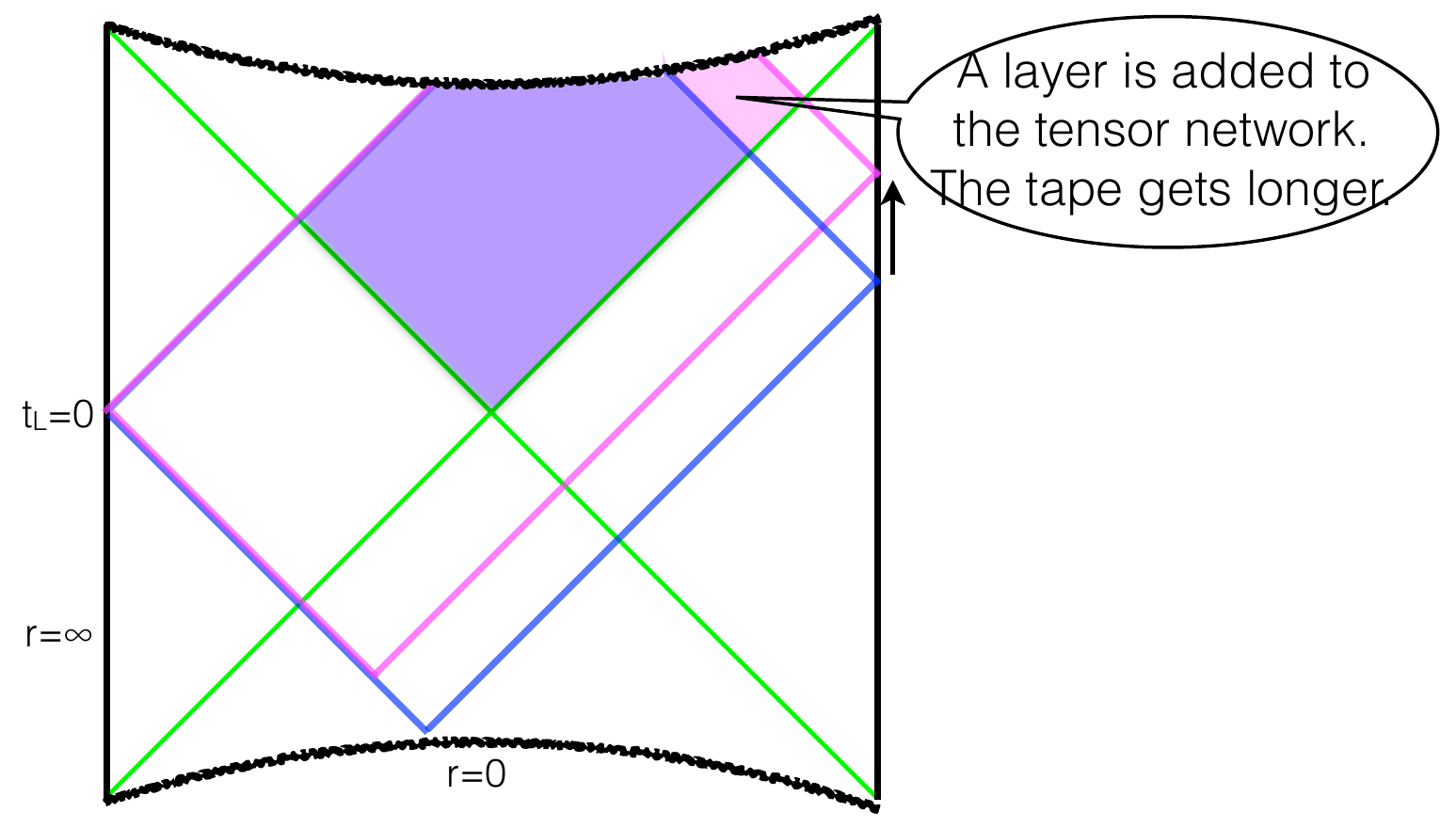}
\caption{The growth of the future tape}
\label{futuretape}
\end{center}
\end{figure}

Tensor networks support spacetime. As pointed out by Hartman and Maldacena \cite{Hartman:2013qma}, we can consider a tensor network laid near the horizon. As time grows, more layers are added to the tensor network, recording the action of the Hamiltonian. It was pointed out by Susskind that the minimal quantum circuit preparing a state is stored in a tape behind the future horizon (Figure \ref{futuretape}) \cite{Susskind:2014moa}. It's called a tape because it keeps a record of the past actions of making the state. This picture is also the basic reason the complexity-volume duality works. For more details, see section 3.3 and Figure 10, 17, 19 in \cite{Susskind:2014moa}. 

In this section, we'll see that, the future and past interiors are like two tapes storing different parts of the minimal circuit preparing the state\footnote{What we mean by future and past interiors should be clear from the context in the following discussions. Mathematically, we can consider the part of spacetime inside the future apparent horizon as the future interior, while the part inside the past apparent horizon as the past interior.}. In the following, we'll sometimes call the future interior the future tape, and the past interior the past tape.

\begin{figure}[H]
\begin{center}
    \begin{subfigure}[b]{0.4\textwidth}
    \begin{center}
    \hspace{0cm}
    \includegraphics[scale=0.45]{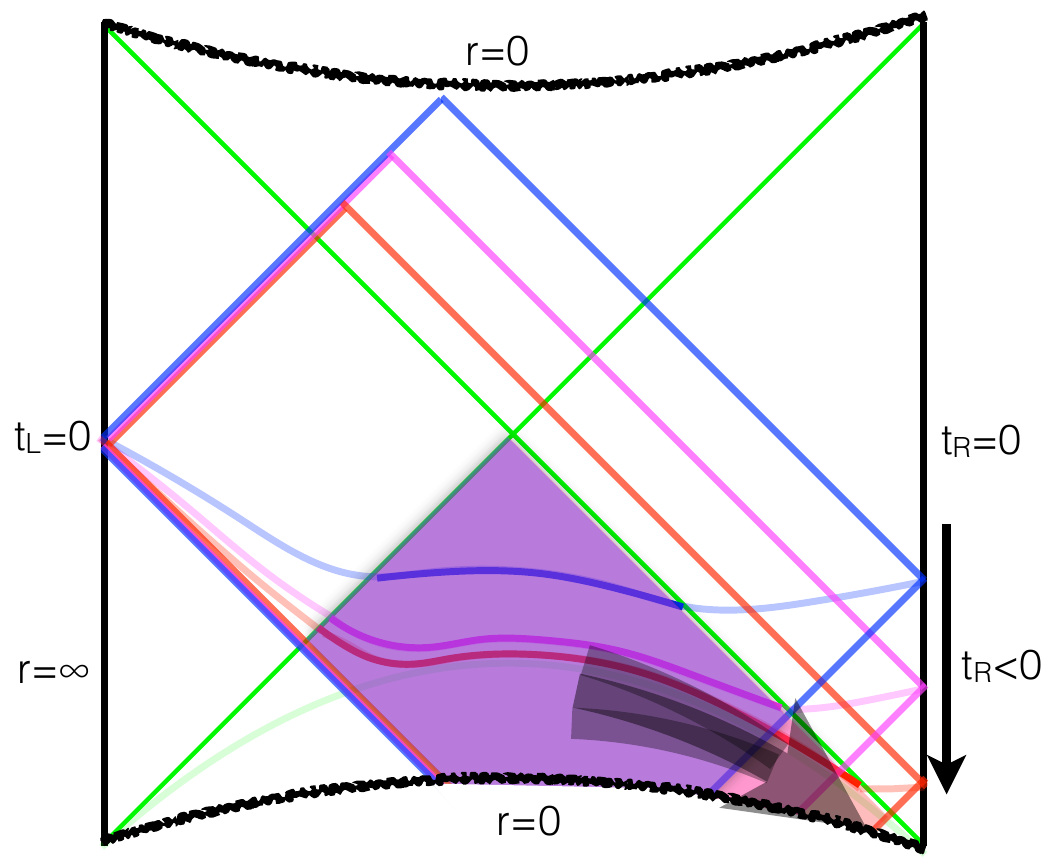}
    \caption{The past tape}
    \label{TwotapeTwosidedBH1}
    \end{center}
  \end{subfigure}
  \hspace{0cm}
  \begin{subfigure}[b]{0.4\textwidth}
  \begin{center}
    \includegraphics[scale=0.45]{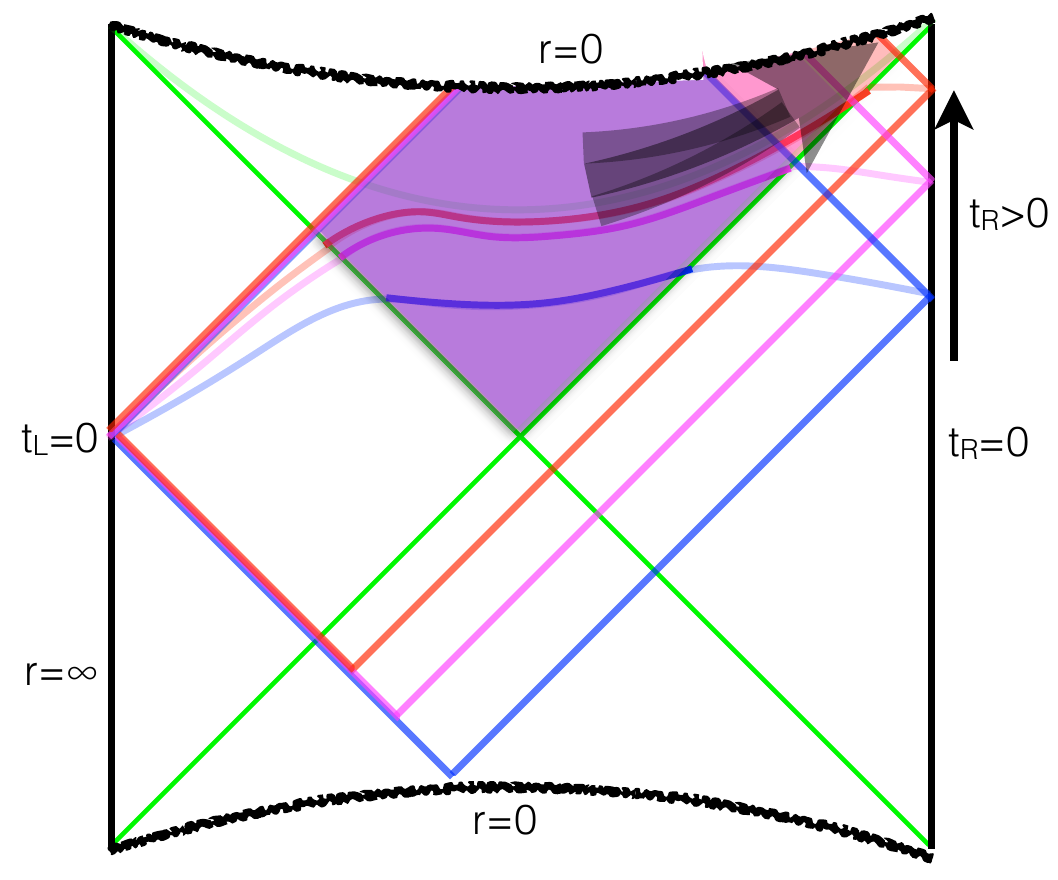}
    \caption{The future tape}
    \label{TwotapeTwosidedBH2}
    \end{center}
  \end{subfigure}
   \caption{The past and future tapes}
  \label{TwotapeTwosidedBH}
  \end{center}
\end{figure}

Here is a simplest example. Consider the time evolution of thermofield double. We fix $t_L=0$, and draw the WDW patches as well as the maximal surfaces at different $t_R$ (Figure \ref{TwotapeTwosidedBH}). When $t_R<0$, the past interior grows to the past (Figure \ref{TwotapeTwosidedBH1}). This is a white hole state, and the minimal circuit preparing this state is stored in the past tape. When $t_R>0$, the future interior expands to the future (Figure \ref{TwotapeTwosidedBH2}). This is a black hole state, and the minimal circuit is stored in the future tape. 

Where are the tapes? There is only AdS distance in the radial direction of the interior. When we talk about tapes, we mean this AdS thickness thing (colored regions in Figure \ref{TwotapeTwosidedBH}), which gets longer along spacelike Schwarzschild time direction. We don't consider the tensor network to be on any specific slice. In complexity-volume duality \cite{Stanford:2014jda} \cite{Roberts:2014isa}, the maximal volume surface is used as a gauge-invariant way to represent the wormhole. As in Figure \ref{TwotapeTwosidedBH}, when the future/past interior expands, the maximal surface also goes through the future/past interior. But it does not mean that particular slice is physically special. In fact, we can't localize the tensor network to distances smaller than AdS scales.

\subsection{Two tape picture: Why are there both future and past horizons?}
\label{twohorizon}

Let's make a more detailed comparison between the picture of the quantum circuit preparing a state and the dual black hole geometry. We'll see that the minimal circuit preparing a state can contain both forward and backward Hamiltonian evolutions (for example, left and right legs in Figure \ref{1precursormincircuit} for a precursor). The forward Hamiltonian evolution part is stored in the future tape, while the backward Hamiltonian evolution part is stored in the past tape. 

To avoid complications from the other side, in this section we'll use a one-sided black hole as an example. We start from an AdS black hole formed from a collapsing shell at $t=0$:

\begin{figure}[H]
\begin{center}
\includegraphics[scale=.4]{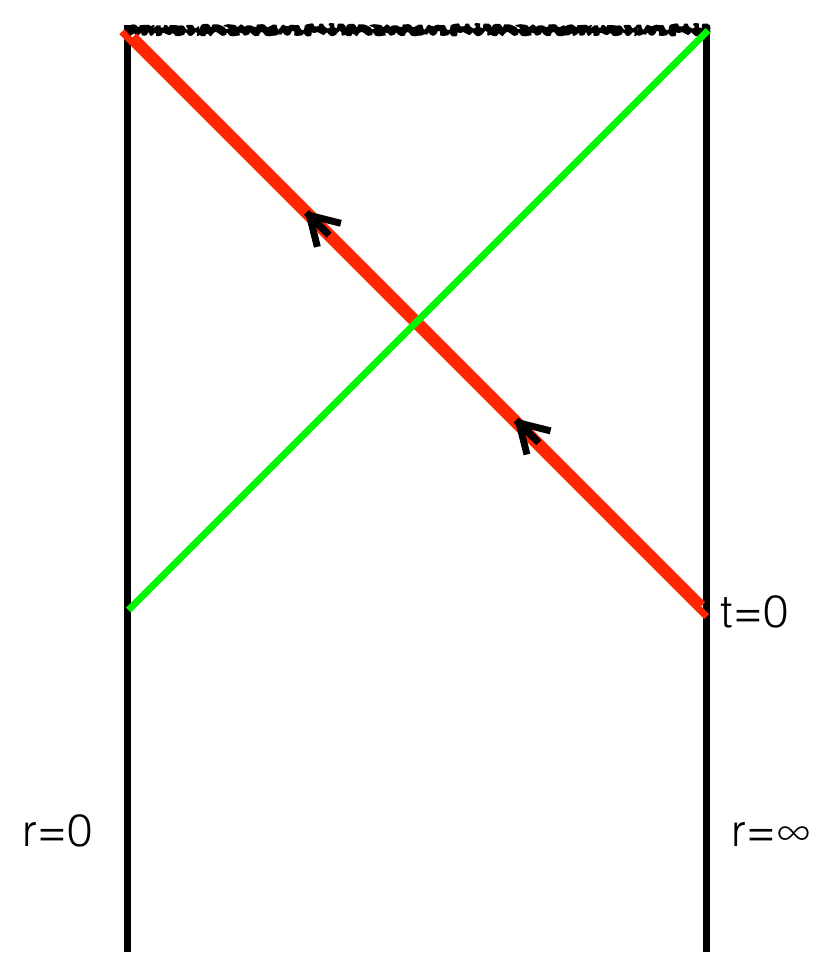}
\caption{A one-sided black hole formed by a collapsing shell.}
\label{onesidedBHunperturbed}
\end{center}
\end{figure}

Now we perturb the state by inserting a precursor: $W(t_w)=e^{iHt_w}We^{-iHt_w}$ with $t_w>t_*$. Figure \ref{1precursor} illustrates such a precursor composed of three parts: forward time evolution $e^{-iHt_w}$, perturbation $W$, and backward time evoluiton $e^{iHt_w}$. There will be cancellations right after the perturbation, and in Figure \ref{1precursormincircuit} the blue line represents the minimal circuit after the cancellations. See \cite{Stanford:2014jda} for more explanations.
\begin{figure}[H]
\begin{center}
    \begin{subfigure}[b]{0.4\textwidth}
    \begin{center}
    \hspace{0cm}
    \includegraphics[scale=0.4]{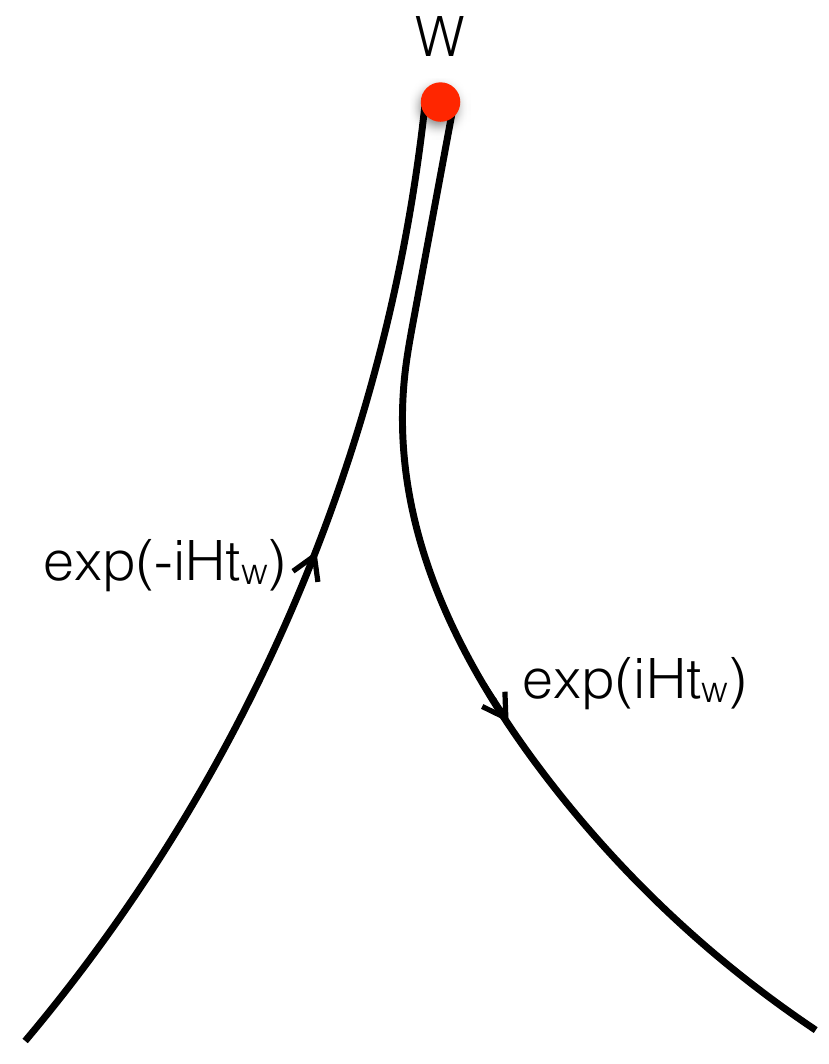}
    \caption{One precursor}
    \label{1precursor}
    \end{center}
  \end{subfigure}
  \hspace{0cm}
  \begin{subfigure}[b]{0.4\textwidth}
  \begin{center}
    \includegraphics[scale=0.4]{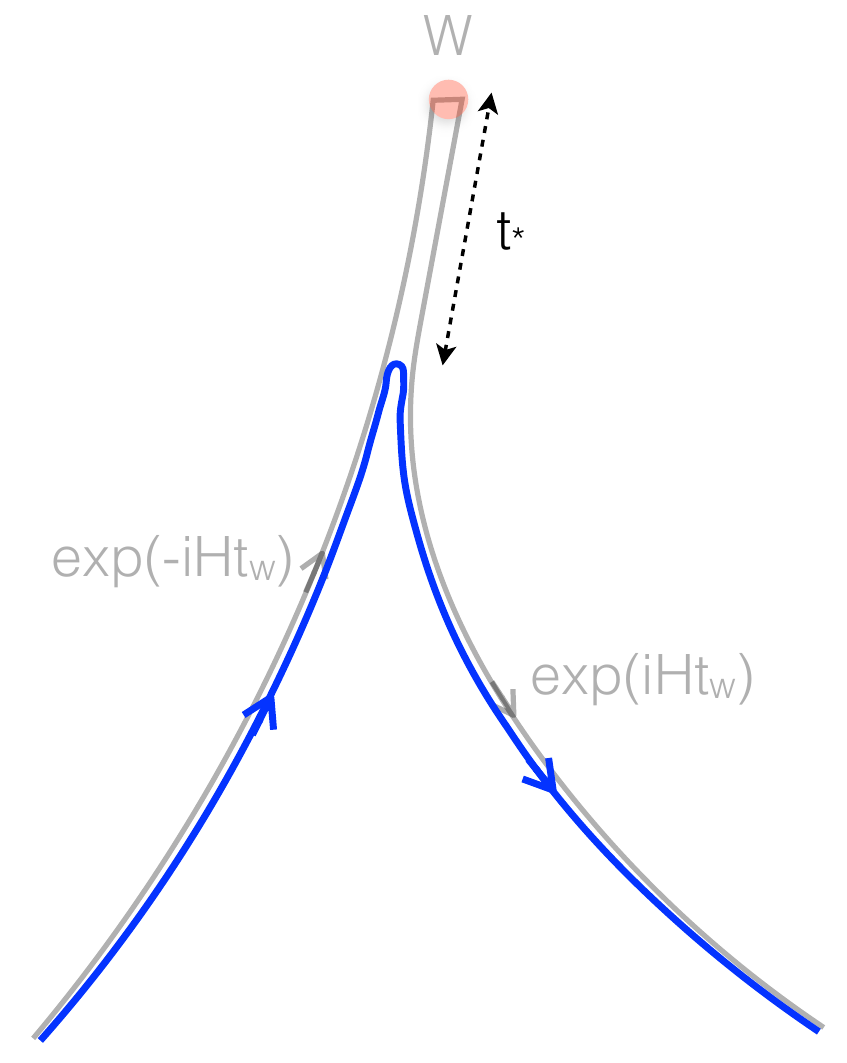}
    \caption{Minimal circuit for one precursor}
    \label{1precursormincircuit}
    \end{center}
  \end{subfigure}
   \caption{One precursor}
  \label{precursorcircuit}
  \end{center}
\end{figure}

With the perturbation, a past singularity will form, and the dual geometry looks like
\begin{figure}[H]
\begin{center}
\includegraphics[scale=.4]{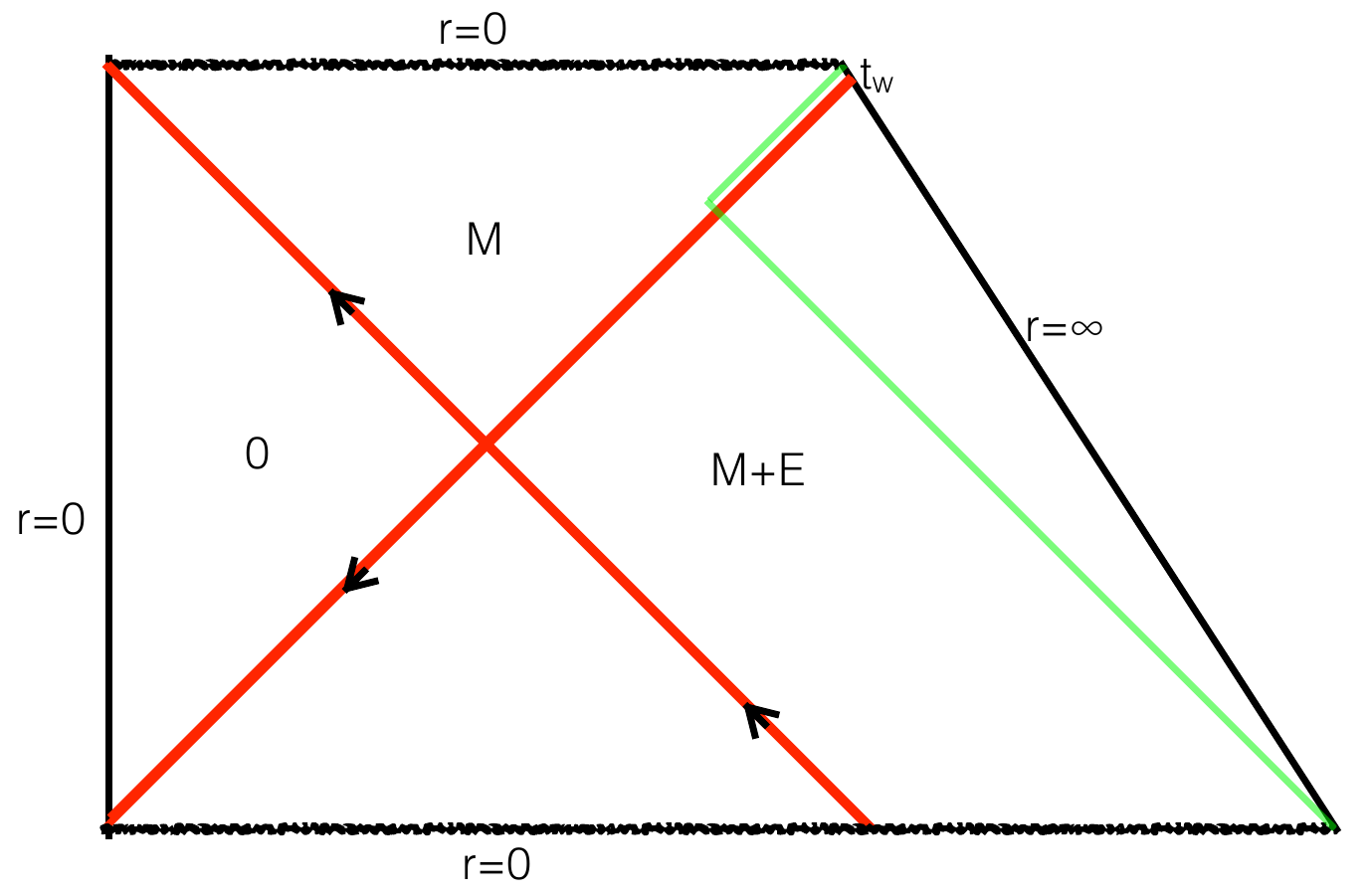}
\caption{A one-sided black hole perturbed by a precursor}
\label{onesidedBHperturbed}
\end{center}
\end{figure}

\begin{figure}[H]
\begin{center}
\hspace{-1cm}
 \begin{subfigure}[b]{0.4\textwidth}
    \begin{center}
    \includegraphics[scale=0.4]{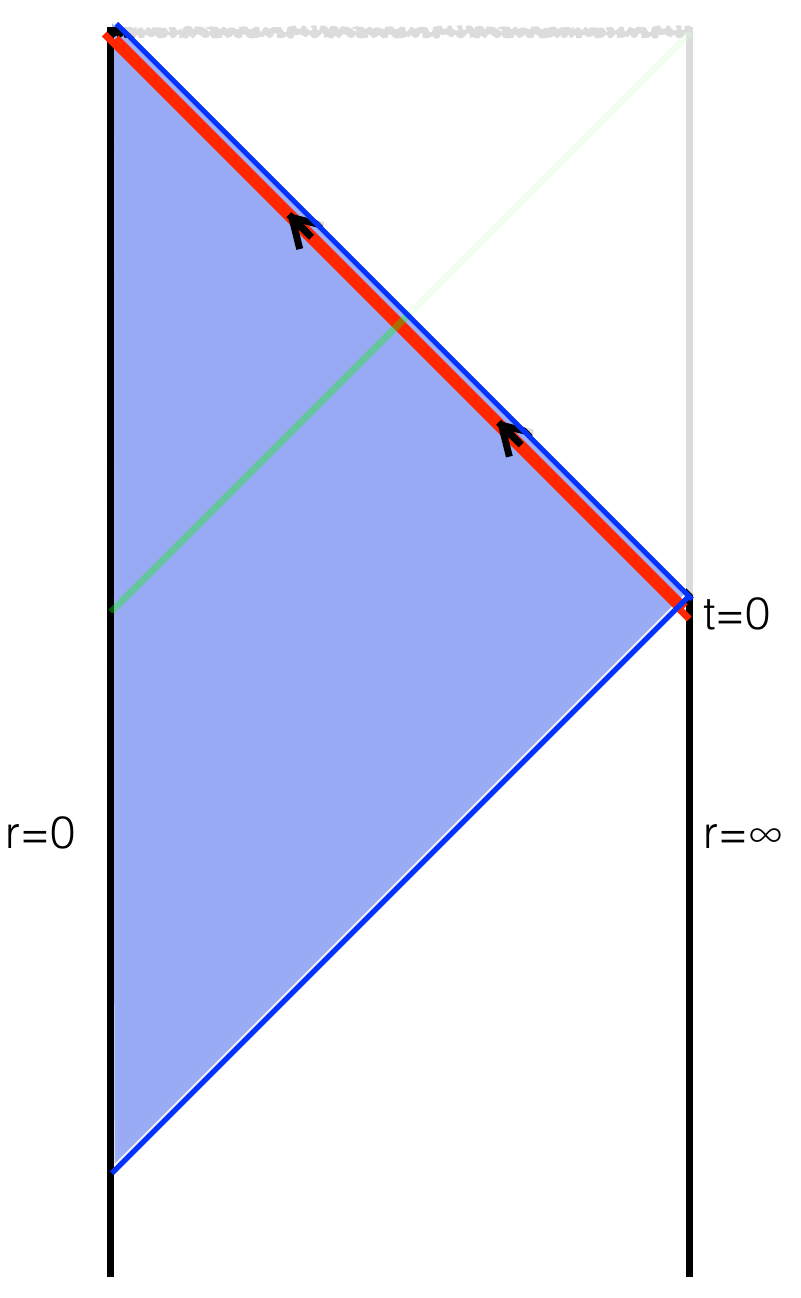}
    \caption{}
    \label{onesidedphase1a}
    \end{center}
  \end{subfigure}
  \hspace{-1cm}
  \begin{subfigure}[b]{0.3\textwidth}
  \begin{center}
    \includegraphics[scale=0.4]{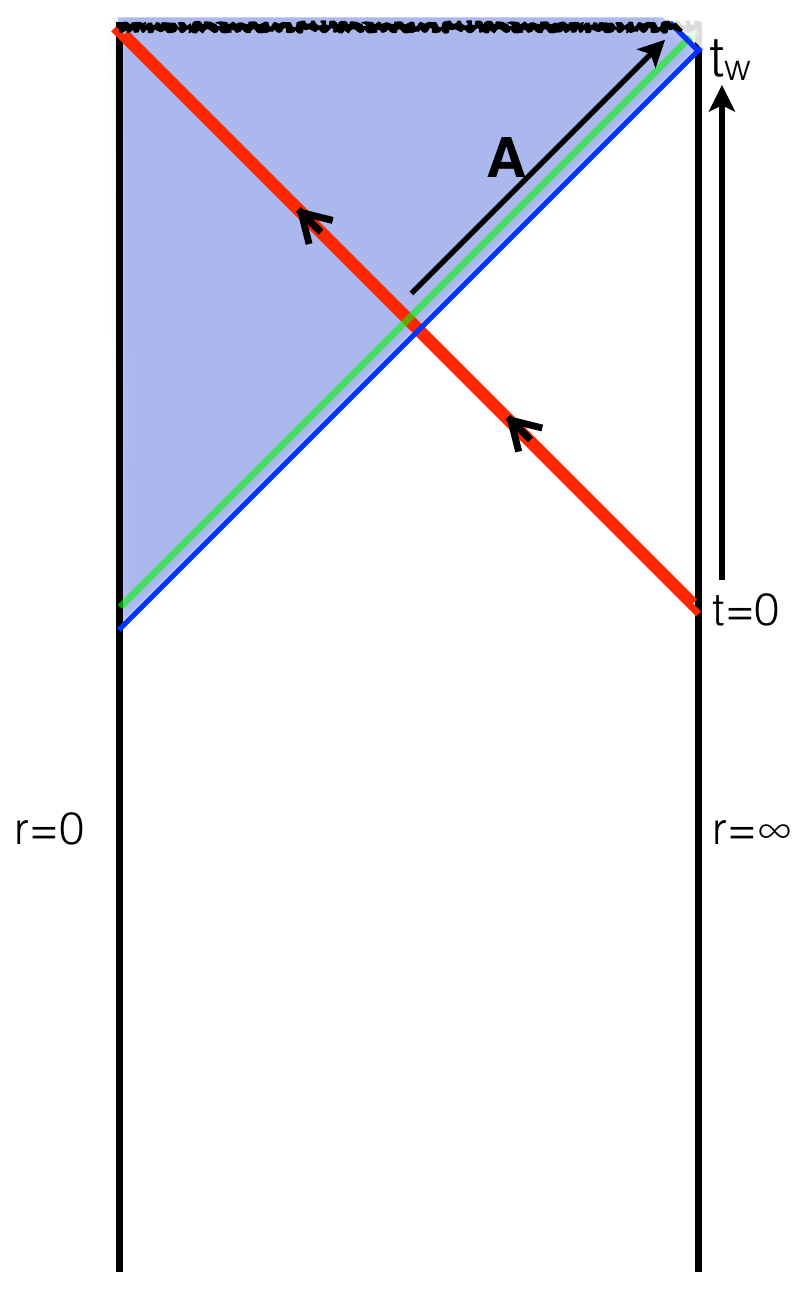}
    \caption{}
    \label{onesidedphase1b}
    \end{center}
  \end{subfigure}
  \hspace{-1cm}
   \begin{subfigure}[b]{0.4\textwidth}
    \begin{center}
        \includegraphics[scale=0.4]{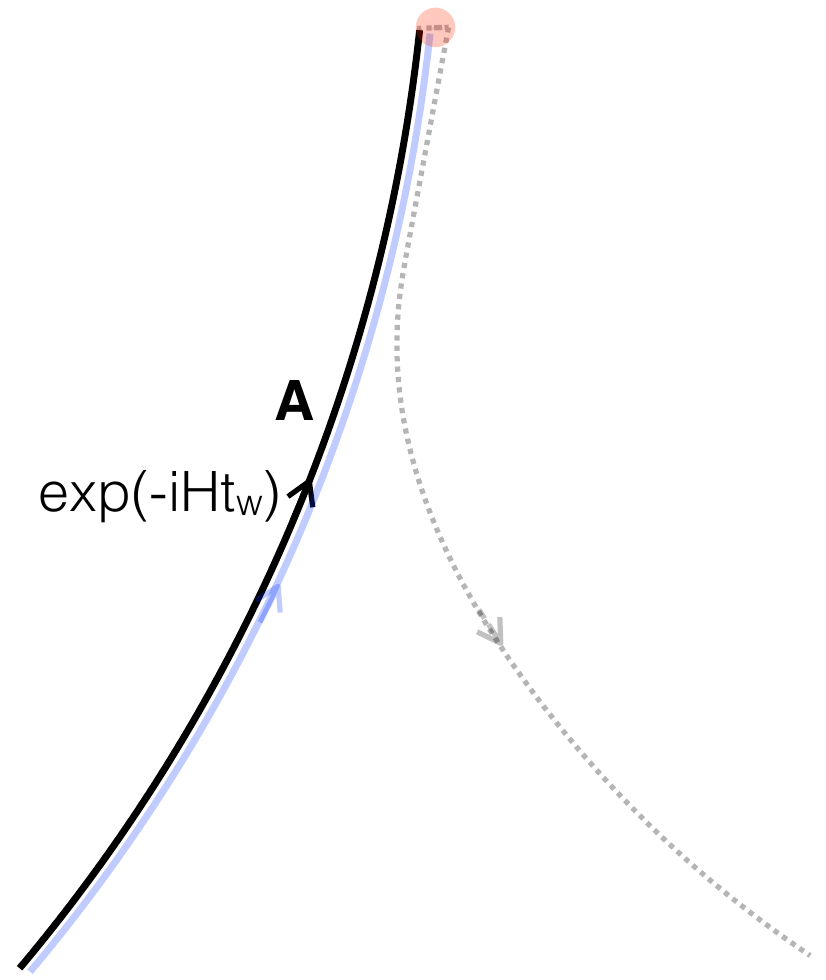}
    \caption{}
    \label{onesidedphase1c}
    \end{center}
  \end{subfigure}
   \caption{A black hole forms}
  \label{}
  \end{center}
\end{figure}

To match the quantum circuit picture with the black hole geometry, we start at $t=0$ when the black hole just formed (Figure \ref{onesidedphase1a}), and time evolve to $t=t_w$ (Figure \ref{onesidedphase1b}). The minimal circuit of this forward time evolution is shown as A in Figure \ref{onesidedphase1c}. It is stored in the future interior (A in Figure \ref{onesidedphase1b}). In this regime, the white hole has not formed yet, and the future tape is working. 

\begin{figure}[H]
\begin{center}
    \begin{subfigure}[b]{0.4\textwidth}
    \begin{center}
    \hspace{0cm}
    \includegraphics[scale=0.4]{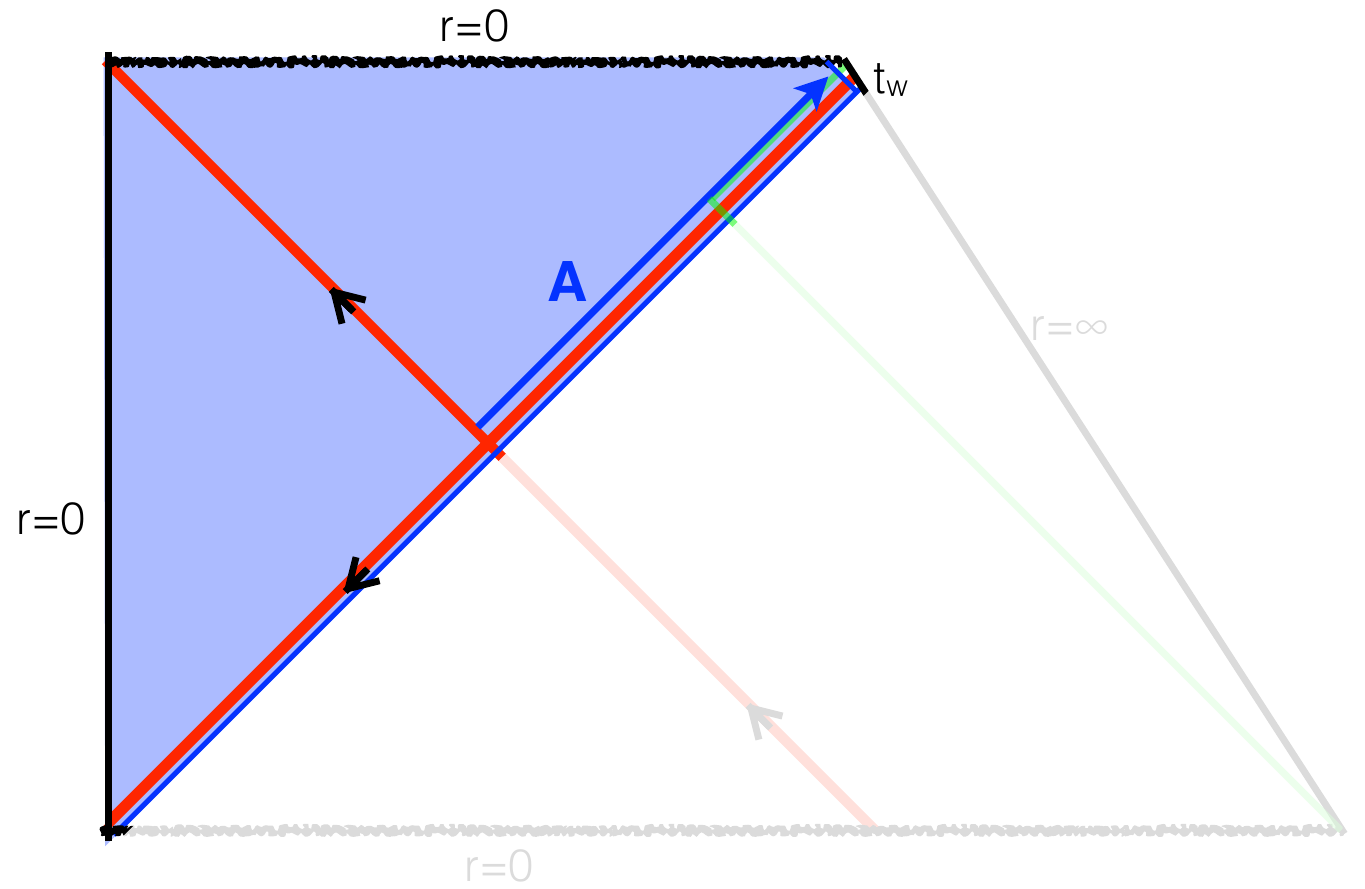}
    \caption{Bulk geometry right after the perturbation}
    \label{onesidedphase2aa}
    \end{center}
  \end{subfigure}
  \hspace{0cm}
  \begin{subfigure}[b]{0.4\textwidth}
  \begin{center}
    \includegraphics[scale=0.4]{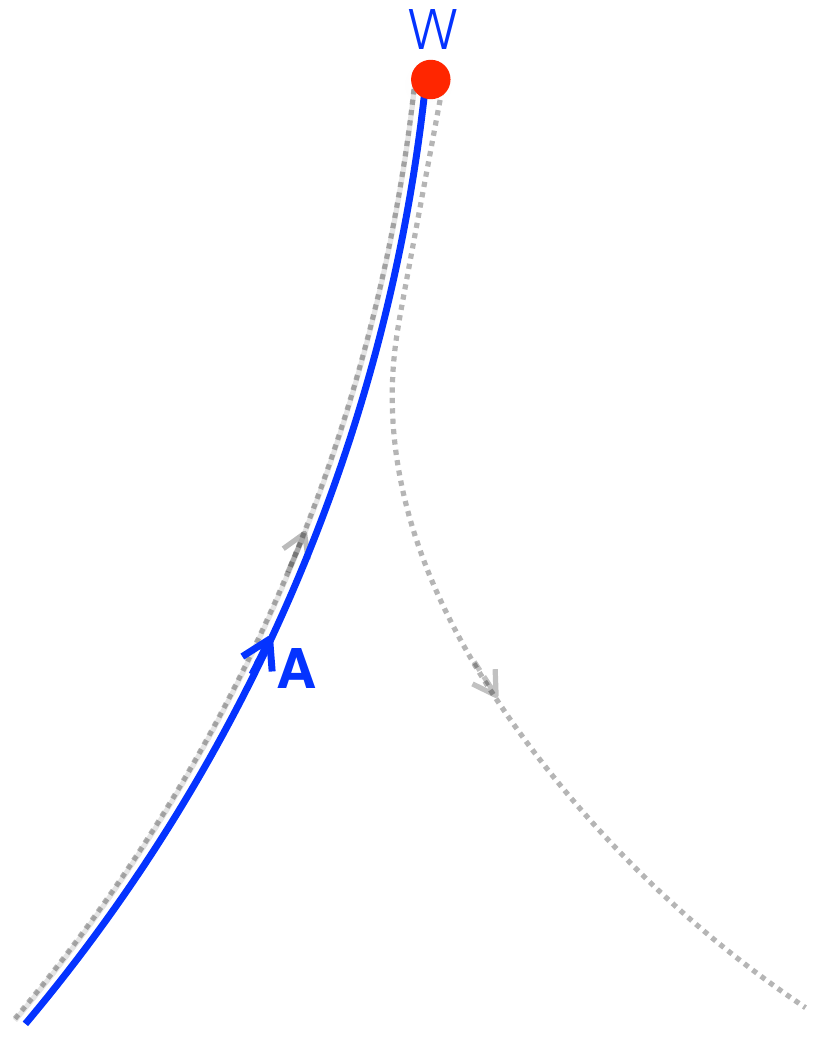}
    \caption{Quantum circuit right after the perturbation}
    \label{onesidedphase2ab}
    \end{center}
  \end{subfigure}
   \caption{A white hole forms}
  \label{onesidedphase2a}
  \end{center}
\end{figure}

Next, we perturb the state by $W$ at $t=t_w>t_*$ and then evolve backward. We start right after the perturbation at $t=t_w$ (Figure \ref{onesidedphase2aa}), where the minimal circuit A (Figure \ref{onesidedphase2ab}) is stored in the future interior (A in Figure \ref{onesidedphase2aa}).  A white hole forms.

\begin{figure}[H]
\begin{center}
    \begin{subfigure}[b]{0.4\textwidth}
    \begin{center}
    \hspace{0cm}
    \includegraphics[scale=0.45]{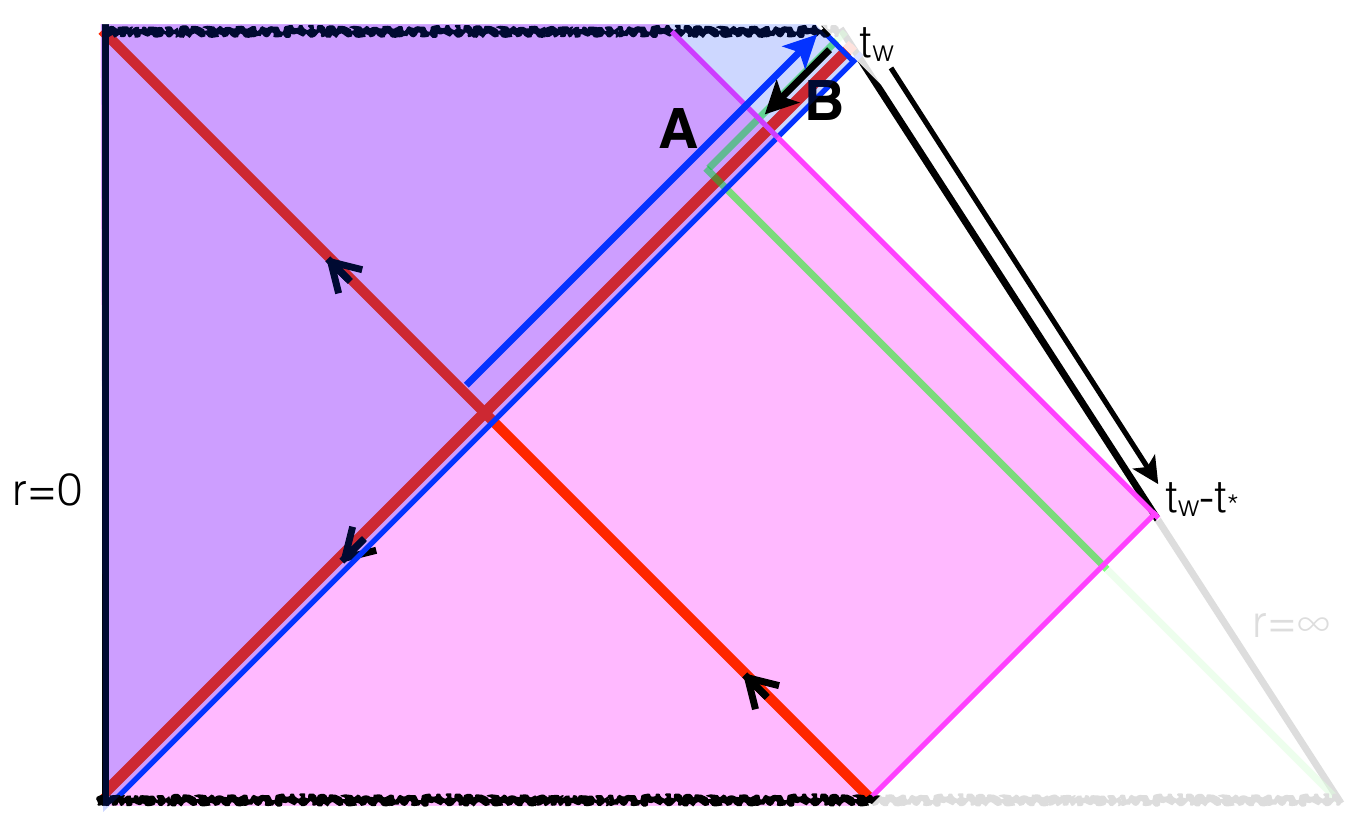}
    \caption{}
    \label{onesidedphase2ba}
    \end{center}
  \end{subfigure}
  \hspace{2cm}
  \begin{subfigure}[b]{0.4\textwidth}
  \begin{center}
    \includegraphics[scale=0.45]{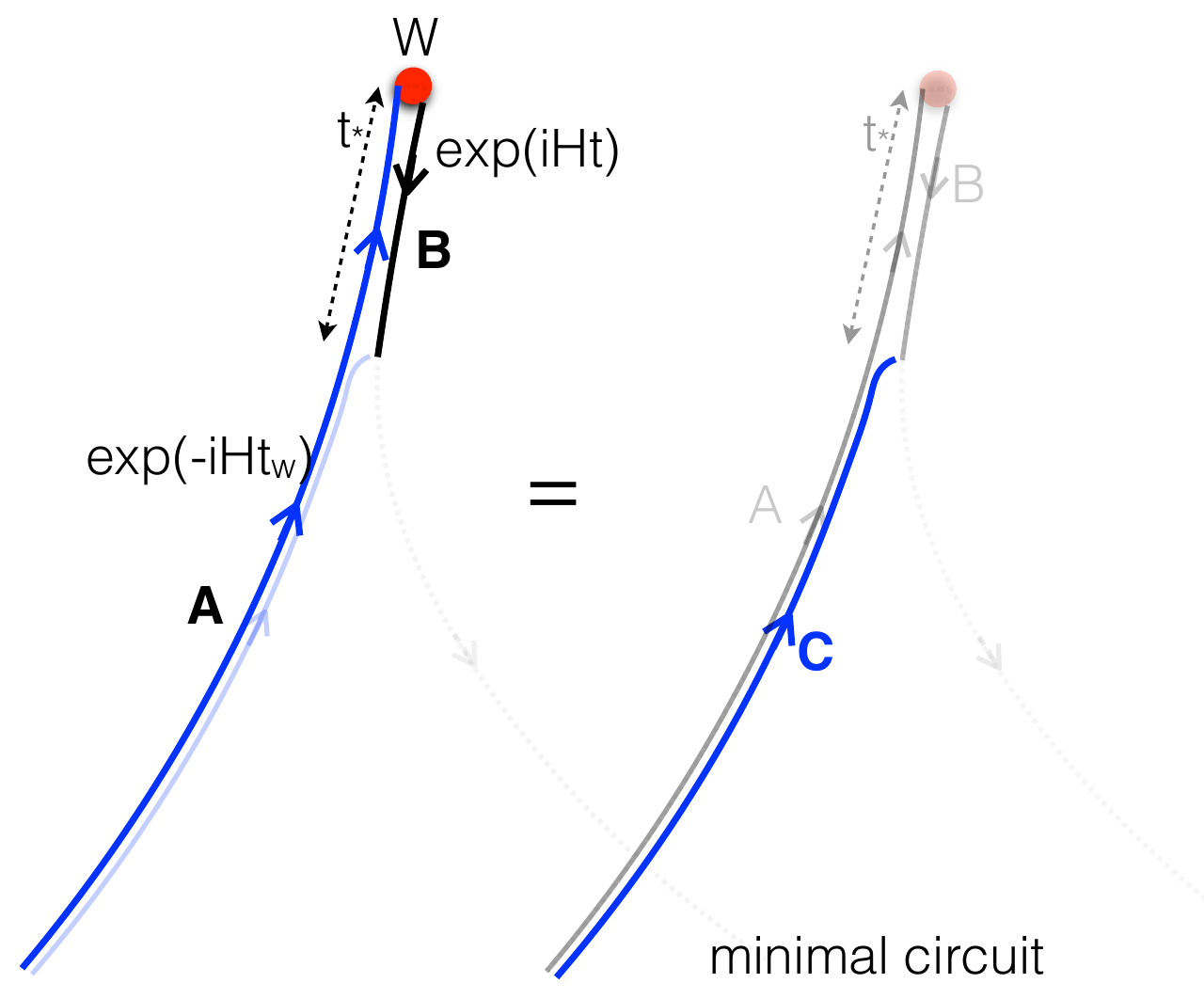}
    \caption{}
    \label{onesidedphase2bb}
    \end{center}
  \end{subfigure}
   \caption{Making the state by forward Hamiltonian evolution}
  \label{onesidedphase2b}
  \end{center}
\end{figure}

We evolve backward, until $t = t_w - t_*$. When $t_w-t<t_*$, the backward time evolution will cancel part of earlier forward time evolution (In Figure \ref{onesidedphase2bb}, the circuit in B cancels part of circuit in A, resulting in minimal circuit C) \cite{Stanford:2014jda}. As a result, the minimal circuit gets smaller, and still only contains forward Hamiltonian evolution (C in Figure \ref{onesidedphase2bb}). On the bulk side, the future interior shrinks (Figure \ref{onesidedphase2ba}). We ignore the change in the past interior, since the volume inside the post collision region is small. 
So we see that the minimal circuit making the state at this time ($t=t_w-t_*$), which contains only forward Hamiltonian evolution, is still stored in the future interior.



\begin{figure}[H]
\begin{center}
    \begin{subfigure}[b]{0.4\textwidth}
    \begin{center}
    \hspace{0cm}
    \includegraphics[scale=0.45]{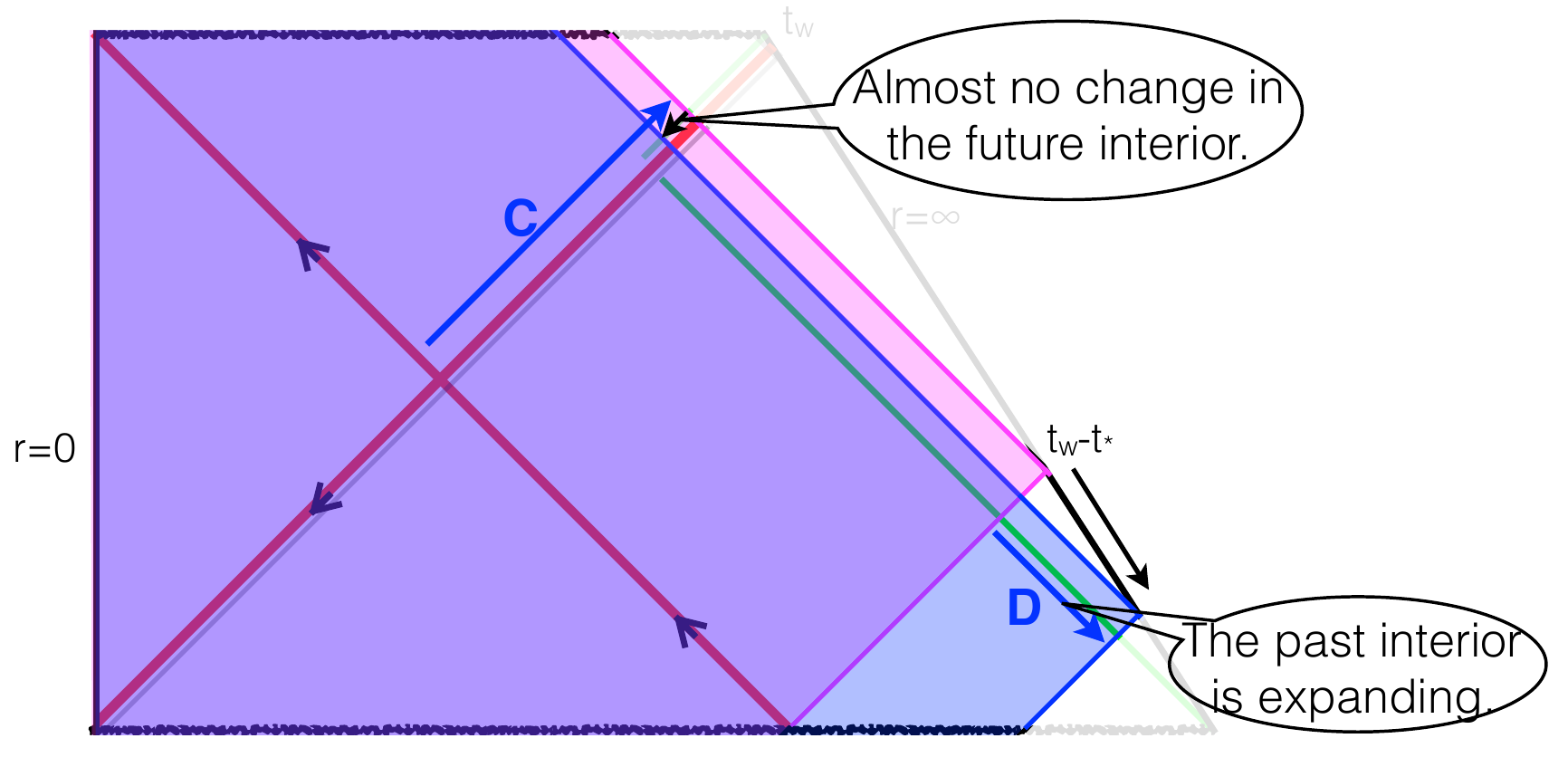}
    \caption{}
    \label{onesidedphase3ba}
    \end{center}
  \end{subfigure}
  \hspace{1.6cm}
  \begin{subfigure}[b]{0.4\textwidth}
  \begin{center}
    \includegraphics[scale=0.45]{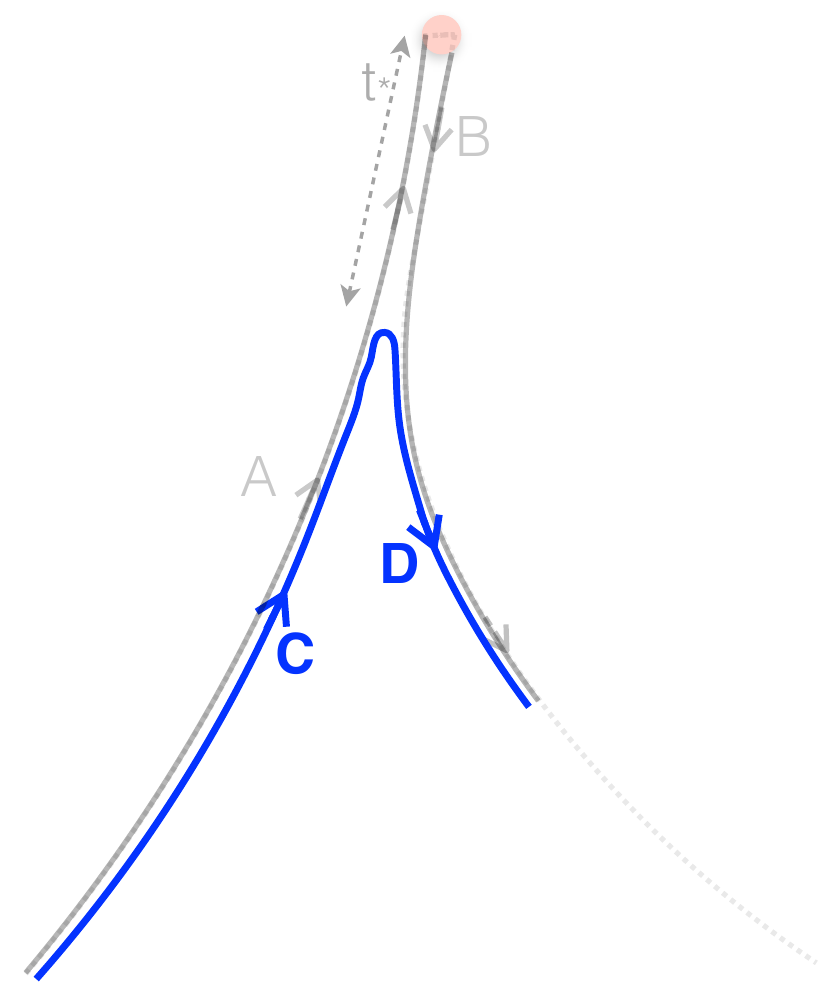}
    \caption{}
    \label{onesidedphase3bb}
    \end{center}
  \end{subfigure}
   \caption{Making the state by backward Hamiltonian evolution}
  \label{onesidedphase3b}
  \end{center}
\end{figure}

Next, we further decrease time $t$. By now ($t<t_w-t_*$) the perturbation has spread to the entire system and there are no more cancellations between the backward and forward Hamiltonian evolutions. The minimal circuit is growing from backward Hamiltonian evolution (D in Figure \ref{onesidedphase3bb}). On the bulk side, the future interior does not change because there is significant time dilation across the firewall (Figure \ref{onesidedphase3ba}). The past interior is expanding, and the circuit from backward Hamiltonian evolution is stored there (D in Figure \ref{onesidedphase3ba}). 

\begin{figure}[H]
\begin{center}
    \begin{subfigure}[b]{0.4\textwidth}
    \begin{center}
    \hspace{0cm}
    \includegraphics[scale=0.45]{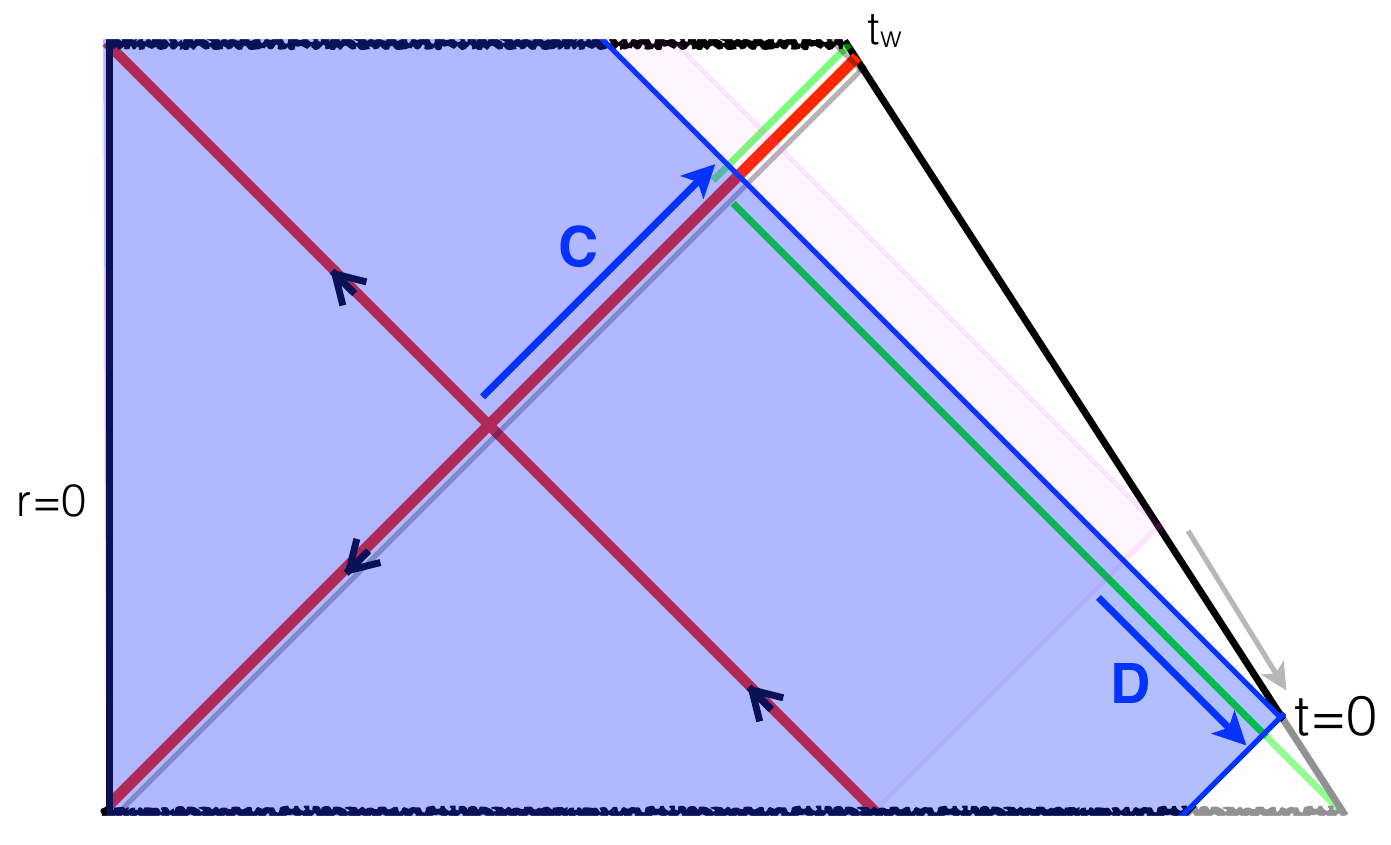}
    \caption{}
    \label{onesidedphase3ca}
    \end{center}
  \end{subfigure}
  \hspace{1.6cm}
  \begin{subfigure}[b]{0.4\textwidth}
  \begin{center}
    \includegraphics[scale=0.45]{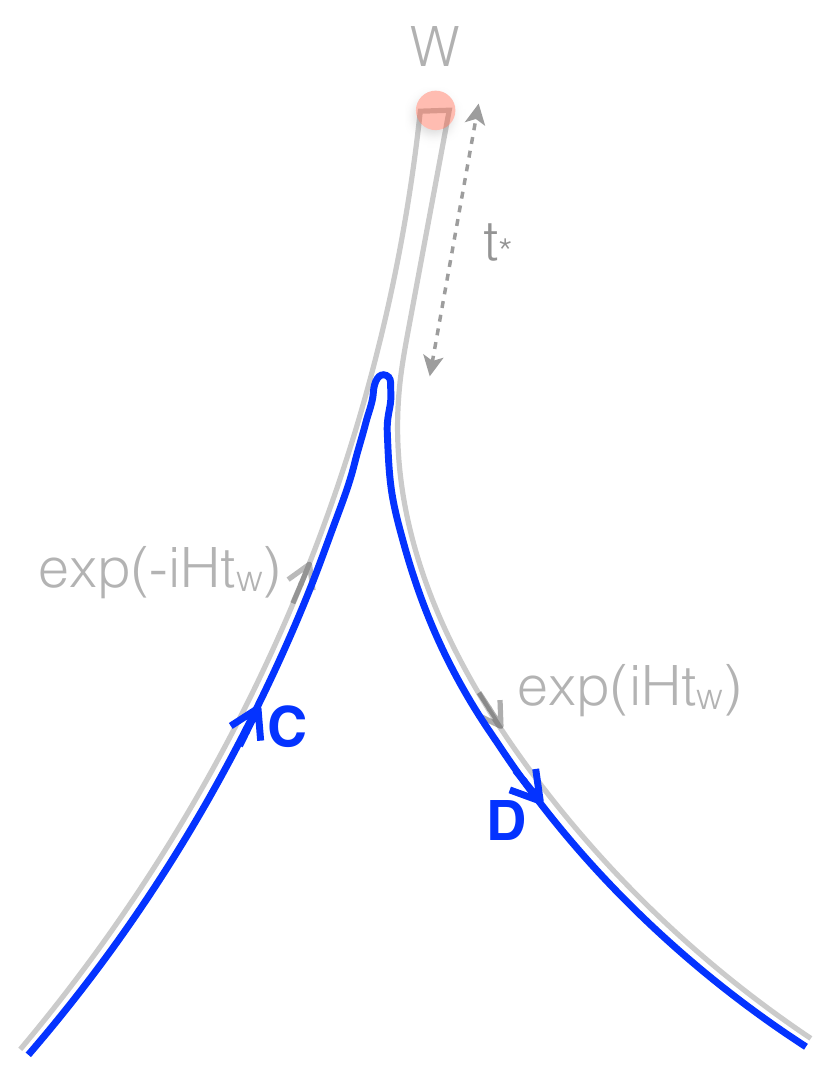}
    \caption{}
    \label{onesidedphase3cb}
    \end{center}
  \end{subfigure}
   \caption{Future and past interiors store different parts of the circuits}
  \label{onesidedphase3c}
  \end{center}
\end{figure}

Eventually, we get to $t=0$, and we have made the state perturbed by the precursor $W(t_w)$. In the minimal circuit making this state, the forward Hamiltonian evolution (C in Figure \ref{onesidedphase3cb}) is stored in the future interior (C in Figure \ref{onesidedphase3ca}), while the backward Hamiltonian evolution (D in Figure \ref{onesidedphase3cb}) is stored in the past interior (D in Figure \ref{onesidedphase3ca}).

We see the roles played by future and past horizons. 
While the stretching of the future interior is responsible for the increase of complexity to the future, to have complexity decrease it seems necessary to have past horizons . Complexity is stored in interiors. The existence of past horizons means the complexity can also increase to the past, and so is a manifestation of time reversibility.


\subsection{Tape working and the smoothness of horizons}
\label{TapeLocking}

There is one confusing point about the relation between the time dependence of complexity and the smoothness of horizons. The time dependence of complexity is not a conventional property of states in general. There are properties of states described by linear operators, like position, momentum, e.t.c.. Entanglement is also a property of states, though may not always be linear \cite{Harlow:2016vwg}. These properties are independent of dynamics. But given a state, without knowing the Hamiltonian, in general one cannot talk about the time dependence of its complexity. If the smoothness of horizon is a state property, then how is it related to the time dependence of complexity? The two-tape picture gives an answer to this question. 

\begin{figure}[H]
\begin{center}
\hspace{-1cm}
    \begin{subfigure}[b]{0.4\textwidth}
    \begin{center}
    \hspace{0cm}
    \includegraphics[scale=0.46]{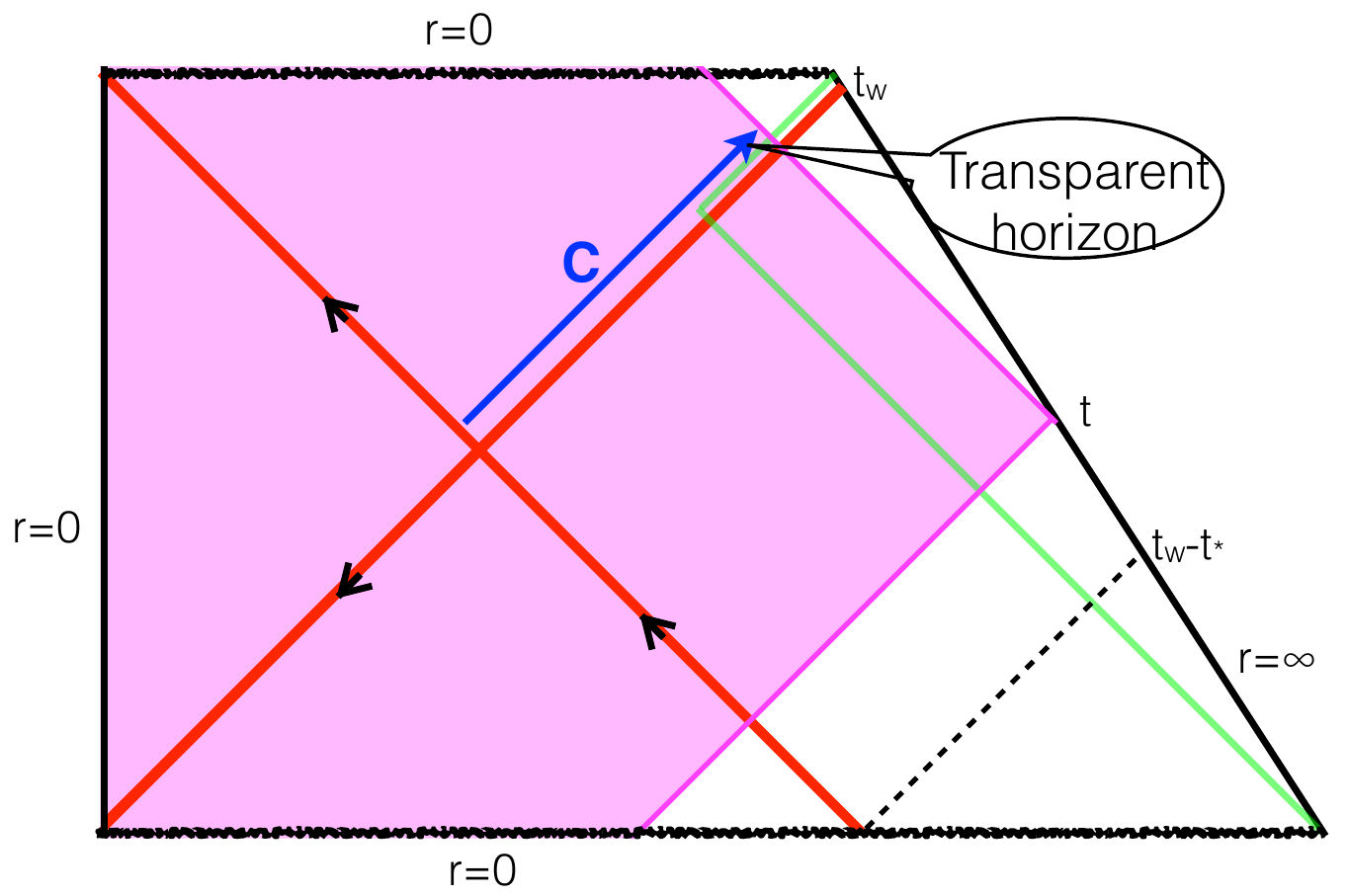}
    \caption{}
    \label{onesidedsmoothBH}
    \end{center}
  \end{subfigure}
  \hspace{1.8cm}
  \begin{subfigure}[b]{0.4\textwidth}
  \begin{center}
    \includegraphics[scale=0.46]{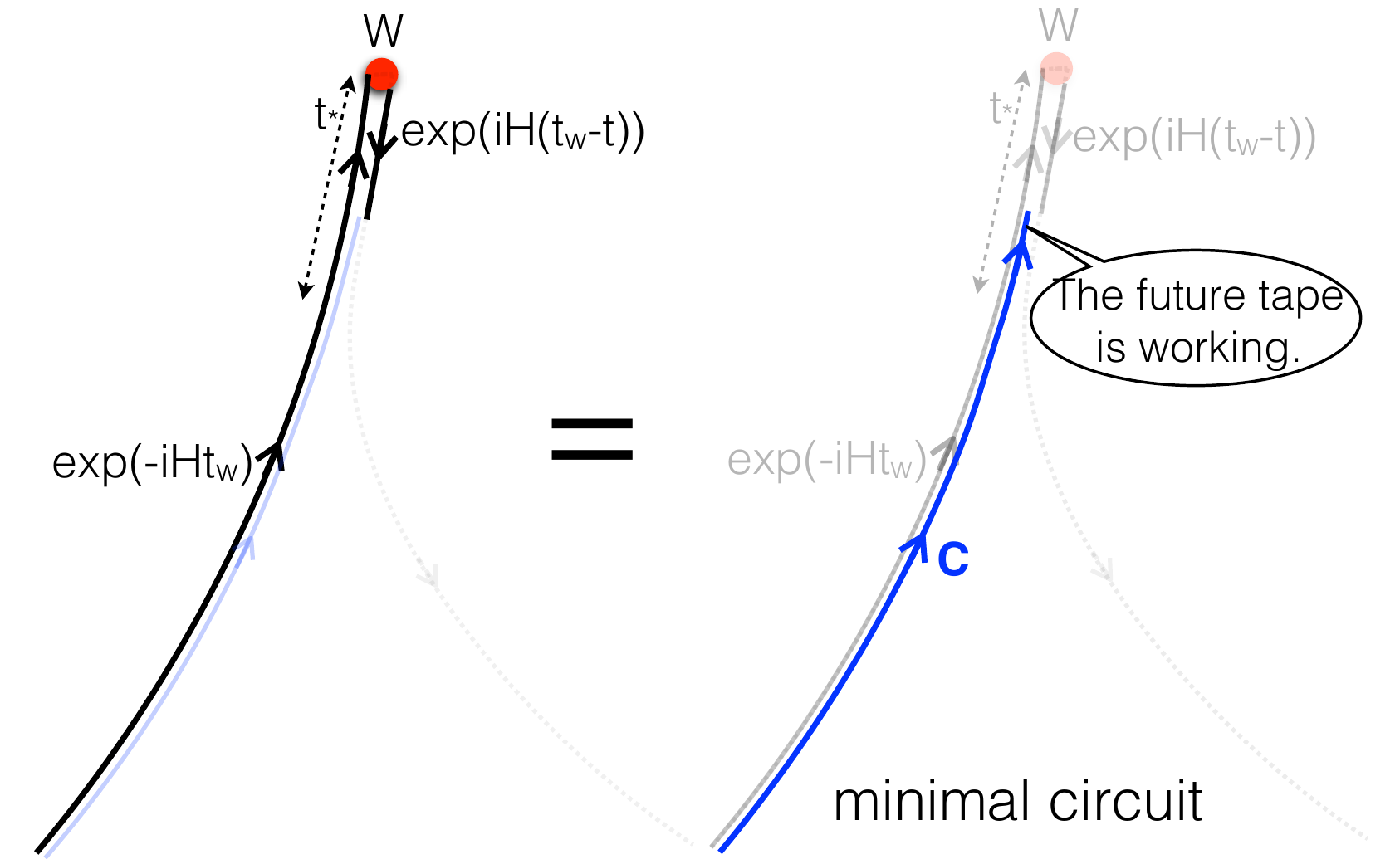}
    \caption{}
    \label{onesidedsmoothcircuit}
    \end{center}
  \end{subfigure}
   \caption{Future tape working}
  \label{onesidedsmooth}
  \end{center}
\end{figure}

Let's look at the perturbed one-sided black hole at a particular time $t>t_w-t_*$ (Figure \ref{onesidedsmoothBH}). We've seen that its complexity increases with time. The perturbation is not too close to the horizon.
The minimal circuit preparing this state is shown as blue line C in Figure \ref{onesidedsmoothcircuit}. At the end of the minimal circuit it is forward Hamiltonian evolution. The quantum gates making the state are being laid on the future tape, so we say the future tape is working. This is why the complexity will increase as we increase the time. Because tensor network supports spacetime, this also implies the future interior is growing.
The future horizon is transparent here. The past tape is not working, and inside the past interior there is high energy collision.\\

If we look at the state at another time $t<t_w-t_*$, the minimal circuit will look completely different:
\begin{figure}[H]
\begin{center}
\hspace{-2cm}
    \begin{subfigure}[b]{0.4\textwidth}
    \begin{center}
    \hspace{0cm}
    \includegraphics[scale=0.46]{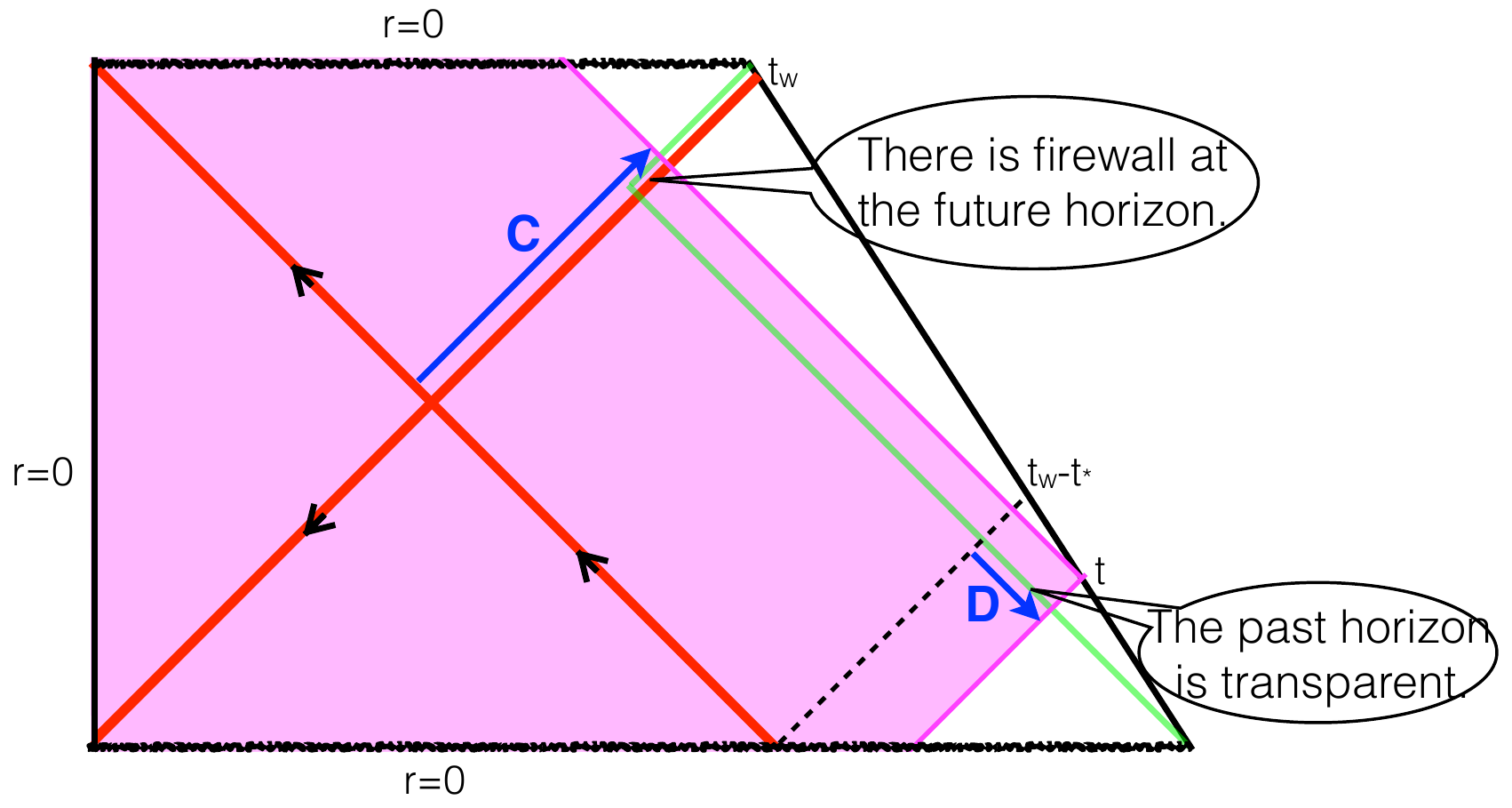}
    \caption{}
    \label{onesidednonsmoothBH}
    \end{center}
  \end{subfigure}
  \hspace{1.8cm}
  \begin{subfigure}[b]{0.4\textwidth}
  \begin{center}
    \includegraphics[scale=0.46]{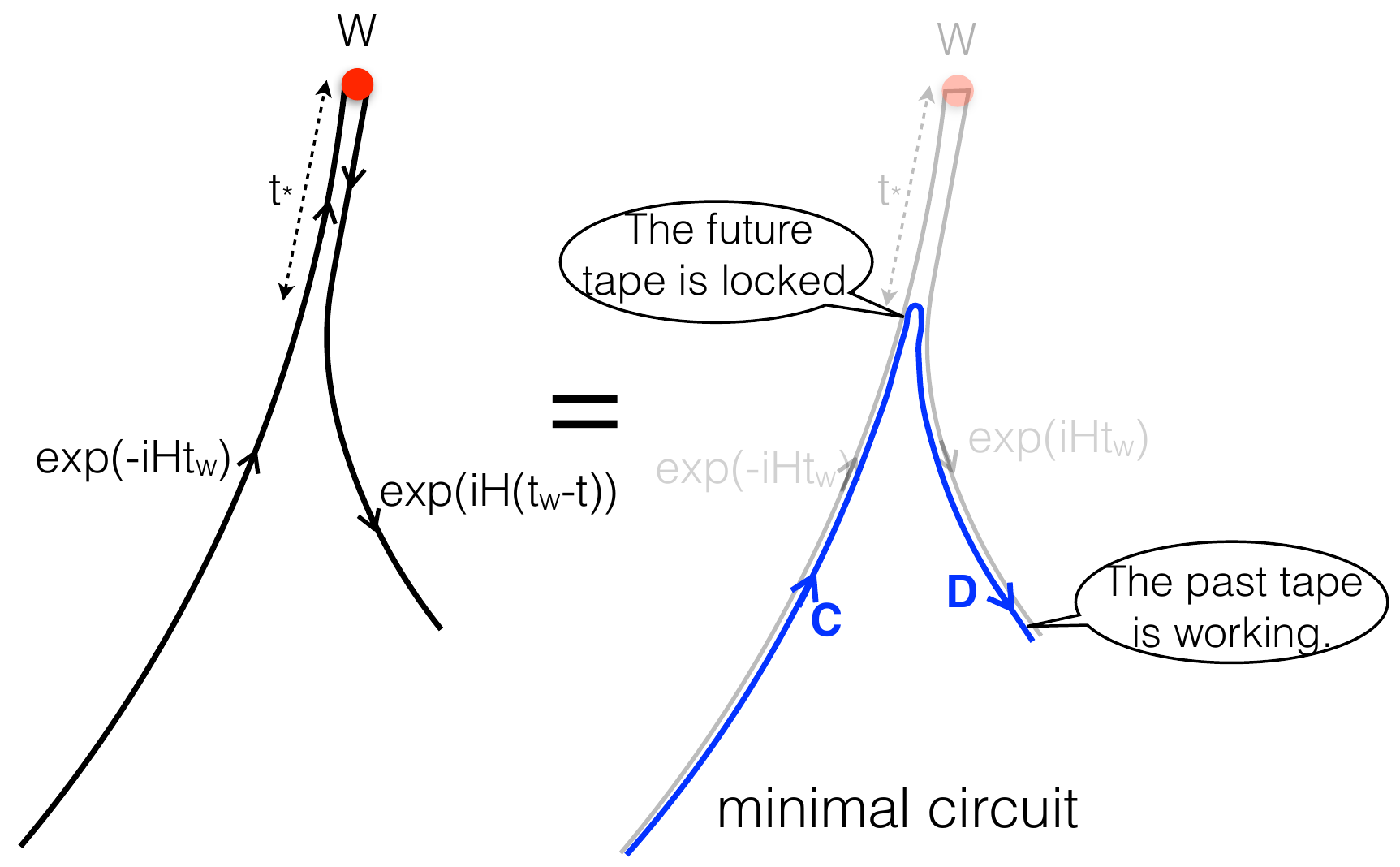}
    \caption{}
    \label{onesidednonsmoothcircuit}
    \end{center}
  \end{subfigure}
   \caption{Past tape working}
  \label{onesidednonsmooth}
  \end{center}
\end{figure}
Now the minimal circuit (blue line in Figure \ref{onesidednonsmoothcircuit}) contains both forward Hamiltonian evolution (C in Figure \ref{onesidednonsmoothcircuit}) and backward Hamiltonian evolution (D in Figure \ref{onesidednonsmoothcircuit}). The forward evolution part is stored in the future tape (C in Figure \ref{onesidednonsmoothBH}), while the backward evolution part is stored in the past tape (D in Figure \ref{onesidednonsmoothBH}). At the end of the minimal circuit it is backward Hamiltonian evolution, i.e., it is the past tape working. No quantum gates are being laid on the future tape at this moment, and we say the future tape is locked. 
From the complexity point of view, we see the complexity decrease because the past tape working means the complexity is increasing towards the past. 


We see that the smoothness of horizons have a simple interpretation in this picture. The future and past interiors are like two tapes storing different parts of the minimal circuit. Each tape can be in two states:  working, when there are quantum gates being laid on it, or locked, when no more gates are being laid on it. When a tape is working, it's getting longer from those newly added quantum gates. So the corresponding interior steadily grows with time. At this moment the interior is open for someone to enter. When a tape is locked, no change happens at the end of it. As a result the corresponding interior ceases to grow. Nothing can fall in at this moment.

\section{Conclusion}

We showed that the time dependence of complexity is closely related to the boost symmetry across horizons.  
The Rindler-like nature of horizons guarantees that the boost symmetry is broken in simple and universal forms, which gives rise to the various robust features of the time dependence of complexity. 


Identifying the black / white hole interiors as tapes storing different parts of the minimal circuit preparing the state, we see the various ideas fit together: the tape working, means the interior growing, the boost symmetry staying good, and transparent horizon. On the other hand, a tape being locked, means the interior stops growing, the boost symmetry being broken, and any infalling object will experience high-energy collision.


It would be interesting to understand this boost symmetry and symmetry breaking in the boundary field theory. Intuitively, without perturbations, there is uniform dynamics. Right after the perturbation, the perturbation is distinguishable and it's not very symmetric. After it is scrambled, locally everywhere is the same, and certain symmetry is regained.  A steady increase of complexity is a kind of time translation symmetry. 

In our analysis we ignored various regions of AdS size and considered them as giving rise to transient behaviors. However, they serve as the ``seeds" for later steady expansion of the interior. On the quantum circuit side, they correspond to the most interesting behavior of switching from one tape to another. But without an understanding of sub-AdS locality, we don't have tools to address them.

Another problem is about linearity of firewalls. If we want to see whether the superposition of two states has a transparent horizon, we need to know something about the minimal circuit preparing the superposition. Unfortunately so far we know very little about the complexity of superpositions without using ancilla qubits.\footnote{We thank Scott Aaronson for explanations on this.} Maybe a more basic question to ask is, on what states can we talk about superpositions? It may not make sense to talk about superpositions of states with large relative complexity. One possibility is that the existence of firewalls behaves like an ordinary property only if we restrict to certain subspace of states, like entanglement entropy as shown by Harlow recently in \cite{Harlow:2016vwg}. Another possibility is that to describe black hole interiors we indeed need some violations of quantum mechanics, but no observer should be able to see it. After all, testing if there exists a firewall or not is not some experiment that can be repeated by a single observer. 


A key assumption we used throughout this paper is the validity of classical general relativity, at least until very late time. This assumption might break down earlier. But it's interesting to see how far we can go with these assumptions. We see that all these ideas: complexity, minimal circuit, and firewalls fit together, at least in the examples we studied. However, we do keep in mind that these examples can be special. Furthermore, it may be a very strong constraint to require that a state determines the minimal circuit making it. To further address this, a better understanding of complexity and minimal circuit is needed.

\section*{Acknowledgements}
I am grateful to my advisor Lenny Susskind for tremendous help in various aspects. I thank Scott Aaronson for explanations on complexity of superposition. 

\appendix

\section{Complexity of a precursor}\label{complexityoneprecursor}

\begin{figure}[H]
 \begin{center}
  \begin{subfigure}[b]{0.3\textwidth}
  \begin{center}
    \includegraphics[scale=0.33]{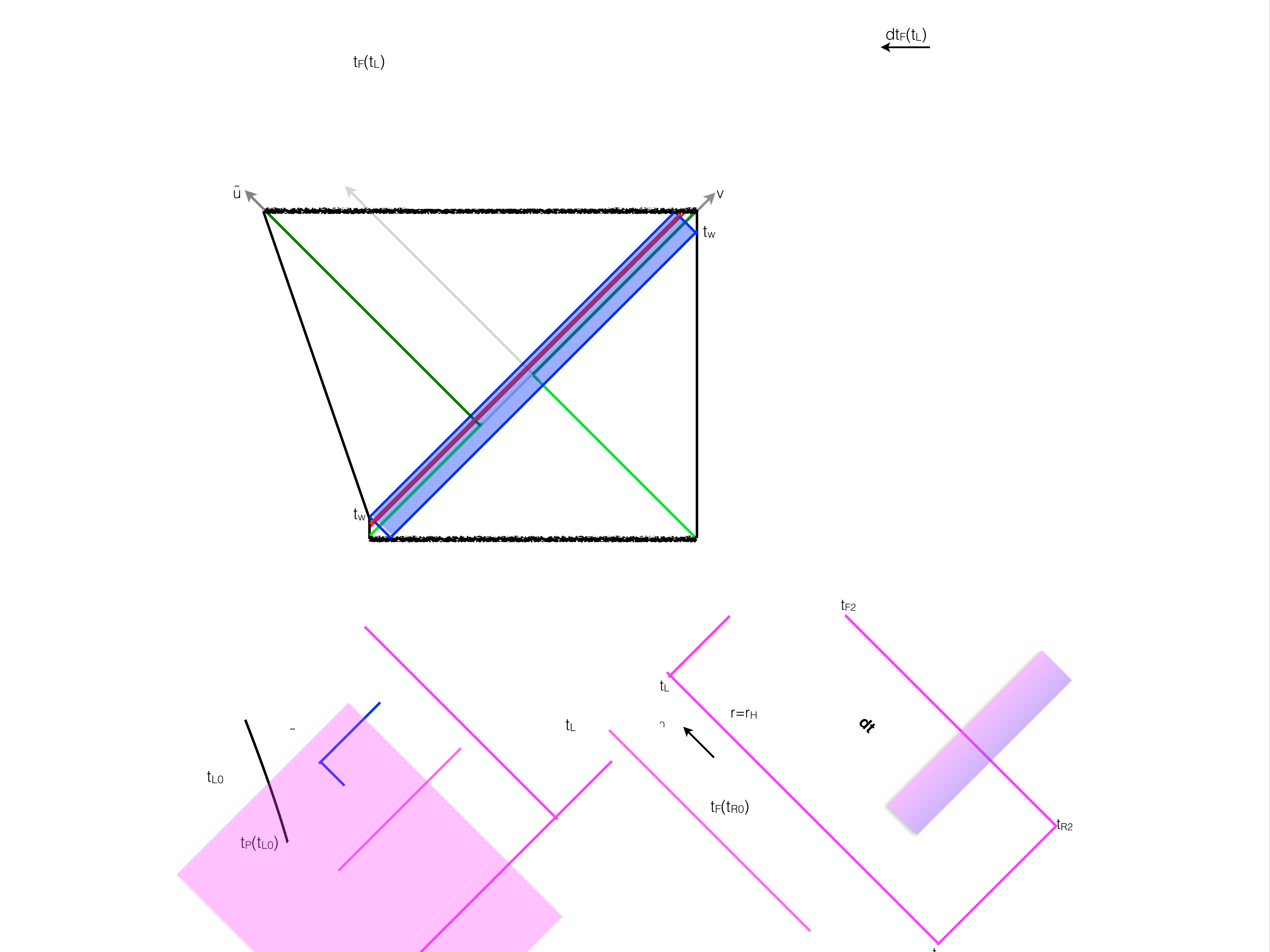}
    \caption{$t_R=t_L=t_w$}
    \label{oneprecursorpicturea}
    \end{center}
  \end{subfigure}
  \hspace{-0.64cm}
  \begin{subfigure}[b]{0.4\textwidth}
  \begin{center}
    \includegraphics[scale=0.33]{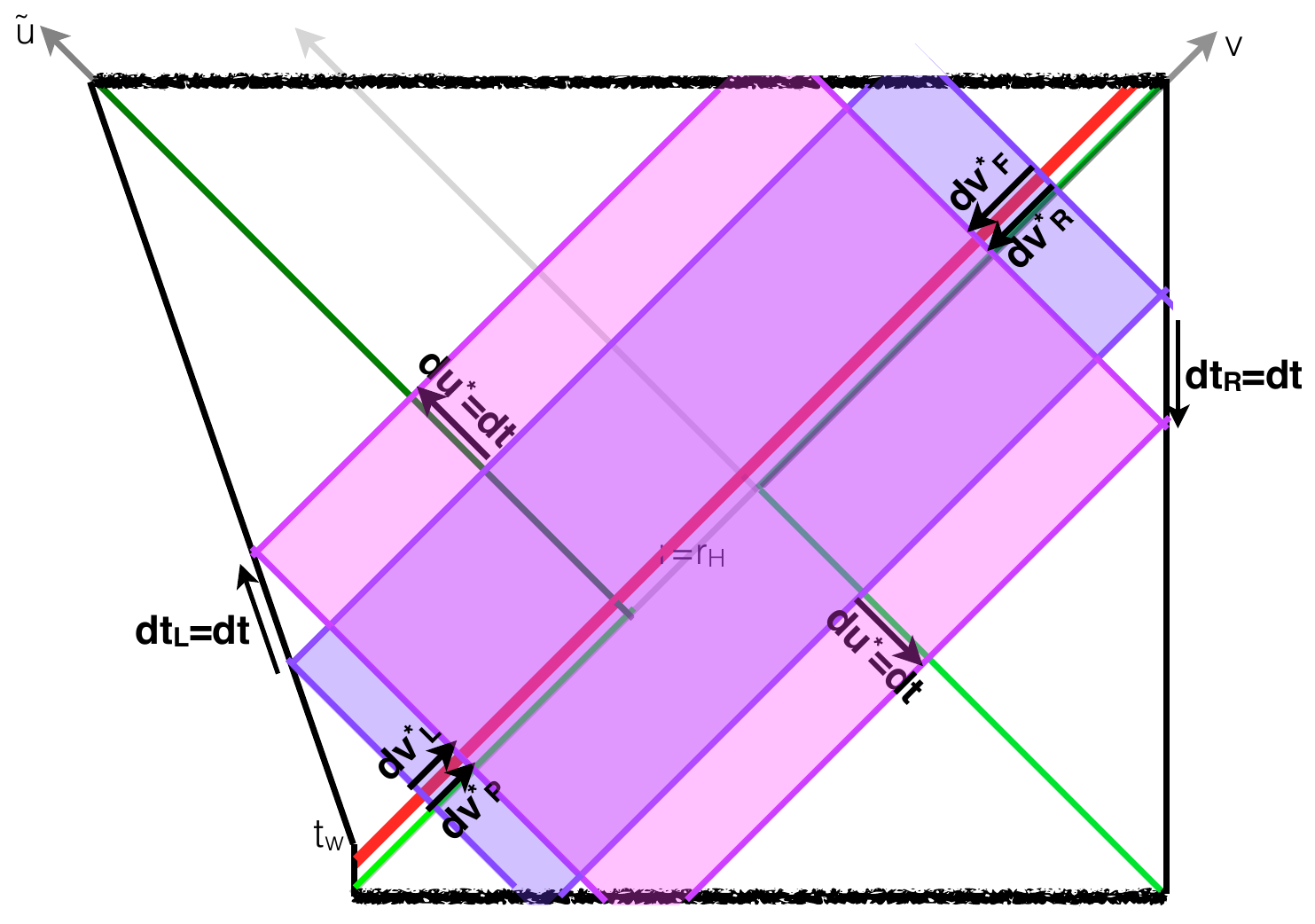}
    \caption{Change both $t_R$ and $t_L$ by $dt$.}
    \label{oneprecursorpictureb}
    \end{center}
  \end{subfigure}
  \hspace{-1.46cm}
    \begin{subfigure}[b]{0.4\textwidth}
  \begin{center}
    \includegraphics[scale=0.33]{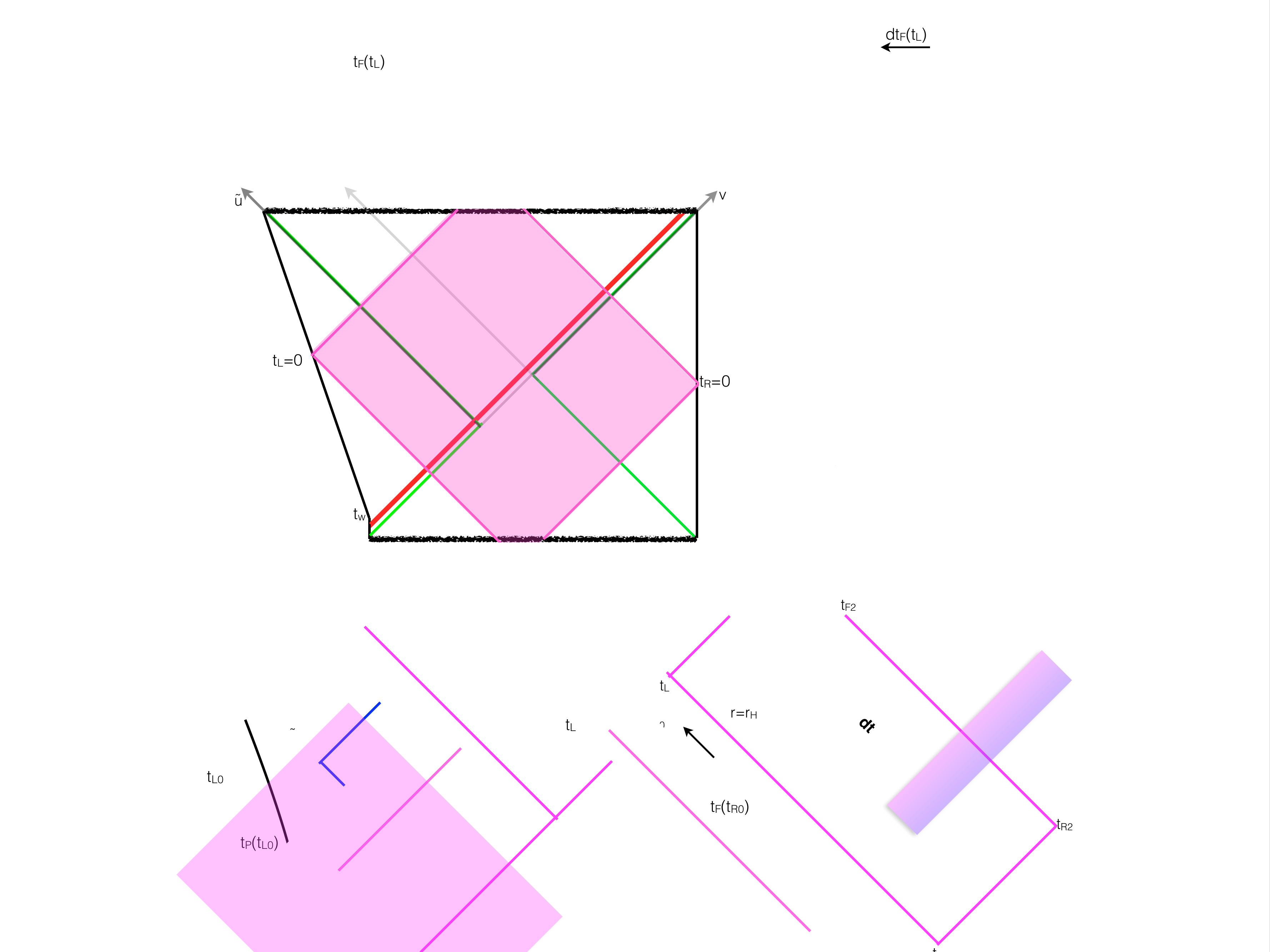}
    \caption{$t_R=t_L=0$}
    \label{oneprecursorpicturec}
    \end{center}
  \end{subfigure}
  \end{center}
  \caption{Complexity of a precursor}
  \label{oneprecursorpicture}
\end{figure}

Our goal is to calculate the complexity of the precursor $W(t_w)=e^{iHt_w}We^{-iHt_w}$. Consider the complexity of $W_L(t_w)|\text{TFD}\rangle$. Geometrically, we start from $t_R=t_L=t_w$ (Figure \ref{oneprecursorpicturea}), decrease $t_R=t_L=t$ (Figure \ref{oneprecursorpictureb}) until $t=0$ (Figure \ref{oneprecursorpicturec}). During this process the tip of WDW patch never leaves the singularity, so the symmetry argument always works. 
%
We have
\begin{align*}
\frac{d\mathcal{C}_1(t,t)}{dt}&=C\left(\frac{dv^*_F}{dv^*_R}-1+\frac{dv^*_P}{dv^*_L}-1\right)
=-2C\frac{ce^{\frac{2\pi}{\beta}(t_w-t_*-t)}}{1+ce^{\frac{2\pi}{\beta}(t_w-t_*-t)}}\\
\mathcal{C}_1(t,t)&=\mathcal{C}(|\text{TFD}\rangle)+2C\frac{\beta}{2\pi}\log\left(1+ce^{\frac{2\pi}{\beta}(t_w-t_*-t)}\right)
\end{align*}
With $t=0$, we get the complexity of a precursor:
\begin{align}
\label{oneprecursorcomplexity}
\mathcal{C}(W(t_w))=2C\frac{\beta}{2\pi}\log\left(1+ce^{\frac{2\pi}{\beta}(t_w-t_*)}\right).
\end{align}
This has exactly the same functional form as geodesic distance in the BTZ black hole \cite{Shenker:2013pqa}. 
The symmetry argument will be accurate when $t_w$ is a fraction of scrambling time (when Rindler approximation is good). Since this calculation only depends on the boost symmetry and the Rindler nature of horizons, it must reflect some universal property of horizon dynamics. It's interesting to see the exact functional from also come out of a simple epidemic model \cite{Susskind:2014jwa}.

\section{Switchback from multiple precursors}
\label{multipleperturbation}

\begin{figure}[H]  
      \includegraphics[width=2.5in]{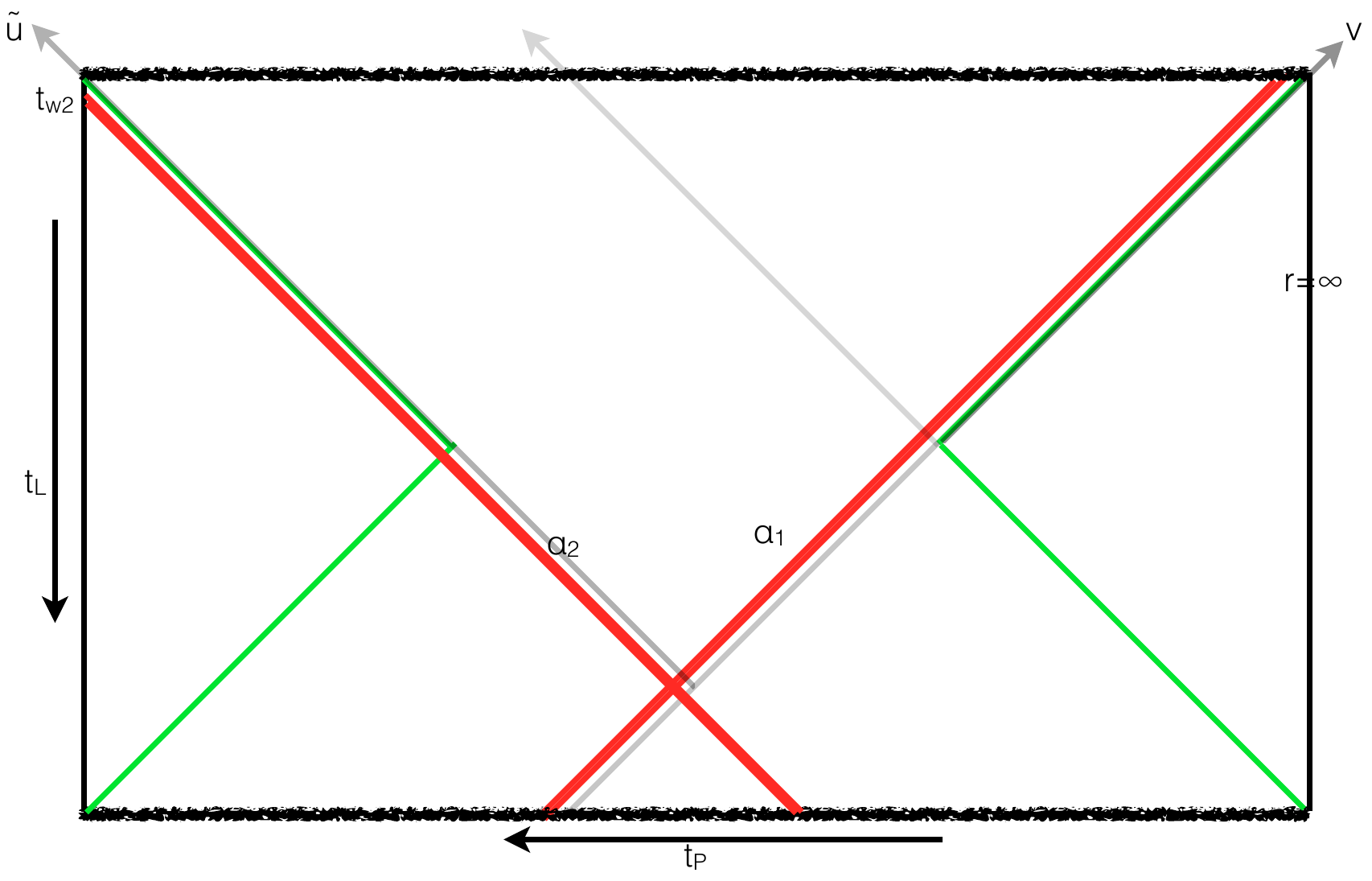}
  \caption{Thermofield double perturbed by two precursors separated by large time}
  \label{sphericaltwoshock}
\end{figure}

In section \ref{homogeneousperturbation} we studied in detail the case of one precursor, and saw that the time dependence of complexity is determined by the behavior of boost symmetry across the horizon. Let's see how this works with multiple perturbations separated by large time: $W(t_{w2})W(t_{w1})|\text{TFD}\rangle$. We assume $t_{w1}>t_*$, and $t_{w2}<-t_*$. See Figure \ref{sphericaltwoshock}. \cite{Shenker:2013yza}

\begin{figure}[H]
 \begin{center}
  \begin{subfigure}[b]{0.3\textwidth}
  \begin{center}
    \includegraphics[scale=0.27]{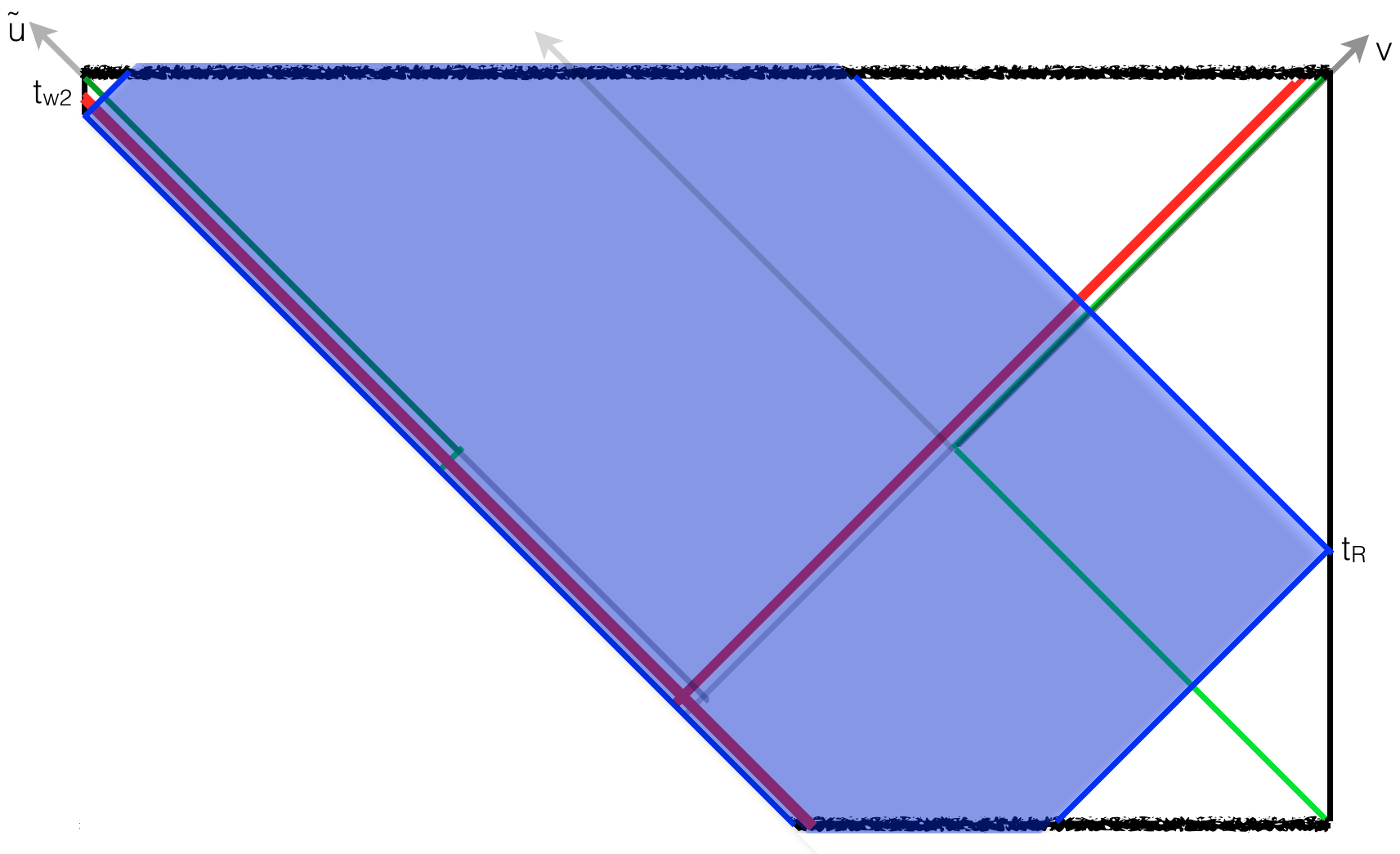}
    \caption{Right after the second perturbation}
    \label{sphericaltwoshocktime1a}
    \end{center}
  \end{subfigure}
  \hspace{-0.51cm}
  \begin{subfigure}[b]{0.4\textwidth}
  \begin{center}
    \includegraphics[scale=0.27]{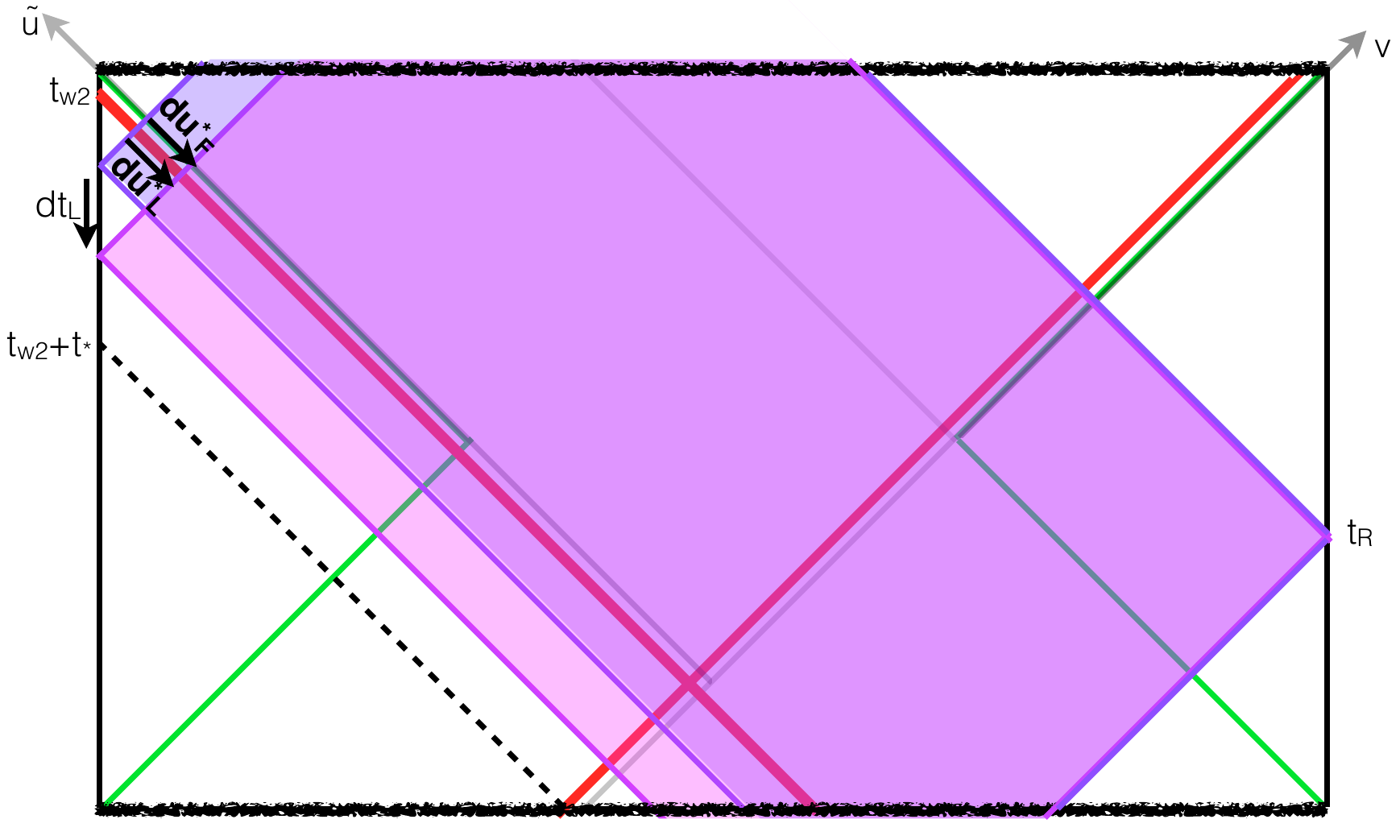}
        \caption{Decrease $t_L$}
        \vspace{0.38cm}
    \label{sphericaltwoshocktime1b}
    \end{center}
  \end{subfigure}
  \hspace{-1.59cm}
    \begin{subfigure}[b]{0.4\textwidth}
  \begin{center}
    \includegraphics[scale=0.27]{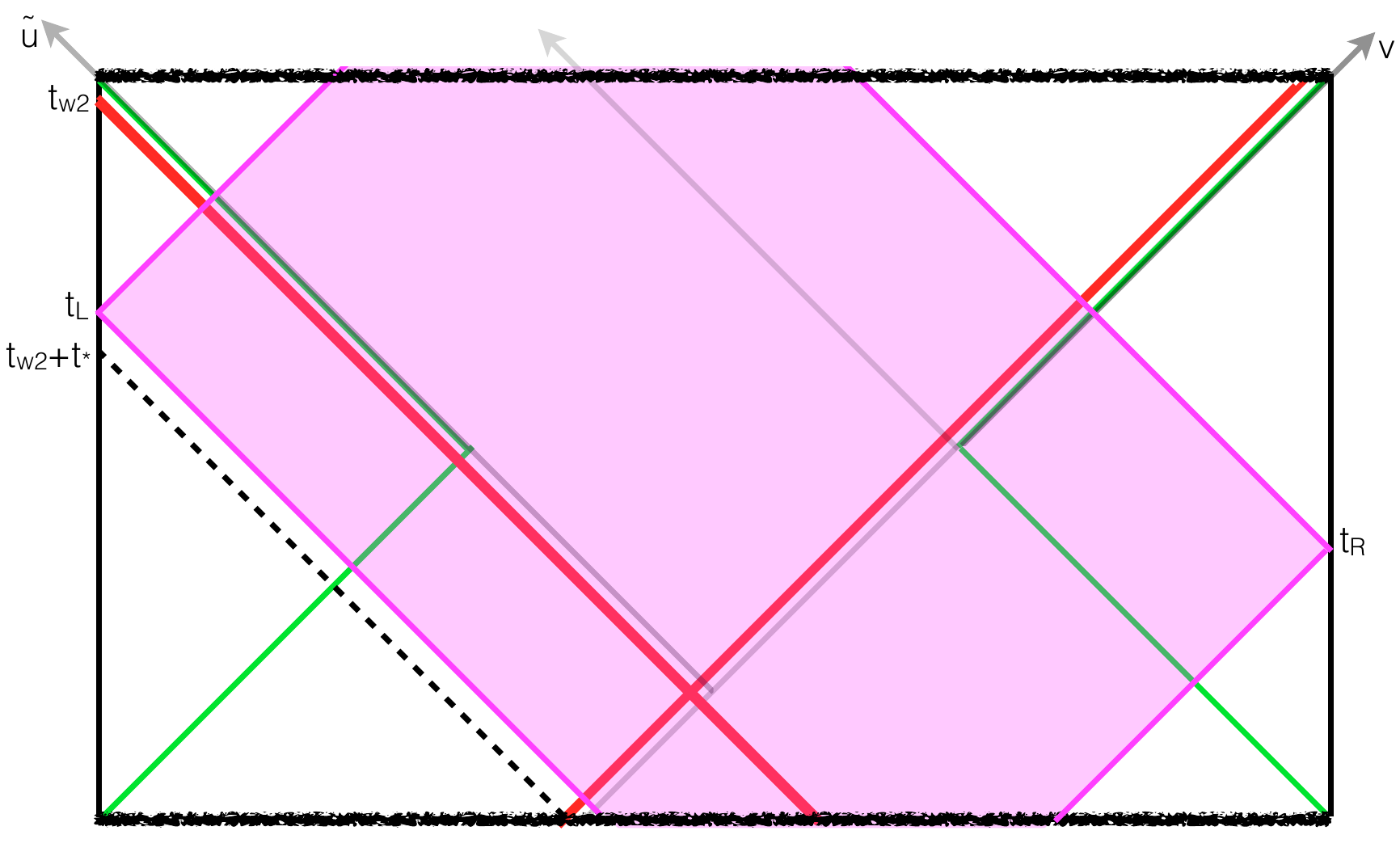}
    \caption{$t_L-t_{w2}<t_*$}
     \vspace{0.38cm}
    \label{sphericaltwoshocktime1c}
    \end{center}
  \end{subfigure}
   \end{center}
  \caption{Less than scrambling time after the second perturbation}
  \label{sphericaltwoshocktime1}
 \end{figure}

At $t_L=t_{w2}$, the effect of the second perturbation is small and we essentially have a state with one shockwave (Figure \ref{sphericaltwoshocktime1a}). We have its complexity from previous section \eqref{complexitytime1}, \eqref{complexitytime2}, and \eqref{abnormalregime}. 

Now we evolve $t_L$ downward. The black hole interior shrinks, but inside the white hole, we'll encounter a post collision region (Figure \ref{sphericaltwoshocktime1b}). The collision energy increases exponentially with $t_{w1}-t_{w2}$. With $t_{w1}-t_{w2}>2t_*$, it will be comparable to the mass of the black hole, but the size of spacetime region there is exponentially small. Here, we use assumption \ref{assumption4} to ignore its contribution.\footnote{We are aware that the high energy here could lead to complications. The circuit picture indicates that there are no significant additional contributions.} It takes scrambling time for the first shockwave to fall into the singularity. In this regime ($t_{L}-t_{w2}<t_*$),
\begin{align}
\label{twoshockcancellation}
&\frac{d\mathcal{C}_2(t_L, t_R)}{dt_L}=-C\frac{du^*_F}{du^*_L}\nonumber\\
&\mathcal{C}_2(t_L, t_R)=\mathcal{C}_1(t_L=t_{w2}, t_R)-C(t_{L}-t_{w2})
\end{align}

Notice that during this regime when $0<t_L-t_{w2}<t_*$, as we evolve $t_L$ down, the white hole interior does not expand much due to the collision, and the black hole interior shrinks because the new perturbation is still mild. This will give rise to $2t_*$ switchback. If we ever want to compare black hole geometry with quantum circuit picture, this regime corresponds to the cancellations between forward and backward time evolutions. \\

\begin{figure}[H]
 \begin{center}
  \begin{subfigure}[b]{0.3\textwidth}
  \begin{center}
    \includegraphics[scale=0.28]{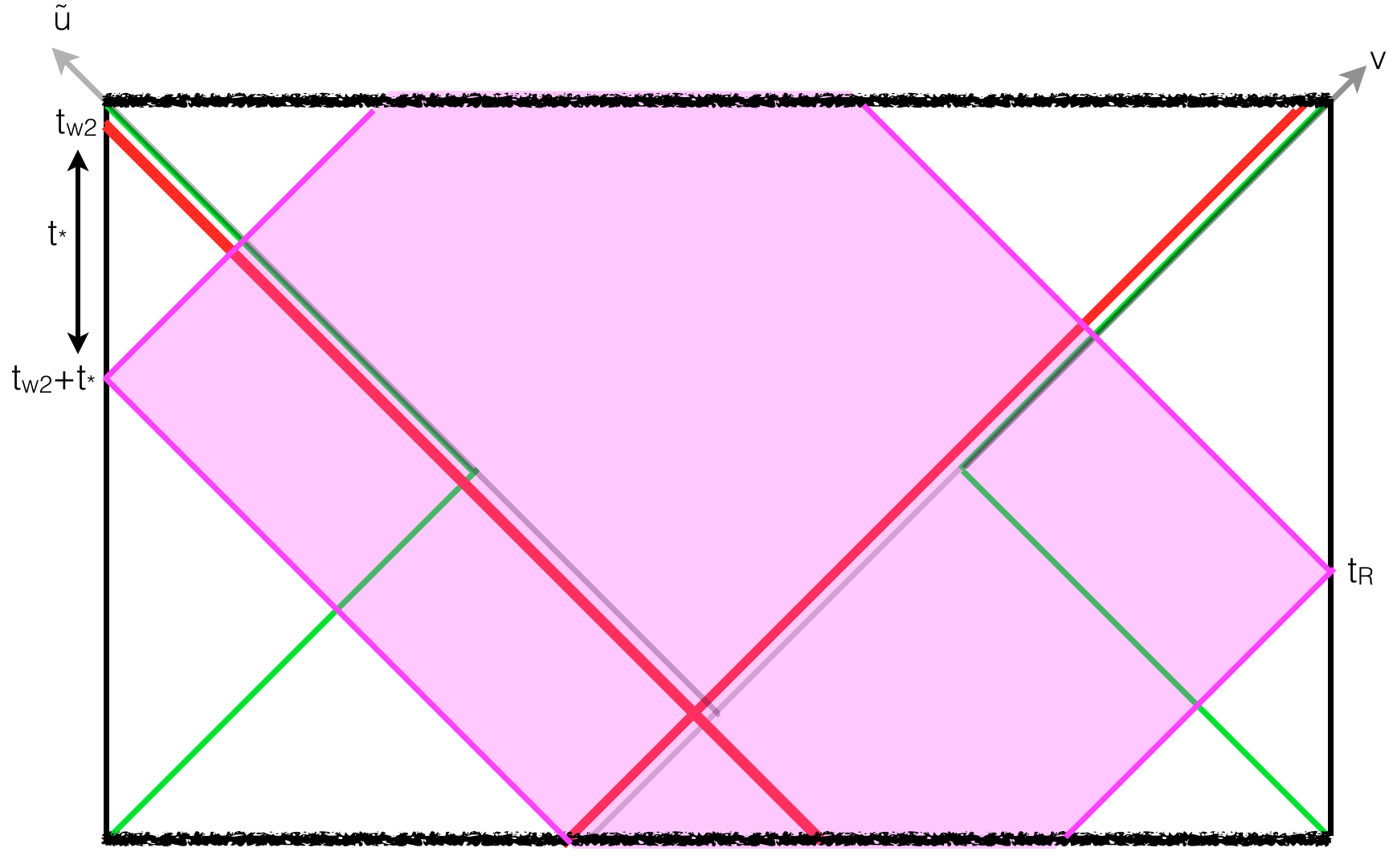}
    \caption{Scrambling time after the second perturbation}
    \label{sphericaltwoshocktime2a}
    \end{center}
  \end{subfigure}
  \hspace{-0.51cm}
  \begin{subfigure}[b]{0.4\textwidth}
  \begin{center}
    \includegraphics[scale=0.28]{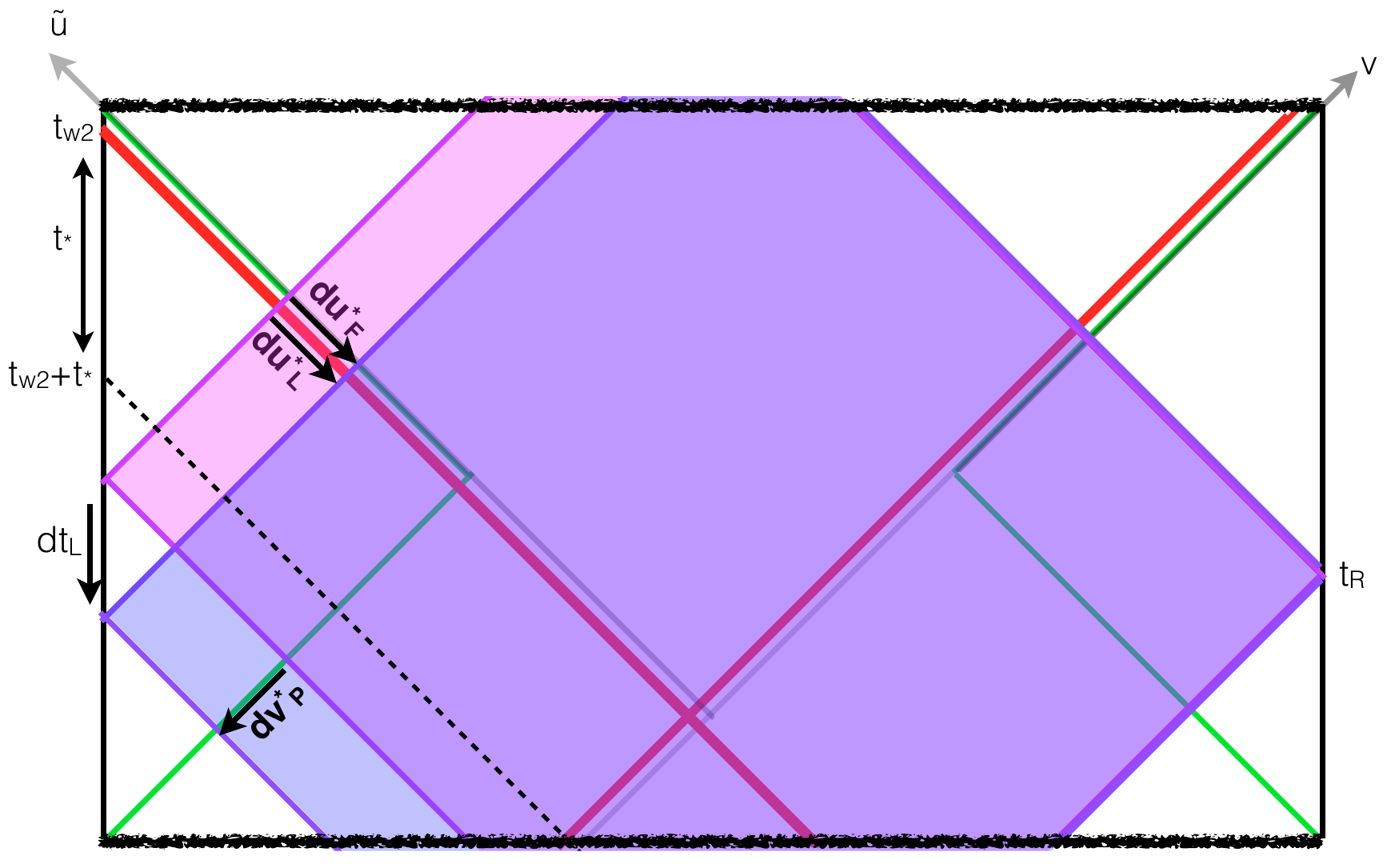}
    \caption{Further decrease $t_L$}
    \vspace{0.4cm}
    \label{sphericaltwoshocktime2b}
    \end{center}
  \end{subfigure}
  \hspace{-1.59cm}
    \begin{subfigure}[b]{0.4\textwidth}
  \begin{center}
    \includegraphics[scale=0.28]{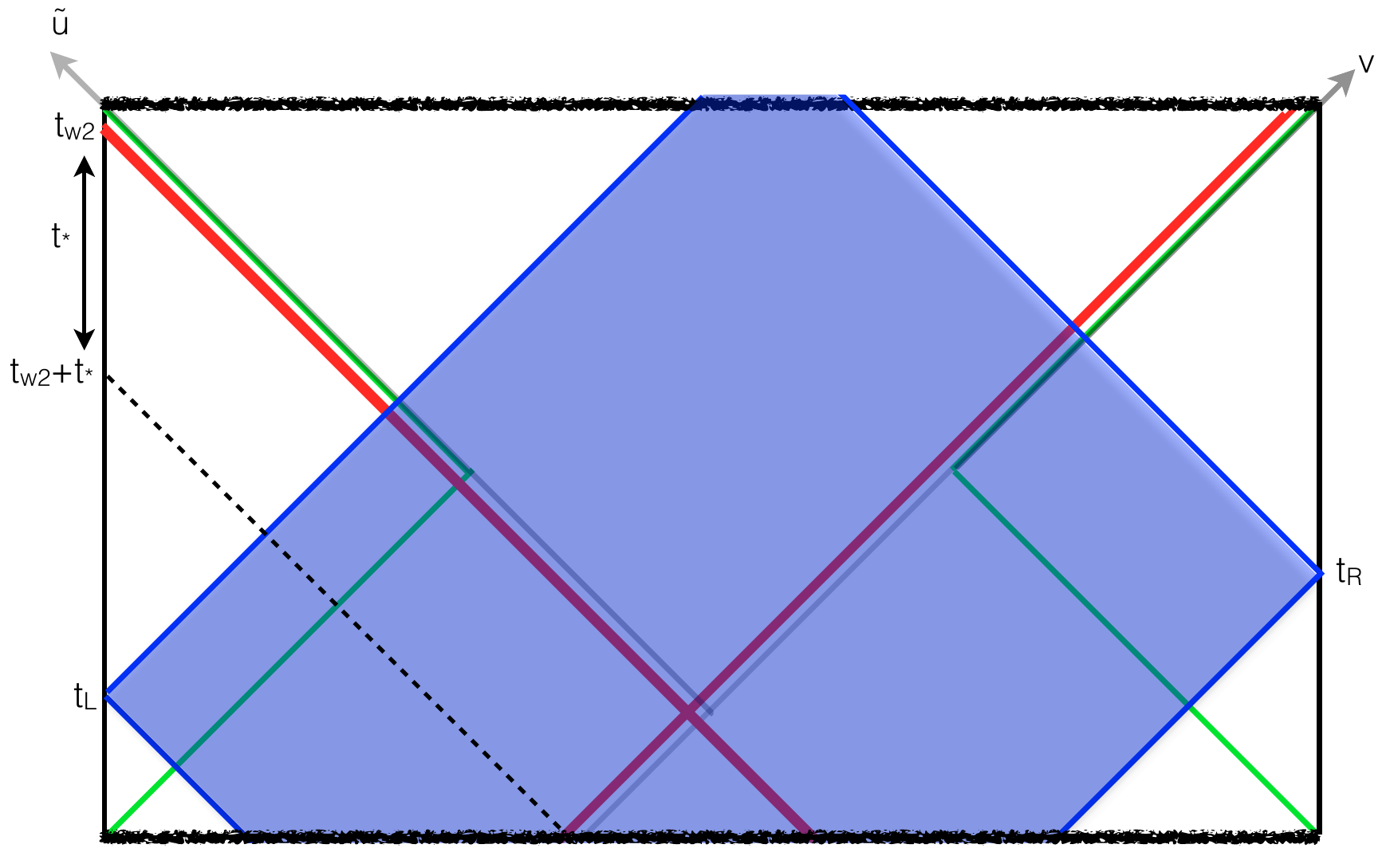}
    \caption{$t_L-t_{w2}>t_*$}
    \vspace{0.4cm}
    \label{sphericaltwoshocktime2c}
    \end{center}
  \end{subfigure}
   \end{center}
  \caption{More than scrambling time after the second perturbation}
  \label{sphericaltwoshocktime2}
 \end{figure}

As we further push $t_L$ down, once $t_L-t_{w2}>t_*$, we leave the collision region (Figure \ref{sphericaltwoshocktime2}). The white hole starts to expand, while the shrinking of the black hole is significantly delayed due to the second shockwave.

We have
\begin{align}
&\frac{d\mathcal{C}_2(t_L, t_R)}{dt_L}=C\left(1-\frac{du^*_F}{du^*_L}\right)=C\frac{ce^{\frac{2\pi}{\beta}(t_L-t_{w2}-t_*)}}{1+ce^{\frac{2\pi}{\beta}(t_L-t_{w2}-t_*)}}\nonumber\\
\label{decreasetwoshock}
&\mathcal{C}_2(t_L, t_R)=\mathcal{C}_2(t_{w2}+t_*, t_R)+C\frac{\beta}{2\pi}\log\left(1+ce^{\frac{2\pi}{\beta}(t_L-t_{w2}-t_*)}\right)\nonumber\\
&\ \ \ \ \ \ \ \ \ \ \ \ \ \ \ \ \ -C\frac{\beta}{2\pi}\log\left(1+ce^{\frac{2\pi}{\beta}(t_{w2}+t_*-t_{w2}-t_*)}\right)\nonumber\\
&\ \ \ \ \ \ \ \ \ \ \ \ =\mathcal{C}_1(t_L=t_{w2}, t_R)+C(t_L-t_{w2}-2t_*)
\end{align}

We notice that in this regime, as we push $t_L$ up, the complexity will decrease, which reflects the fact that the left future horizon is not transparent. \\

Now we look at the regime when $t_L-t_{w2}>t_*$, $t_{w1}-t_R>t_*$. Plug in $\mathcal{C}_1(t_L=t_{w2}, t_R)$ from \eqref{abnormalregime}, we have
\begin{align}
\label{multishock}
\mathcal{C}_2(t_L, t_R)=\mathcal{C}(|\text{TFD}\rangle)+C\left[2(t_{w1}-t_*)-t_R+2(-t_{w2}-t_*)+t_L\right]
\end{align}

There are two switchbacks, exactly as in \eqref{multiswitchback}. 

\section{Breaking spatial translation symmetry: Localized shocks}
\label{localperturbation}

We can also look at localized perturbations. Their holographic duals are black holes with localized shocks \cite{Roberts:2014isa}. 
Assume a planer black hole without perturbations: 
\begin{align*}
ds^2=l^2\left(-f(r)dt^2+\frac{dr^2}{f(r)}+r^2 dx^i dx^i\right)
\end{align*}
where $f(r)=r^2-r^{3-D}$.

With localized perturbation at an early time, $W_{x_w}(t_w)=e^{iHt_w}W_{x_w}e^{-iHt_w}$, null lines of constant $v^*$ and $x$ again will be pushed forward,\footnote{Contrary to the case with homogeneous perturbations, they are no longer null geodesics.} but this time in a $x$-dependent way. It was shown in \cite{Roberts:2014isa} that at large $|x-x_w|$, when crossing the horizon, the null lines will be pushed forward in Kruskal coordinates by the amount
\begin{align}
\label{shift}
h(x)=&\frac{\exp\left[\frac{2\pi}{\beta}\left(t_w-\frac{|x-x_w|}{v_B}-t_*\right)\right]}{|x-x_w|^{\frac{D-3}{2}}}\nonumber\\=&\exp\left[\frac{2\pi}{\beta}\left(t_w-\frac{|x-x_w|}{v_B}-t_*\right)-\frac{D-3}{2}\log|x-x_w|\right],
\end{align}
where $v_B=\sqrt{\frac{(D-1)}{2(D-2)}}$. 

This again will cause a mismatch between Eddington-Finkelstein time inside and outside the horizon, in a $x$-dependent way. 
If we look at \eqref{shift}, when $|x-x_w|\gg 1$, inside the exponential the linear term dominates over the logarithmic term, and the perturbation to spatial location $x$ effectly starts at time $t_w-\frac{|x-x_w|}{v_B}$, so there will be $x$-dependent time delay. If we look at a fixed value $x$, the situation is similar to the homogeneous case as discussed in earlier sections once we replace $t_w$ by $t_w-\frac{|x-x_w|}{v_B}$. 

Now let's consider the complexity of this perturbed state. We can first consider the contribution from fixed $x$ slice, which we know how it looks like from earlier discussions, then do an integration in $x$. Again, when we do this we implicitly used assumption \ref{assumption2}, i.e., locality of tensor network down to AdS scale. Here, a fixed $x$ slice has width $l_{\text{AdS}}$. So far we recovered all time dependence of complexity as discussed in section \ref{knownresults}.

\section{Collision energy and time dilation factors}

\subsection{Raychaudhuri equation}\label{Raychaudhuriequation}
We've seen that 
we can use a bundle of ingoing light rays to detect the change of Eddington-Finkelstein time across the horizon, and hence detect the time dependence of complexity. 
Raychaudhuri equation describes the behaviors of a congruence of geodesics  \cite{Hawking:1973uf}, in particular, how they respond to stress-energy tensors. 

Consider a bundle of ingoing light rays. This bundle can either expand, or shrink as they travel. To quantitatively describe this, consider a small area element $A$ in transverse directions. The expansion rate $\theta$ is defined as
\begin{align}
\label{expansionrate}
\frac{dA}{d\tau}=\theta A
\end{align}
Raychaudhuri equation gives:
\begin{align}
\label{Raychaudhuri}
\frac{d\theta}{d\tau}=-R_{ab}E_0^aE_0^b-2\sigma^2+2\omega^2-\frac{1}{D-2}\theta^2,
\end{align}
where $\omega$ reflects the rotation of neighboring light rays, $\sigma$ reflects the distortion of shapes, and $\theta$ reflects the area expansion. $E_0=\partial_{\tau}$ is the vector field along proper time. From the term $R_{ab}E_0^aE_0^b=8\pi G_N T_{ab}E_0^aE_0^b$, we see that, the presence of energy causes the bundle of light rays to focus. 

With a shell of matter present, the expansion rate $\theta$ can suddenly jump, while there will be time dilation $\frac{d\tilde v^*}{dv^*}$ across the shell. In case of homogeneous perturbations, radial light rays stay on constant transverse positions as they travel. We have
\begin{align}
\label{changeexpansionspherical}
&\frac{d\tilde v^*}{dv^*}=\frac{\theta}{\tilde\theta}=\frac{\theta}{\theta+\Delta\theta},\ \ \ 
\text{i.e.,}\ \ \ \frac{\Delta\theta}{\theta}=\left(\frac{d\tilde v^*}{dv^*}\right)^{-1}-1.
\end{align}
With localized perturbations, light rays will be refracted, and the time dilation factors will depend on transverse positions. This will give additional contributions to the expansion rate:
\begin{align}
\label{changeexpansionlocalized}
\frac{\Delta\theta}{\theta}=\left[\left(\frac{d\tilde v^*}{dv^*}\right)^{-1}-1\right]-\left(\frac{\beta v_B}{2\pi}\right)^2\nabla^2\left[\left(\frac{d\tilde v^*}{dv^*}\right)^{-1}-1\right],
\end{align}
where the Laplacian in \eqref{changeexpansionlocalized} is in transverse $x^i$ directions.\footnote{\eqref{changeexpansionspherical}, \eqref{changeexpansionlocalized} are exact in the limit of thin matter shell on the horizon. Away from horizon, $T_{ab}E_0^aE_0^b$ won't get too large and matter won't cause significant jump in expansion rate. }

On the other hand, from \eqref{expansionrate}, \eqref{Raychaudhuri}, we have: 
\begin{align}
\label{changeexpansionRaychaudhuri}
\frac{d\theta}{\theta}=-\frac{1}{D-2}\frac{d A}{A}\left[1+(D-2)\left(\frac{8\pi G_N T_{ab}E_0^aE_0^b+2\sigma^2-2\omega^2}{\theta^2}\right)\right].
\end{align}

In the limit of thin shell, $\frac{\Delta A}{A}\rightarrow 0$, and there won't be singularities in $\omega$ and $\sigma$ since they reflect rotations and separations of neighboring light lays. So the only potential contribution to the sudden jump of expansion rate comes from the stress-energy tensor term $\frac{\Delta A}{A}\frac{8\pi G_N T_{ab}E_0^aE_0^b}{\theta^2}$. This term is proportional to the collision-energy-squared. Near horizons, the large relative boost between the matter and the detecting light beam makes it grow exponentially in time. Combining with \eqref{changeexpansionspherical}, we see that with homogeneous perturbations, the collision-energy-squared with matter is inversely proportional to the time dilation factor.

With localized perturbations \cite{Roberts:2014isa}, $T_{ab}$ has localized support in transverse spatial directions: $T\propto E e^{\frac{2\pi}{\beta}t_w}\delta^{D-2}(x-x_w)$. The solution of equation \eqref{changeexpansionlocalized} shows that the time dilation factors $\frac{d\tilde v^*}{dv^*}$ can still be proportional to $Ee^{\frac{2\pi}{\beta}t_w}$, even at positions where $x\neq x_w$ and $T$ vanishes. Someone falling in there will not encounter matters, but will still be hit hard by gravitational shockwaves. In next section, we'll estimate the collision energy by evaluating the Landau-Lifshitz pseudotensors. 

This shows how the presence of stress-energy tensor affects boost symmetry across horizons. \eqref{changeexpansionspherical}, \eqref{changeexpansionlocalized} relate change of expansion rate to Eddington-Finkelstein time dilation, which controls time dependence of complexity, while Raychaudhuri equation \eqref{changeexpansionRaychaudhuri} shows that a sudden change of expansion rate is related to high collision energy. 

\subsection{Landau-Lifshitz pseudotensor}\label{Landau-LifshitzTensor}

To see the energy carried by gravitons, we look at the Landau-Lifshitz pseudotensor \cite{Landau:1982dva}:
\begin{equation}
\label{LandauLifshitz}
\begin{aligned}
t^{\mu\nu}=\frac{1}{16\pi G_N}\Big[&(g^{\mu a}g^{\nu b}-g^{\mu\nu}g^{ab})\left(2\Gamma^{c}_{ab}\Gamma^d_{cd}-\Gamma^{c}_{ad}\Gamma^d_{bc}-\Gamma^c_{ac}\Gamma^d_{bd}\right)\\
&+g^{\mu a}g^{bc}\left(\Gamma^{\nu}_{ad}\Gamma^d_{bc}+\Gamma^{\nu}_{bc}\Gamma^d_{ad}-\Gamma^{\nu}_{cd}\Gamma^d_{ab}-\Gamma^{\nu}_{ab}\Gamma^{d}_{cd}\right)\\
&+g^{\nu a}g^{bc}\left(\Gamma^{\mu}_{ad}\Gamma^d_{bc}+\Gamma^{\mu}_{bc}\Gamma^d_{ad}-\Gamma^{\mu}_{cd}\Gamma^d_{ab}-\Gamma^{\mu}_{ab}\Gamma^{d}_{cd}\right)\\
&+g^{ab}g^{cd}\left(\Gamma^{\mu}_{ac}\Gamma^{\nu}_{bd}-\Gamma^{\mu}_{ab}\Gamma^{\nu}_{cd}\right)\\
&-2\Lambda g^{\mu\nu}\Big]
\end{aligned}
\end{equation}

Landau-Lifshitz pseudotensor is only a tensor under linear coordinate transformations. To describe the experience of an infalling observer, we evaluate this tensor in the local inertial frame right before the observer crosses the horizon. 

Near horizons, assume a near-Rindler metric of the form\footnote{This metric is not exactly the Rindler metric. What we need here is an approximation of horizon metric to first order in $uv$, while in the Rindler metric, the transverse size is a constant.}:
\begin{align}
ds^2&=-\frac{4\pi}{\beta}(r-r_H)dt^2+\frac{dr^2}{\frac{4\pi}{\beta}(r-r_H)}+r^2dx^idx^i\nonumber\\
&=-\frac{\tilde c\beta r_H}{\pi} dudv+r_H^2(1-\tilde cuv)^2dx^idx^i\label{horizonmetric}
\end{align}
where $\tilde c$ is an order one constant depending on the details of the geometry. 
With localized perturbations,
\begin{align}
T_{uu}=E e^{\frac{2\pi}{\beta}t_w}\frac{\beta}{2\pi r_H^{D-2}}\delta(u)\delta^{(D-2)}(x-x_w),
\end{align}
the metric changes to
\begin{align}
\label{metricperturbed}
ds^2=-\frac{\tilde c\beta r_H}{\pi} du\left[dv-\delta(u)h(x)du\right]+r_H^2(1-\tilde cuv)^2dx^idx^i.
\end{align}
At large $x-x_w$ when the delta function perturbation is a good approximation, 
\begin{align*}
h(x)=\tilde c Er_H\frac{e^{\frac{2\pi}{\beta}\left(t_w-t_*-\frac{|x-x_w|}{v_B}\right)}}{|x-x_w|^{\frac{D-3}{2}}}
\end{align*}
where $\tilde c$ is a dimension-dependent constant. 

Kruskal coordinates $u,v$ in \eqref{horizonmetric} are not local inertial frames at $u=0, v=v_0, x=x_0$. We do coordinate transformations:
\begin{align*}
&\tilde u = u\\
&\tilde v = v-v_0-\frac{\pi r_H}{\beta}v_0(1-\tilde c u v_0)^2(x-x_0)^2\\
&\tilde x=(1-\tilde cuv_0)(x-x_0).
\end{align*}
$\tilde u, \tilde v, \tilde x$ give local initial frame right before an observer crosses the horizon at $\tilde u = 0, \tilde v = 0, \tilde x=0$. In these coordinates, the shockwave metric \eqref{metricperturbed} becomes\footnote{Note that here we ignored higher than first order terms in $\tilde u$, $\tilde x$, but we keep $\tilde u\tilde v$ term, because $\tilde v$ is not continuous across the horizon. }
\begin{align}
ds^2 = -\frac{\tilde c\beta r_H}{\pi}d\tilde u\left[d\tilde v - \delta(\tilde u)h\left(x_0+\frac{\tilde x}{1-\tilde c\tilde u v_0}\right)d\tilde u\right]+r_H^2 (1-2\tilde c\tilde u\tilde v)d\tilde x^2 +\mathcal{O}(\tilde u, \tilde x)^2.
\label{shockinertial}
\end{align}
Evaluating the Landau-Lifshitz pseudotensor \eqref{LandauLifshitz} in this frame, we see that, apart from the cosmological constant term,
\begin{align}
\label{LandauLifshitzinertial}
t_{uu} = \frac{1}{8\pi G_N}\tilde c \delta(u)h(x).
t_{\tilde u\tilde u} = \frac{1}{8\pi G_N}c \delta(\tilde u)h(x_0+\frac{\tilde x}{1-c\tilde u v_0})
\end{align}

Assume we also send in a thermal scale quantum from the right hand side. To consider the collision energy, we can look at the quantity appearing in Raychaudhuri equation $-8\pi G_N\frac{\Delta A}{A}\frac{t_{ab}E_0^aE_0^b}{\theta^2}$. 
From the definition of stress-energy tensor $T^{ab}=p^a\frac{dx^b}{d\tau}\int d\tau \frac{1}{\sqrt{-g}}\delta^{(D)}(x-x(\tau))$, one can verify that\footnote{In fact, one can already see this from the calculations of spherical perturbations \eqref{collisionenergydilation}, \eqref{changeexpansionspherical}, \eqref{changeexpansionRaychaudhuri}.}
\begin{align*}
\frac{E_c^2}{m_p^2}\sim \frac{r_H}{\beta} S^{\frac{D-4}{D-2}}8\pi G_N\left(\frac{-\Delta A}{A}\right)\frac{t_{ab}E_0^aE_0^b}{\theta^2}.
\end{align*}
On the other hand, a direct calculation using \eqref{LandauLifshitzinertial} shows that\footnote{The cosmological constant term does not contribute to this.}
\begin{align*}
-8\pi G_N\frac{\Delta A}{A}\frac{t_{ab}E_0^aE_0^b}{\theta^2}\sim\frac{h(x)}{v}=\left(\frac{d\tilde v^*}{dv^*}\right)^{-1}-1
\end{align*}
So the collision-energy-squared with gravitons is also inversely proportional to the time dilation factor.

\bibliographystyle{utphys}

\bibliography{reference}

\providecommand{\href}[2]{#2}\begingroup\raggedright\begin{thebibliography}{10}

\bibitem{Susskind:2014rva}
L.~Susskind, ``{``Computational Complexity and Black Hole Horizons"},''
  \href{http://dx.doi.org/10.1002/prop.201500092}{{\em Fortsch. Phys.}
  {\bfseries 64} (2016) 24--43},
\href{http://arxiv.org/abs/1403.5695}{{\ttfamily arXiv:1403.5695 [hep-th]}}.

\bibitem{Swingle:2009bg}
B.~Swingle, ``{Entanglement Renormalization and Holography},''
  \href{http://dx.doi.org/10.1103/PhysRevD.86.065007}{{\em Phys. Rev.}
  {\bfseries D86} (2012) 065007},
\href{http://arxiv.org/abs/0905.1317}{{\ttfamily arXiv:0905.1317
  [cond-mat.str-el]}}.

\bibitem{Swingle:2012wq}
B.~Swingle, ``{Constructing holographic spacetimes using entanglement
  renormalization},''
\href{http://arxiv.org/abs/1209.3304}{{\ttfamily arXiv:1209.3304 [hep-th]}}.

\bibitem{Hartman:2013qma}
T.~Hartman and J.~Maldacena, ``{Time Evolution of Entanglement Entropy from
  Black Hole Interiors},''
  \href{http://dx.doi.org/10.1007/JHEP05(2013)014}{{\em JHEP} {\bfseries 05}
  (2013) 014},
\href{http://arxiv.org/abs/1303.1080}{{\ttfamily arXiv:1303.1080 [hep-th]}}.

\bibitem{Roberts:2014isa}
D.~A. Roberts, D.~Stanford, and L.~Susskind, ``{``Localized shocks"},''
  \href{http://dx.doi.org/10.1007/JHEP03(2015)051}{{\em JHEP} {\bfseries 03}
  (2015) 051},
\href{http://arxiv.org/abs/1409.8180}{{\ttfamily arXiv:1409.8180 [hep-th]}}.

\bibitem{Brown:2015lvg}
A.~R. Brown, D.~A. Roberts, L.~Susskind, B.~Swingle, and Y.~Zhao,
  ``{``Complexity, action, and black holes"},''
  \href{http://dx.doi.org/10.1103/PhysRevD.93.086006}{{\em Phys. Rev.}
  {\bfseries D93} no.~8, (2016) 086006},
\href{http://arxiv.org/abs/1512.04993}{{\ttfamily arXiv:1512.04993 [hep-th]}}.

\bibitem{Stanford:2014jda}
D.~Stanford and L.~Susskind, ``{``Complexity and Shock Wave Geometries"},''
  \href{http://dx.doi.org/10.1103/PhysRevD.90.126007}{{\em Phys. Rev.}
  {\bfseries D90} no.~12, (2014) 126007},
\href{http://arxiv.org/abs/1406.2678}{{\ttfamily arXiv:1406.2678 [hep-th]}}.

\bibitem{Brown:2015bva}
A.~R. Brown, D.~A. Roberts, L.~Susskind, B.~Swingle, and Y.~Zhao,
  ``{``Holographic Complexity Equals Bulk Action?"},''
  \href{http://dx.doi.org/10.1103/PhysRevLett.116.191301}{{\em Phys. Rev.
  Lett.} {\bfseries 116} no.~19, (2016) 191301},
\href{http://arxiv.org/abs/1509.07876}{{\ttfamily arXiv:1509.07876 [hep-th]}}.

\bibitem{Susskind:2014moa}
L.~Susskind, ``{Entanglement is not enough},''
  \href{http://dx.doi.org/10.1002/prop.201500095}{{\em Fortsch. Phys.}
  {\bfseries 64} (2016) 49--71},
\href{http://arxiv.org/abs/1411.0690}{{\ttfamily arXiv:1411.0690 [hep-th]}}.

\bibitem{Susskind:2015toa}
L.~Susskind, ``{``The Typical-State Paradox: Diagnosing Horizons with
  Complexity"},'' \href{http://dx.doi.org/10.1002/prop.201500091}{{\em Fortsch.
  Phys.} {\bfseries 64} (2016) 84--91},
\href{http://arxiv.org/abs/1507.02287}{{\ttfamily arXiv:1507.02287 [hep-th]}}.

\bibitem{Zhao:2017iul}
Y.~Zhao, ``{Complexity, boost symmetry, and firewalls},''
\href{http://arxiv.org/abs/1702.03957}{{\ttfamily arXiv:1702.03957 [hep-th]}}.

\bibitem{Shenker:2013pqa}
S.~H. Shenker and D.~Stanford, ``{``Black holes and the butterfly effect"},''
  \href{http://dx.doi.org/10.1007/JHEP03(2014)067}{{\em JHEP} {\bfseries 03}
  (2014) 067},
\href{http://arxiv.org/abs/1306.0622}{{\ttfamily arXiv:1306.0622 [hep-th]}}.

\bibitem{Shenker:2013yza}
S.~H. Shenker and D.~Stanford, ``{``Multiple Shocks"},''
  \href{http://dx.doi.org/10.1007/JHEP12(2014)046}{{\em JHEP} {\bfseries 12}
  (2014) 046},
\href{http://arxiv.org/abs/1312.3296}{{\ttfamily arXiv:1312.3296 [hep-th]}}.

\bibitem{tHooft:1984kcu}
G.~'t~Hooft, ``{On the Quantum Structure of a Black Hole},''
\href{http://dx.doi.org/10.1016/0550-3213(85)90418-3}{{\em Nucl. Phys.}
  {\bfseries B256} (1985) 727--745}.

\bibitem{Maldacena:2001kr}
J.~M. Maldacena, ``{Eternal black holes in anti-de Sitter},''
  \href{http://dx.doi.org/10.1088/1126-6708/2003/04/021}{{\em JHEP} {\bfseries
  04} (2003) 021},
\href{http://arxiv.org/abs/hep-th/0106112}{{\ttfamily arXiv:hep-th/0106112
  [hep-th]}}.

\bibitem{Atia:2016sax}
Y.~Atia and D.~Aharonov, ``{Fast-forwarding of Hamiltonians and Exponentially
  Precise Measurements},''
\href{http://arxiv.org/abs/1610.09619}{{\ttfamily arXiv:1610.09619
  [quant-ph]}}.

\bibitem{Susskind:2014jwa}
L.~Susskind and Y.~Zhao, ``{``Switchbacks and the Bridge to Nowhere"},''
\href{http://arxiv.org/abs/1408.2823}{{\ttfamily arXiv:1408.2823 [hep-th]}}.

\bibitem{Brown:2016wib}
A.~R. Brown, L.~Susskind, and Y.~Zhao, ``{Quantum Complexity and Negative
  Curvature},''
\href{http://arxiv.org/abs/1608.02612}{{\ttfamily arXiv:1608.02612 [hep-th]}}.

\bibitem{Dray:1984ha}
T.~Dray and G.~'t~Hooft, ``{The Gravitational Shock Wave of a Massless
  Particle},''
\href{http://dx.doi.org/10.1016/0550-3213(85)90525-5}{{\em Nucl. Phys.}
  {\bfseries B253} (1985) 173--188}.

\bibitem{Sfetsos:1994xa}
K.~Sfetsos, ``{On gravitational shock waves in curved space-times},''
  \href{http://dx.doi.org/10.1016/0550-3213(94)00573-W}{{\em Nucl. Phys.}
  {\bfseries B436} (1995) 721--745},
\href{http://arxiv.org/abs/hep-th/9408169}{{\ttfamily arXiv:hep-th/9408169
  [hep-th]}}.

\bibitem{Maldacena:2013xja}
J.~Maldacena and L.~Susskind, ``{Cool horizons for entangled black holes},''
  \href{http://dx.doi.org/10.1002/prop.201300020}{{\em Fortsch. Phys.}
  {\bfseries 61} (2013) 781--811},
\href{http://arxiv.org/abs/1306.0533}{{\ttfamily arXiv:1306.0533 [hep-th]}}.

\bibitem{Lehner:2016vdi}
L.~Lehner, R.~C. Myers, E.~Poisson, and R.~D. Sorkin, ``{Gravitational action
  with null boundaries},''
  \href{http://dx.doi.org/10.1103/PhysRevD.94.084046}{{\em Phys. Rev.}
  {\bfseries D94} no.~8, (2016) 084046},
\href{http://arxiv.org/abs/1609.00207}{{\ttfamily arXiv:1609.00207 [hep-th]}}.

\bibitem{Landau:1982dva}
L.~D. Landau and E.~M. Lifschits, {\em {The Classical Theory of Fields}},
  vol.~Volume 2 of {\em Course of Theoretical Physics}.
\newblock Pergamon Press, Oxford,
1975.
\newblock

\bibitem{Harlow:2016vwg}
D.~Harlow, ``{The Ryu-Takayanagi Formula from Quantum Error Correction},''
\href{http://arxiv.org/abs/1607.03901}{{\ttfamily arXiv:1607.03901 [hep-th]}}.

\bibitem{Hawking:1973uf}
S.~W. Hawking and G.~F.~R. Ellis,
  \href{http://dx.doi.org/10.1017/CBO9780511524646}{{\em {The Large Scale
  Structure of Space-Time}}}.
\newblock Cambridge Monographs on Mathematical Physics. Cambridge University
  Press,
2011.
\newblock

\end{thebibliography}\endgroup

\end{document}